\documentclass[a4paper,12pt]{article}
\usepackage{graphicx}
\usepackage{lnfprep}
\usepackage[top=2.5cm, bottom=3cm, left=3cm, right=3cm]{geometry}
\newcommand{\Header}{
  \resizebox{15cm}{!}{
  \begin{tabular}{rl}
  \includegraphics[width=5cm, trim={50 100 0 0}]{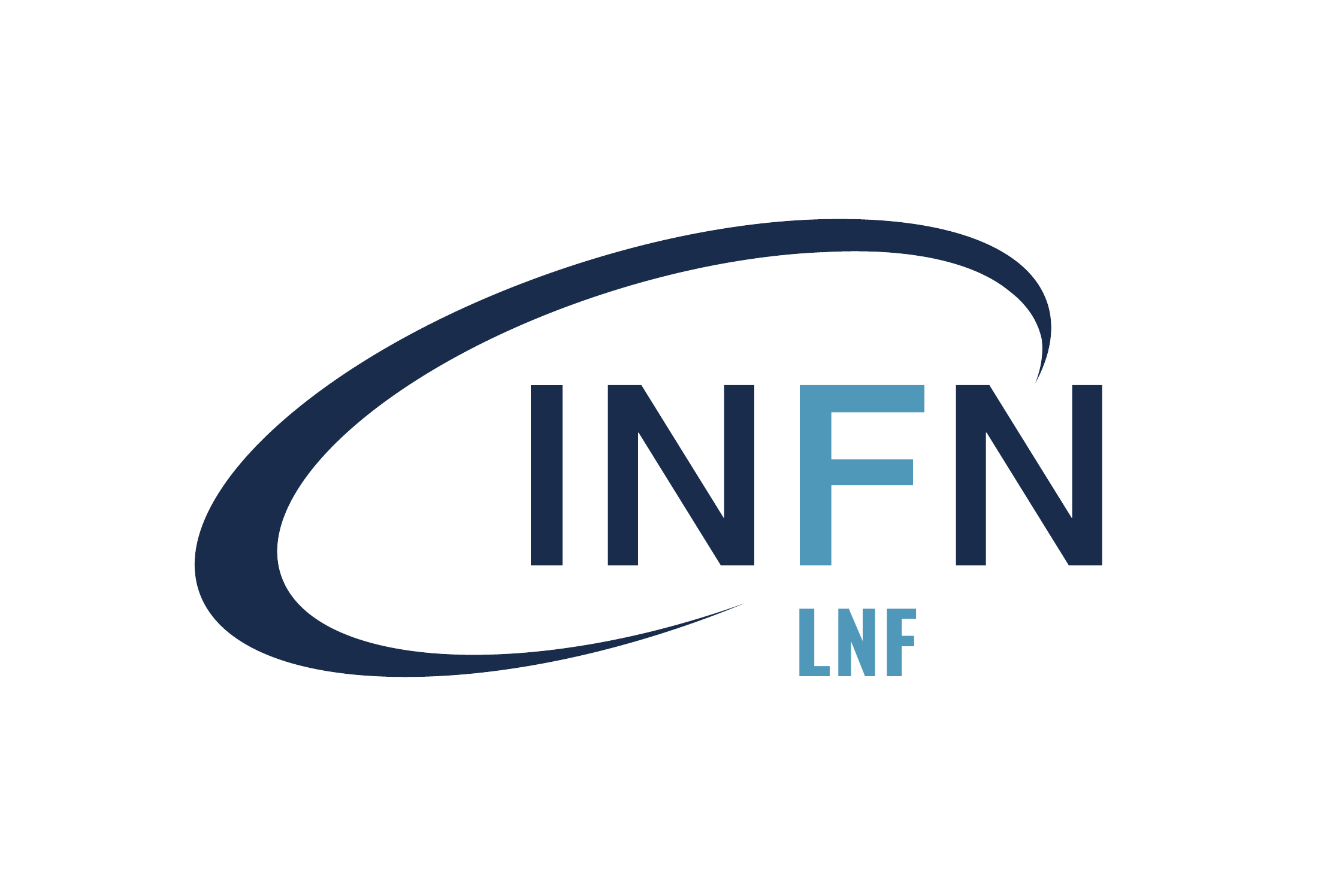} & {\LARGE\sffamily ISTITUTO NAZIONALE DI FISICA NUCLEARE}\\
      \\
  \end{tabular}
  }
\begin{center}
      {\large\sffamily Laboratori Nazionali  di Frascati}\\
\end{center}
    \renewcommand{\arraystretch}{1}
\vskip 0.5cm
\rule{15.0cm}{0.09mm}
  \begin{flushright}
      {\underline{\bf INFN-18/12/LNF}}\\    
      {\small\bf December 31, 2018} \\      
  \end{flushright}
}
\usepackage[normalem]{ulem} 
\usepackage {ulem} 

\usepackage{wrapfig}
\usepackage[english]{babel}
 \usepackage{hyperref}
\usepackage{cleveref}
\usepackage{natbib}
\bibpunct[;]{(}{)}{,}{a}{}{;}
 \usepackage{type1cm}   
 \usepackage{array} 
\hypersetup{ colorlinks=true,  linkcolor=blue,  filecolor=magenta,       urlcolor=blue, citecolor=black}
\usepackage{mathptmx}
\usepackage[intlimits]{amsmath}
\usepackage{helvet}
\usepackage{courier}
\usepackage{type1cm}  
\usepackage[font={small}]{caption}
\usepackage{makeidx}         
\usepackage{multicol}        
\usepackage[bottom]{footmisc}
\def\BT{Bruno Touschek}
\def\RW{Rolf Wider\o e}
\def\CB{Carlo Bernardini}
\def\PM{Pierre Marin}
\def\LAL{Laboratoire de l'Acc\'el\'erateur Lin\'eaire}
\def\FJ{Fr\'ed\'eric Joliot}

\def\IJC{Ir\`ene Joliot-Curie}

\def\FL{Fran\c cois Lacoste}
\def\JH{Jacques Ha\"issinski}
\def\ABL{Andr\'e Blanc-Lapierre}

\def\ENS{\'Ecole Normale Sup\'erieure}

\def\TwAO{{\it Touschek with AdA in Orsay}}
\def\WW2{World War II}
\def\JHT{Jacques Ha\"issinski's thesis}
\def\ACO{Anneau de Collision d'Orsay}

\setcitestyle{notesep={, }}
\begin{document}
\begin{titlepage}
\title
 {\Header {\Large \bf Bruno Touschek with AdA in Orsay:}
  \\{\bf The first direct observation of electron-positron collisions}
 } 

\author{ Giulia Pancheri$^1$,
   Luisa Bonolis$^2$\\
{\it ${}^{1)}$INFN, Laboratori Nazionali di Frascati, P.O. Box 13,
I-00044 Frascati, Italy}\\
{\it ${}^{2)}$Max Planck Institute for the History of Science, Boltzmannstra\ss e 22, 14195 Berlin, Germany}
} 
\maketitle
\vskip 0.7 cm
\baselineskip=14pt
\begin{abstract}
We describe how the first 
 direct observation of  electron-positron collisions took place 
   in 1963-1964  at the {\it Laboratoire de l'Acc\'el\'erateur Lin\'eaire d'Orsay},
 in France, with the storage ring AdA, which had been proposed and constructed in the Italian National Laboratories of Frascati in 1960, under the
 guidance of Bruno Touschek.  The obstacles and  successes of the two and a half years during which the feasibility of electron-positron colliders was  proved will be illustrated  using   archival  and forgotten documents, in addition to  transcripts  from
  interviews with  Carlo Bernardini, Peppino Di Giugno, Mario Fascetti,  Fran\c cois Lacoste, and Jacques Ha\"issinski.
\end{abstract}
\vskip 0.8cm
\begin{figure}[htb]
\centering
\includegraphics[scale=0.2]{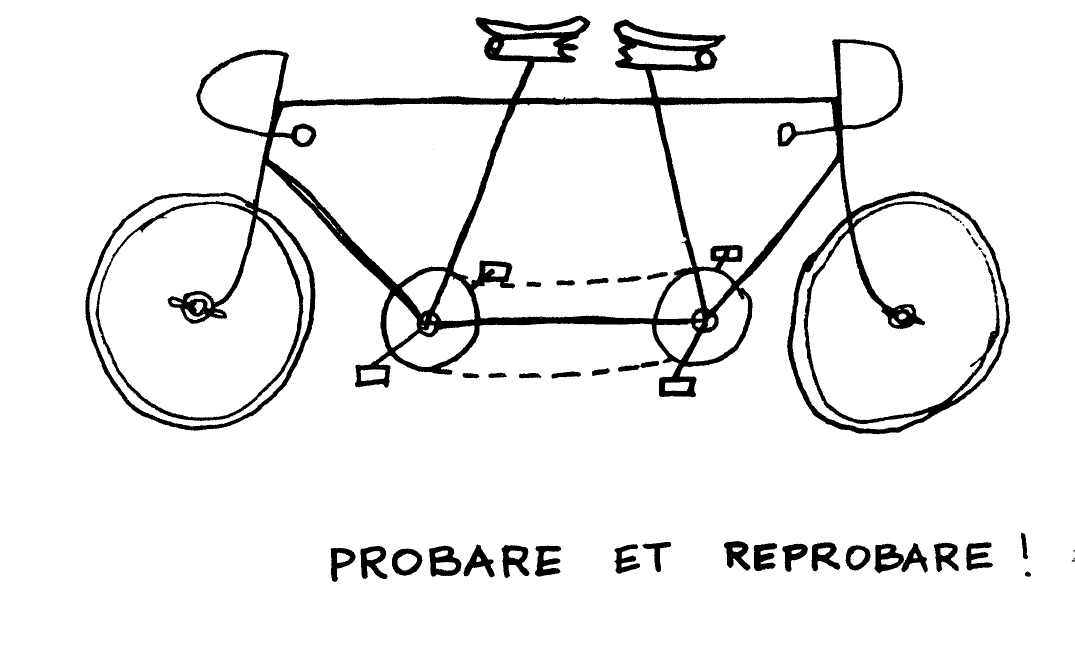}
\end{figure}
\vspace{-1cm}
\begin{center}
\rm Drawing by Bruno Touschek \citep{Amaldi:1981}.
\end{center}
\vskip 1cm
\rule{15.0cm}{0.09mm}\\
\begin{footnotesize}
 Authors' ordering in this and related works alternates to reflect that this work is part of a joint collaboration project with no principal author.
\end{footnotesize}
\end{titlepage}
\pagestyle{plain}

\newpage
\tableofcontents

\section{Introduction}
\begin{figure}[ht]
\centering
\includegraphics[scale=0.53]{
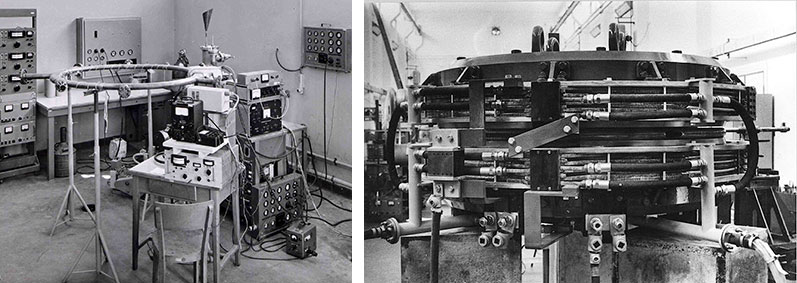}
\caption{At left, AdA's first vacuum chamber is shown  under construction in Frascati in 1960 \citep{Corazza:2008aa}. At right,  the ring AdA ready to start functioning  in January 1961.}
\label{fig:AdA}
\end{figure}

In July 1962, two trucks arrived at the {\it \LAL} \ in Orsay, South of Paris, in France. They came from Frascati, a small town near Rome, in Italy, and had crossed the Alps, in a trip which became legendary in the memory of its protagonists.    The biggest of the two trucks carried AdA, a small particle accelerator  of a completely new design and concept.  AdA was formed by an 8.5 ton iron magnet, shaped like a pancake  160 cm across   in diameter  
and 1 meter high,
which  housed a thin walled steel doughnut --- an empty  chamber, whose preliminary version under construction in Frascati is seen   in Fig.~\ref{fig:AdA} 
together with the image of AdA  in  1961, just before it started operating.
In the chamber,    electrons and positrons would be injected to be accelerated by the radio-frequency cavities, bent and focused by  magnetic fields into closed paths along opposite directions.  In due time, these particles would  be shown to collide and thus make history in particle physics.

AdA, in the first truck, was accompanied by  its support system, a set of heavy batteries needed  to power the vacuum pumps which kept the space inside the doughnut as empty of any trace of matter as it could be possible. The second truck carried other equipment needed to install AdA in Orsay, and which was  too big to be sent by plane.  AdA was arriving  from  the Frascati National Laboratories in Italy, where it  had been built and started to function in February 1961. By itself this had been a remarkable feat: when the team of Frascati scientists registered the first electrons or positrons to circulate in AdA, it had been  the first time ever that these  particles had been stored  for such a long time. 

AdA had been designed and built in Frascati by a  brilliant team of scientists and technologists, led and inspired by Bruno Touschek  \citep{Amaldi:1981}, the Austrian-born theoretical physicist, who had learnt the art of making electron accelerators from the Norwegian  \RW ~\citep{Wideroe:1994},  when  working in Germany, during \WW2, on a secret project financed by the Ministry of Aviation of the Reich, the Reichsluftfahrtministerium (RLM).\footnote{Touschek's  past experience was fundamental, both in the effect which would ultimately define his  personal reaction to later events, but also for the intimate connection with the working of an accelerator, Wider\o e's betatron, which he had helped to  build. At the end of the war, when the Allied Forces presented a report on the betatron, his contribution had been examined carefully. Between 1944 and about 1949, British-American investigations  were held on various German Science and Industrial Institutions. An on-line data base is available at \url{http://www.cdvandt.org/fiat-cios-bios.htm}, but not all the reports are included for download. These reports  subjects  cover a wide variety of  German scientific and industrial Institutions. Authored by officers from B.I.O.S. (British Intelligence Objective Sub-Committee), C.I.O.S. (Combined Intelligence Objectives Sub-Committee) and F.I.A.T (Field Information Agency Technical, United States Group Control for Germany), they also include reports where the Wider\o e's betatron is described and Touschek's contribution acknowledged. In particular, see B.I.O.S MISC.77 and  B.I.O.S 148.  In B.I.O.S MISC.77, page 6, the following is said abut Touschek's work: ``In collaboration with the design work of Wider\o e, a considerable amount of work was carried out by Touschek. This is known to have been of invaluable aid in the development of the 15 MeV accelerator. Further theoretical work has also been done by Touschek in the starting of electrons in the accelerator. Some of the work is along the lines initiated by Kerst and Serber which were known to Touschek." In the same page 6, it is also said that ``Wider\o e and the group that came to be associated with him in the war-time German betatron work were not in sympathy with the Nazi-cause, and were persuaded to continue their work for purely scientific considerations." It should also however be noted that the relevant betatron work is also mentioned in B.I.O.S. 201, \url{http://www.cdvandt.org/BIOS-201.pdf},  where it says  ``the M\"uller factory had built a 15 million volt betatron in conjunction with a Norwegian scientist" and that   "Dr. Fehr stated that the project had been experimented for  the Luftwaffe with the hope of obtaining a  death-ray  for anti-aircraft work." A similar statement appears in B.I.O.S., Final Report N. 148, Item No.1, ``German Betatrons", dated 24.10.1945, on page 8. In the same report, at page 11, it is also recommended that ``Mr. Touschek  [\dots]  be taken to U.K. for theoretical work."} 

\begin{figure}[ht]
\centering
\includegraphics[scale=0.55]{
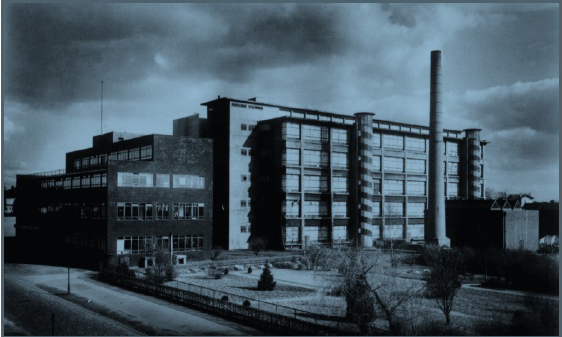}
\caption{The X-ray tube factory C.H.F. M\"uller  in Hamburg, where \RW\  with Bruno Touschek, Rudolf Kollath and Gerhard Schumann built the 15-MeV German betatron commissioned by the RLM.} 
\label{fig:mueller}
\end{figure}

In Fig.~\ref{fig:mueller} we show the C.H.F. M\"uller factory building in Hamburg, where  Wider\o e's  betatron had been constructed and had started to function at the end of 1944 \citep{KollathSchumann:1947}.
  
\begin{wrapfigure}{R}{0.5\textwidth}
  \centering
    \includegraphics[width=0.45\textwidth]{
    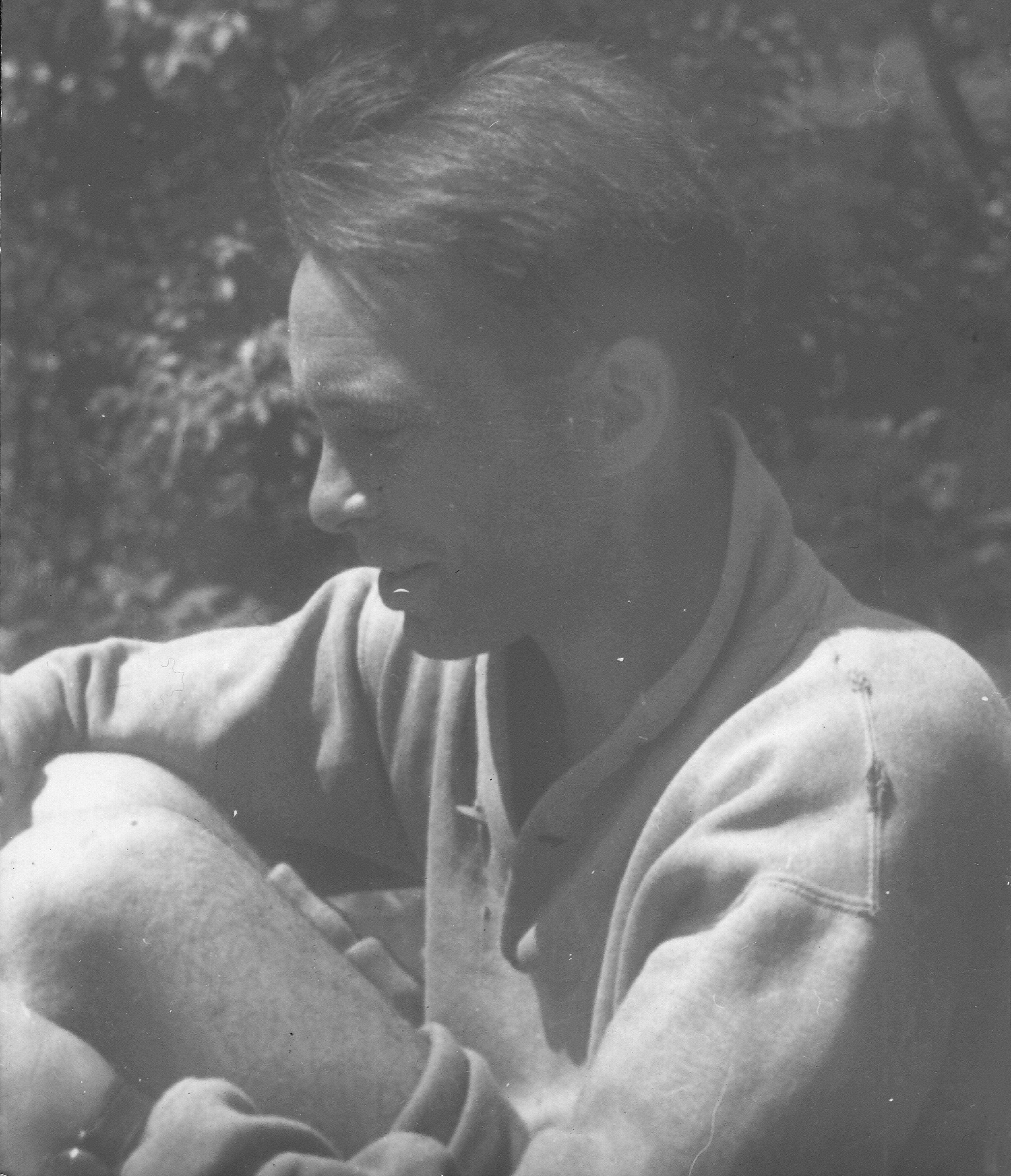}
  \caption{Bruno Touschek on the shore of the  Lake Albano, a small volcanic crater lake in the Alban Hills near Rome, where he loved to go fishing and swimming. Courtesy of Elspeth Yonge Touschek.}
  \label{fig:touschekalbano52}
\end{wrapfigure}

  After the war, Touschek completed his degree in Physics at the University of G\"ottingen  with a dissertation on the theory of the betatron \citep{Bonolis:2019qqh} and in early 1947 he moved to the University of Glasgow, where he obtained his 
  doctorate and began his career as theoretical physicist, also contributing to the development of the first electron synchrotron in Europe, a project for a 350 MeV machine led by Philip Dee. By the time he moved to Rome, in 1952, ``he was arguably the world leader on all aspects of circular electron accelerators.''\footnote{\href{http://www.worldchanging.glasgow.ac.uk/article/?id=61}{World Changing Project. ``Inventing the storage ring for high-energy elementary particles''}, University of Glasgow, 2010.}   In Fig.~\ref{fig:touschekalbano52} we show Bruno Touschek a short time after his arrival in Rome. 

  
The construction of Ada in Frascati in 1960 benefited from the enthusiasm and capabilities provided by a new Laboratory,  which  had been built in mid 1950s as part of the post-war reconstruction of Italian physics, and  housed an electron synchrotron, a first rate particle accelerator.\footnote{
See the special issue of \textit{Il Nuovo Cimento} dedicated to a complete overview of the Frascati electron synchrotron, also including notes on the history of the project \citep{Bernardini_Gilberto:1962ab}.} But the final goal of AdA was beyond the still remarkable technological and scientific know-how of a single national laboratory in Europe. This goal was  to make electrons and positrons collide in the same ring with sufficient intensity to prove the feasibility of this type of accelerators as powerful tools  in  reaching  deeper and deeper into the world of elementary particles. 
{A machine such as AdA had never been built before and its success was far from being assured.}

What transformed a partial success into a world first was the joining of forces with another national laboratory, which had also been built after the war, in France. 
The {\it \LAL  \ d'Orsay}, where AdA arrived in July 1962,   had been founded 
{five years earlier}, not unlike the  Frascati Laboratories in Italy, which AdA had just left,  with {\it tout son attirail} \citep{Marin:2009}, namely  its travelling outfit of working pumps, batteries, oscillograph, movie camera, and more. 
Both laboratories 
 had come to life 
 after the war, as  the decision to found CERN was taken and, with it, the need to train new generations of European scientists in accelerator science arose. It also sprang
   from  the dreams
  of three great pre-war 
  scientists, Enrico Fermi  on one side of the Alps,
   \  \FJ,\ and \IJC\ on the other \citep{Pinault:2000aa,Bimbot:2007aa}. 
   They shared a vision, one of open spaces for students and researchers,  to lodge new equipments, and overcome the strictures of University settings.
 
 In 1962, in Orsay, the parallel roads followed by France and Italy in the development of particle accelerators, met  when AdA was downloaded from the truck and placed in Salle 500, next to the Linear accelerator.  
 The Italian road had its origin  from Enrico Fermi and his pre-war dream of a national laboratory for nuclear physics.\footnote{For an historical overview on Fermi's frustrated plans to have an accelerator and an institute for nuclear physics in Rome see \citep{Battimelli:1997aa}.}  After the war, when Fermi came to Italy for the last time in 1954, shortly before his untimely death,  the decision to build a national laboratory was taking shape, especially thanks to the joint efforts of Edoardo Amaldi and Gilberto Bernardini, who were the main driving forces in the reconstruction of physics in Italy, also playing a leading role in the birth and development of CERN.  Fermi's dream came true in 1959 when a 1100 MeV  electron synchrotron  started operating in the new  laboratories built near Frascati, an ancient town south-east of Rome,  built on the slopes of the   hills of volcanic origin, which   look down to the city spreading wide along  the Tiber and towards the Tyrrhenian Sea.\footnote{Fermi did not see  his dream  come to life, as he died in Chicago on November 28th  1954, soon after returning from his last visit to Italy, during which he also attended the 2nd Summer School dedicated to elementary particles and accelerators held in Varenna, on the Como Lake, and where he had the opportunity to see his plans for reconstruction of Italian and European science getting on their  way.}
 The Frascati laboratory was officially born in 1957, 
{when offices and buildings housing the synchrotron were ready, an official address was assigned  as {\it Via del Sincrotrone 12, Frascati} and a team of scientists, engineers and technicians arrived from Pisa, where part of the equipment for the synchrotron had been under construction since 1955 \citep{Alberigi:1959}, under the direction of Giorgio Salvini.\footnote{For a recent history of the Frascati Laboratories, see Vincenzo Valente, {\it La strada del sincrotrone km 12. Cinquant'anni di acceleratori e particelle nei laboratori di Frascati}, Imprimenda - INFN 2007, Limena (PD).}}  

  In Orsay, a dream had also come true in the post-war years: the dream of  \FJ\ and \IJC, who had supported the creation of a large  research laboratory outside Paris.  After the war, driven by their vision \citep{Bimbot:2007aa},    the Institut de Physique Nucl\'eaire (IPN) had been created,  in the old municipality of Orsay,  in the Vall\'ee de Chevreuse,  
  which threads its way along the  Yvette river,   next to the 
   Saclay plateau,   south of Paris. 
  In 1942, Fr\'ed\'eric Joliot wrote a letter to the Recteur
de l'Acad\'emie de Paris suggesting to buy a large piece of land
South of Paris, in view of a future extension of
the University of Paris,
not too far from the railway station of Gif-sur-Yvette. The piece
of land that Joliot was thinking of was the one where the CEA is 
located today. 
After the 2nd World War, both Fr\'ed\'eric and \IJC\  were eager 
to extend the Facult\'e des Sciences de l'Universit\'e de Paris, but
Fr\'ed\'eric Joliot was very much involved in the setting up of the
CEA in the vicinity of Saclay, while Ir\`ene Joliot-Curie wanted to set up
an ``Institut de physique nucl\'eaire et radioactivit\'e". 
 
The events leading to the choice of the Orsay site and the birth there of the IPN, were recollected by \FJ  \  in a conference   at the 8th Lindau Nobel Laureate Meeting dedicated to Chemistry, on 1st July 1958, 
   just two months before his death.\footnote{The text of the Lindau conference {\it  Le nouveau Centre de
recherches fondamentales en physique nucl\'eaire d'Orsay et la formation des
chercheurs} was  used in part by Joliot at the inauguration of the
International Conference of Nuclear Physics held in July in Paris and later
published in the journal \textit{L'\^Age Nucl\'eaire}, 11, Juillet-A\^out 1958, p.
183 \citep[154]{Pinault:2000aa}. This version was also reproduced in  \FJ's
collected works, a copy of which was kindly provided by Jacques Ha\"issinski.}
  In his recollections, \FJ\ recalls the r\^ole  he played  together with  his wife, \IJC, after the war, in  establishing    the development of a new institution in an adequate  location  in the surroundings south of Paris:

   \begin{quote}
   \begin{small}
[\dots] en 1950, [les] laboratoires [existants] \'etaient surchag\'es par l'afflux des \'etudiants et chercheurs attir\'es vers les sciences nucl\'eaires [\dots]  il fallait envisager   d'importantes constructions nouvelles dans la banlieue parisienne et les \'equiper des g\'en\'erateurs modernes et puissants [\dots] \IJC\ proposa \`a la Facult\'e des Sciences [de Paris] un project tr\`es \'elabor\'e, d'acquisition d'\'equipment  dont un synchrocyclotron  de 150 MeV (protons) et de construction de locaux dans la r\'egion parisienne [\dots] la Facult\'e fit l'acquisition d'un vaste terrain de plus de 100 hectares, \`a Orsay au sud de Paris, dont une partie fut destin\'ee \`a la construction du laboratoire de physique nucl\'eaire [\dots] \\ En juillet 1955 fut d\'ecid\'e la construction d'une premi\`ere tranche des b\^atiments \`a Orsay. Les avantage de ce lieu sont nombreux. Il se trouve \`a proximit\'e du Centre d'\'Etudes  nucl\'eaires du Commissariat \`a l'\'Energie Atomique (C.E.A.) \'a Saclay et du Centre de Recherches du C.N.R.S. \` a Gif-sur-Yvette, centres \`a la creation desquel j'avais pu contribuer en 1945 et en 1948 [\dots] \\Il fallut environ six mois pour que les b\^atiments commencent \`a peine \`a \'emerger du sol. \IJC, qui avait tant donn\'e des ses forces pour la r\'ealisation  de cette oeuvre, ne put h\'elas!  avoir la joie d'assister \`a cette naissance [\dots] \\Un an plus tard, en ao\^ut 1957, la construction de la premi\`ere tranche des b\^atiments etait achev\'ee et une partie importante du gros et du petit \'equipment \'etait en place.\footnote{[\dots] In 1950, the [existing] laboratories were overcrowded by students and researchers interested in nuclear sciences [\dots] it was [thus] necessary to envisage new constructions  in the suburbs out of Paris and to provide them with modern and powerful power generators [\dots] \IJC \ proposed to the Faculty of Sciences [of Paris University] a very elaborate plan for the acquisition of [new] equipment, including   a synchrocyclotron of 150 MeV (for protons), as well as the building of new office spaces within the Paris region, and [\dots] the Faculty bought a large piece of land of more than 100 hectares, in Orsay, south of Paris, part  of which was to be devoted to the construction of a laboratory for nuclear physics [\dots] In July 1955, the decision to build the first  buildings in Orsay was taken.   This location has many advantages. It is close to the Center for Nuclear Studies of the  Atomic Energy Commission (C.E.A.) in Saclay, as well as to Centre de Recherches of the C.N.R.S. in Gif-sur-Yvette, centers  whose   I had been  able to contribute to  in 1945 and 1948 [\dots] It took about six months for these buildings to  come up just above ground. Unfortunately, \IJC, who had given so much of her forces for the creation of this  project, did not have the joy to see  its birth [\dots] One year later, in August 1957, the construction of the first buildings was completed and an important portion  of the big and small equipment was in place. }
   \end{small}
   \end{quote}
\noindent
  In 1954,
the decision was then taken to build both ``un Synchrotron 
\`a protons" and ``un acc\'el\'erateur lin\'eaire d'\'electrons". The first machine was to 
be installed in Ir\`ene's Institute and its construction began 
first, while the Linac was to be installed in the future LAL. When the decision to build a linear accelerator was taken by Yves Rocard, Directeur du Laboratoire de physique de l'\ENS \ 
 de la rue d'Ulm,   in Paris, the question of  an appropriate location was debated and the decision was taken to build it  in Orsay, near the IPN.  As Pierre Marin says \citep[33]{Marin:2009},  ``Le site: le b\^{a}timents en fond de vall\'ee!''\footnote{The place: the buildings down in  the valley!}
 

In  what follows, we shall tell what happened  during the two and a half years of  experimentation in Orsay with AdA by the team of Italian and French scientists, their  hopes and disappointments, and the final success.   In this note we present 
{an anticipation of} the Orsay chapter of a larger project which will follow the development of electron-positron colliders in Europe as the joining of   different roads which, before and through \WW2,  led to  the birth in Europe of electron-positron colliders, which played a key role in developing  and completing the experimental foundation of the Standard Model of elementary particles.    Bruno Touschek was the main mover for the early development of these machines in Europe. Sadly, he died   young, in 1978, being then only 57 years old, and could not see these developments. The aim of our project is also to rekindle the interest in his genius and that of the colleagues who followed his lead  in the AdA project. 

\subsection{
{Sources and outline}}
In \citep{Bonolis:2011wa}, and more recently  with further detail \citep{Bonolis:2018gpn}, we have described the making of the Franco-Italian collaboration and the transfer of AdA from Frascati to Orsay. In the present  note, and in particular in Sec.~\ref{sec:touschekeffect},   we shall see how the team was able, against all  odds, to prove that such particle accelerators, in which  electrons and positrons   collide, could actually be made to work.  We shall thus  see how Touschek's   understanding of the internal dynamics of a bunch of stored electrons or positrons  circulating in AdA, would be the turning point in  being the  first    {\it ``in realizing  the  impossible and   thinking the unthinkable''}  \citep[57]{Rubbia:2004aa}.

{Concerning the two and a half years  during  which AdA was in Orsay,  there exist  extensive written documentations 
 by the protagonists of the AdA adventure. In addition to the published articles of the AdA collaboration, namely  \citep{Bernardini:1963sc,Bernardini:1964lqa}, we shall rely upon a number of individual contributions and  oral history interviews, which will be highlighted in the course of our narration. One  primary source  is    \BT's    contribution to  a  
  \href{https://lss.fnal.gov/conf/C630610/}{Summer Study on Storage Rings, Accelerators and Experimentation at Super-High Energies} held in June-July 1963 in Upton, NY. Another is   \JH's {\it Th\`ese d'\'Etat}  submitted in 1965 at the end of the experimentation, when AdA had returned to Italy.}\footnote{Excerpts from the thesis, in its original version, will be published   in a  work in progress.   \JH's memories of  AdA's Orsay  period 
  collected 
   in \citep{Haissinski:1998aa}    add to the knowledge of the important milestones  reached by the AdA team in the understanding of how  particle colliders operate,  placing them in 
   the perspective of more than two decades of colliders built after Touschek's death. }

 

 {Another  protagonist of the AdA adventure was \PM,  the developer  and moving force of France's modern accelerators. We owe him  a beautiful description of   the story of AdA in Orsay \citep{Marin:2009}. Published  posthumously  in 2009,  this book is entitled {\it  Un demi-si\`ecle 
d'acc\'el\'erateurs de particules} and   places AdA in the full context of  the development of colliders. \PM \  had barely completed the manuscript,   when he was struck down by a heart attack in 2002. As \JH \ says in his introduction to the book, Marin's intention   had been to tell the passions and the hopes which surrounded the installation and experimentation with AdA. Indeed, in the chapter he dedicates to AdA, we learn how AdA's adventure was a run against time,  a foray into unknown territory, where everything had still to be discovered.}

To these testimonials from the French side, one adds  contributions  by  Carlo Bernardini, Touschek's friend and close collaborator  from the earliest days of AdA's proposal,  
in whose work  one can find  detailed personal memories of both the AdA construction and of the Orsay period,  in addition to the scientific descriptions.
Recent among them are \citep{Bernardini:2004aa}, where many details of AdA's early times in Frascati and the Orsay adventure are  illustrated with attention to  non professional readers, and     \citep{bernardini2006fisica}, where  the Orsay  period is sketched with  all the pathos of a scientist who,  in his youth,  participated in  one of the great adventures of particle physics. \CB, who unfortunately passed away a short time ago, was also an exceptional story teller and has  contributed  many    anecdotes to  AdA's  and  Touschek's story, as in \citep{Bernardini:2015wja}.\footnote{See also the docu-films  \href{https://youtu.be/R2YOjnUGaNY}{\it Bruno Touschek and the art of physics} by E. Agapito and L. Bonolis (Italian version INFN 2003, English version INFN 2005) and \href{http://www.lnf.infn.it/edu/materiale/video/AdA_in_Orsay.mp4}{\it Touschek with AdA in Orsay}, by E. Agapito, L. Bonolis and G. Pancheri,  INFN 2013.}

{ We shall start this paper by  presenting in Section \ref{sec:prequel}  a prequel to the Orsay  events, with our view of    what  took place between  the Conferences of summer   1959, in Kiev  and at CERN,  and  the seminar held on March 7th, 1960 in the Frascati National Laboratories, in Italy, where Touschek's detailed proposal to construct AdA  was presented 
 and immediately approved by the  scientific board} \citep{Amaldi:1981}. To this approval, there followed  the construction of AdA through 1960
and, in December,
  the  proposal to build a bigger and more powerful machine, to be called ADONE, 
  a pun on AdA's name invented by Touschek.\footnote{ In Italian, ADONE can  mean a bigger AdA, but is also the name of Adonis, the lover of the Greek goddess Aphrodite, and a synonym for a handsome man.}
  
  When AdA's  magnet was first turned on in  February 1961,  hopes for an early success arose.  We shall then  recall how, after the  initial elation,  there followed discouragement, as  the overall set up, including the injector, failed to perform  as planned, and how hopes were revived by the unexpected visit by two French scientists. A summary of the events which passed between March 1960 and  
July 1962,  when   AdA was taken to France as detailed  in  \citep{Bonolis:2018gpn}, will be given. 

In  Section \ref{sec:inorsay}, {we shall  present, in sequence, first the installation of AdA in Orsay and then the first experiments, which took place between September and December 1962. Once more, enthusiasm was followed by disappointment, when AdA's beam  life time  was seen not to conform to expectations.}  

In Section \ref{sec:touschekeffect} we shall describe  the  discovery of the Touschek effect, 
{which carried   the bitter acknowledgement of AdA's limits in proving annihilations into new particles,}  and  the reaction to the news on the part of the (still small) world of collider physicists, namely the  American physicists  in   Princeton and Stanford, and the Russian team of Gersh Budker  in Novosibirsk.\footnote{
{ Both the Americans and the Russians had been   building  electron-electron colliding machines, and, in the case of the Russians, had also  started building   an  electron-positron storage ring, as shall be seen later.}}

How Touschek and the Franco-Italian team were able to overcome the bitter discovery that AdA could not prove that electrons and positrons were annihilating into new particles
at a sizable rate, will be the subject of Sections \ref{sec:summer1963} and  \ref{sec:collisions}. The contribution of   two young theoretical physicists from  the University of Rome, under the guidance of both Bruno Touschek and Raoul Gatto, will be outlined here for the first time. 

{In Section \ref{sec:ironcurtain} we shall also comment on the progress which was taking place in Novosibirsk,  how the AdA team learnt of it and their reaction.}

In  Section \ref{sec:afterAdA}   we shall see 
what followed AdA's success in both France and Italy
and how the world of particle physicists and accelerator builders reacted to the news of AdA's successful  experimentation, and give a brief overview of the projects which were put in motion after laboratories around the world learnt of AdA's success. 

\section{Prequel}
\label{sec:prequel}
The story of AdA officially starts   on February 17th 1960, during a meeting of the scientific staff of the (Italian) National Frascati Laboratories when Bruno Touschek proposed to his colleagues at the Frascati Laboratory near Rome to  make  an electron positron experiment.  
On the same exact day when Touschek advanced his proposal,  an article was received by {\it The Physical Review Letters} in Upton, New York, entitled {\it Pion Form Factors from Possible High-Energy Electron-Positron Experiments }  \citep{Cabibbo:1960zza},  while earlier in the month another article from University of Rome had been submitted by  Laurie  M. Brown, then visiting the Physics Institute in Rome,  and Francesco  Calogero \citep{Brown:1960} about {\it Effects of pion-pion interaction in electromagnetic processes}. Both articles deal with the pion form factor, and the interest of studying it in the time-like region through electron positron experiments is mentioned and clearly emphasized  in the Cabibbo and Gatto's article.\footnote{
This article, whose preparation {\it precedes} Touschek's proposal, should not be confused  with a second more complete article on electron positron physics \citep{Cabibbo:1961sz} which was published a year later and became known as {\it the Bible } among AdA's physicists. }

But how did  such  powerful and disruptive idea start? 
In some accounts of the birth of electron positron colliding beams,  Touschek's proposal is often described  as springing from nowhere, 
except  from being related to Wider\o e's war time  idea of colliding oppositely charged particles.\footnote{ 
According to \RW, the idea came  to him by watching clouds collide in the sky, during a vacation he took in late summer 1943 \citep[81]{Wideroe:1994}.}
 He shared the idea with \BT, then his collaborator on the German funded betatron project, but Touschek was not impressed. He considered the energetic advantage in center-of-mass energy provided by colliding beams of relativistic particles too obvious an idea to be patented, as Wider\o e himself recalled in his autobiography \citep[82]{Wideroe:1994}: ``He said that they were obvious, the type of thing that most people would learn at school (he even said `primary school') and that such an idea could not be published or patented.''  But  Wider\o e still wanted to be assured the priority of this idea and submitted a patent which was only issued after the war.\footnote{Ernst Sommerfeld, Arnold Sommerfeld's son, took care of all Wider\o e 's patents. Like all the several patents filed in that period, most of them related to the betatron design, it was given the status of a `secret patent', only recognized and published in 1953 (``Anordnung zur Herbeif\"uhrung von Kernreaktionen'', German patent No. 876279, submitted on Sept. 8, 1943, issued on May 11, 1953). 
 See a reproduction in \citep[179]{Wideroe:1994}.} 
 
 Wider\o e had also considered the possibility that particles to be made to collide could be stored in a ring, even if at the time it was not clear how to accumulate a sufficiently intense beam for such collisions. He not only described electrostatic rings for electrons, but also described storage rings using magnetic fields and wrote about electron-proton, proton-deuteron, and proton-proton collisions.

{After the war, in the mid  1950s, the quest for reaching higher energies in particle scattering  naturally focused on the advantage of  head-on-collisions, even if the practical possibility of colliding beams was to remain  elusive for sometime}.  In April 1956,   Donald Kerst, with Andrew Sessler, Kent Terwilliger and others \citep{Kerst:1997rs,Voss:1996aa} put forward a proposal to attain very high energies by means of intersecting beam of particles, which was followed  a few months later by a paper by  Gerard O'Neill also advocating storage rings for high energy physics research \citep{ONeill:1956aa}.  A few years later, O'Neill with 
W. Carlisle Barber, Bernard Gittelmann and Burton Richter,  proposed to build an electron-electron collider to test quantum electrodynamics \citep{Barber:1959vg}. 

  Thus, the scenario of an idea suddenly coming  in Touschek's mind, based on his conversations with Wider\o e in the early 1940s, is a 
  simplistic approach to the problem of tracing the roots of how  the first electron-positron collider came to be built.
   In this prequel to AdA's successful experimentation in Orsay, we shall try to throw some light on the theoretical 
   and experimental
   origins  of the idea through the published literature, until the day Touschek proposed to construct AdA.

The first calculation of the cross-section for electron positron collisions was done by Homi J. Bhabha, the architect  of the Indian Nuclear program,  in \citep{Bhabha:1935pg,Bhabha:1936zz}.\footnote{
{In a private communication to Yogendra N. Srivastava, Marcello Conversi related the anecdote   that  Bhabha's article was rejected  when first submitted to  a  Danish journal  because the process ``would never be measured".}} A student in Cambridge University  in  1930s,  Bhabha   obtained his doctorate with  Ralph H. Fowler and earlier  studied for this Mathematics Tripos with  Paul M. Dirac, the father of antiparticles, the theorist who   advanced  the concept of a sea of particle and anti-particle states of negative energy.

Dirac's \textit{positive electrons} materialized in 1931, when Carl D. Anderson saw tracks produced by cosmic rays in his cloud chamber and when the first pair production was observed by Patrick Blackett and Giuseppe Occhialini in their counter controlled cloud chamber.\footnote{ It is known that  the name positron for the electron anti-particle was coined by Watson Davis the editor of \textit{Science News Letter} who published the photograph of Anderson's tracks in December 1931.}   It was thus natural that Bhabha's calculation appeared in the context of cosmic ray experiments, the only source of really high energy particles at the time. The reverse process, positron annihilation, was discovered and investigated in 1933-1934 by several physicists by scattering experiments with gamma rays from radioactive sources.\footnote{See Tim Dunker in  \href{https://arxiv.org/abs/1809.04815}{Who discovered positron annihilation}, and references therein \citep{Dunker:2018aa}.} 

In the 1950s, as particle accelerators started taking central stage in the world of nuclear and particle physicists,  an experimental check of   Bhabha's calculation through particle scattering could be envisaged. 

Indeed, in    the talk given  by Wolfgang Panofsky  \citep{Panofsky:1959kz} at  the 9th International Conference on High Energy Physics, held in Kiev from  15th to 25th July 1959 \citep{Kiev:1960} the outcome of such  an experiment was reported.
\subsection{Electron-positron collisions from Kiev to Rome and Frascati}
\label{ssec:kiev}
The Kiev conference, together with  one which followed two months later at CERN,  closed a period of planning and discoveries as  the '50s saw  particle physics overtaking  nuclear physics in terms of interest and new proposals. 
During these years,
  in Europe, CERN launched the construction of  a proton synchrotron, France a linear electron accelerator, Italy a circular electron synchrotron. All three entered into function in 1959, and their working was presented at the Kiev conference. 

In the United States  
{ at Stanford University, 
together with a functioning linear electron accelerator,  the construction of  an electron-electron collider was in advanced stage of development.}\footnote{
{The construction of a multi-GeV two-mile-long linear accelerator (LINAC)
was formally proposed by Stanford researchers in 1957 \citep{Dupen:1966aa}. In the United States,
linear electron accelerators  had been developed after the war,   in the line of  a tradition going back to the 1930s with William V. Hansen. 
The third LINAC of the Stanford  series, the Mark III   and its extensions,  were developed at the High Energy Physics Laboratory directed by Wolfgang Panofsky and towards the mid 1950s the machine began to be used by Robert Hofstadter's group for their pioneering investigations of nuclear structure by means of electron scattering, a remarkable work which Bruno Touschek held in high consideration, and for which Hofstadter was awarded the 1961 Nobel Prize in Physics. Mark III resounding success led  to new ambitious plans. In the meantime, they were  constructing an electron-electron collider formed by two electron storage rings with a common straight section for colliding beam experiments, especially aiming to test current thinking in the field of quantum electrodynamics \citep{Barber:1959vg}.}}
The Russian scientists were also very active in accelerator science. The giant Synchrophasotron for protons designed and constructed under supervision of Vladimir I. Veksler, who had discovered the phase stability principle independently from Edwin McMillan, was operational at the Joint Institute for Nuclear Research in Dubna from 1957 with a proton energy of 10 GeV. At the time it was the largest accelerator in the world and reached the highest energies until the start-up of CERN proton synchrotron in 1959. At the Laboratory of New Acceleration Methods of the Institute of Atomic Energy  in Moscow, Gersh Budker's team  had started since 1957s to develop VEP-1, an electron-electron collider formed by two tangential rings and ideas about an electron-positron collider were beginning to be discussed.\footnote{What was happening in the USSR has  been described by Russian scientists involved at the time in those projects  \citep{Skrinsky:1996aa}, \citep{Baier:2008aa} and  \citep{Levichev:2018aa}.}

\begin{figure}[ht]
\centering
\includegraphics[scale=0.45]{
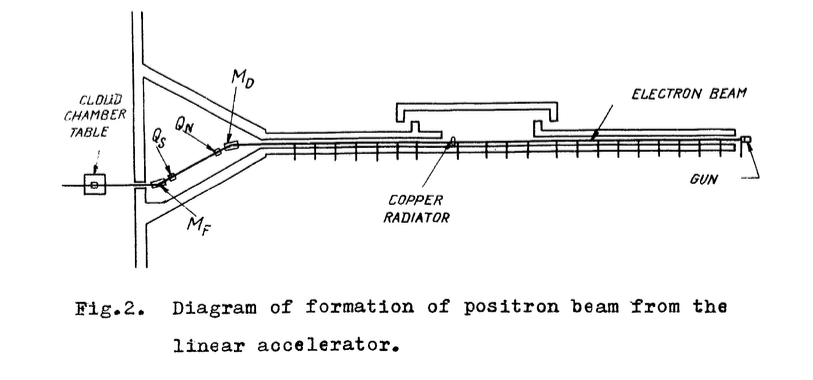}
\caption{Fig.~2 from Wolfgang Panofsky's contribution to the Proceedings of the 1959 Kiev Conference on high energy physics. The figure shows the setup of an experiment at  the Stanford Mark III Linear accelerator  which would produce scattering of electrons against positrons, as described in the  figure caption in Panofsky's contribution \citep{Panofsky:1959kz}.  }
\label{fig:panofsky-kiev-fig2}
\end{figure}

In his talk at the Kiev conference, reporting research work performed at Stanford, Panofsky describes both the electron electron  tangential ring machine under construction and   an 
experiment, performed at the linear accelerator, of the scattering of 200 MeV positron by electrons in a beryllium target.\footnote{
The main goal for setting up a positron beam at Stanford was to compare $e^{-} $ proton and $e^+$ proton
elastic scattering cross sections in order to pin down the interference between the one-photon exchange
and the two-photon exchange Feynman diagrams which has a different sign in the two processes. Comment courtesy of Jacques Ha\"issinski.} 
The experimental set-up for this experiment is shown in Fig.~\ref{fig:panofsky-kiev-fig2} from \citep{Panofsky:1959kz}.
The results of this experiment  indicated good agreement with the Bhabha theory,  in particular noting the contribution of the annihilation term.  Namely, the experimental results   would have been inconsistent with  Bhabha's  theory if the annihilation terms, from direct collisions between electrons and positrons,  were  omitted, as clearly shown in  \citep{PhysRev.117.557}, in an  article submitted shortly after the Conference. The interest of such contribution is unique  for a theorist: here  was the first appearance of the annihilation diagram, to put it in a theorist's jargon, here were Dirac's positrons made to collide with their anti-particles, the electrons,  and annihilate each other in the encounter, as expected. Panofsky's description of the results of this experiment  cannot have escaped the attention  of   three first class Italian theorists from University of Rome, attending the Conference, Marcello Cini, Raoul Gatto and Bruno Touschek.\footnote{The list  of participants  from Italy also includes  the following: Gilberto Bernardini, Paolo Budini, 
 Marcello Conversi, Nicol\`o Dallaporta, Bruno Ferretti, Carlo Franzinetti, Giacomo Morpurgo, Giuseppe Occhialini, Ettore Pancini, Giampiero Puppi, Giorgio Salvini, and Gleb Wataghin. The role of the theorist  Marcello Cini in the  developments around Touschek's proposal in late 1959 and early 1960   and the relevance of his participation to the Kiev Conference is  not to be ignored, as he seems to have inspired the Brown and Calogero work, as acknowledged in \citep{Brown:1960}. He was also close to Touschek at the time, having  coauthored with him  an  article on what was later known  as  Cini-Touschek transformation \citep{CiniTouschek:1958}.}

Touschek presented some {\it Remarks on the neutrino gauge group} \citep{Toushek:1959bza}, a work  which was also related to topics he was discussing in his correspondence with 
{Wolfgang }Pauli (see also the article they published together at that time \citep{Pauli:1959aa}.  It is clear that, until the Kiev's conference, Touschek's scientific  interest had been, and still was, mostly on weak interactions and meson physics, with a keen interest in discrete symmetries.\footnote{
 Touschek's publications during the 1950s were mainly oriented on quantum field theory and research on cosmic ray mesons (in August 1958 he organized and directed the IX Varenna Course on \textit{Pion physics}), but the properties of space-time symmetries were also a main focus of his research interests since 1954, when he wrote \textit{A note on time reversal} with G. Morpurgo and L.A. Radicati \citep{Morpurgo:1954aa}, followed by \textit{Space and time reflection in quantum field theory} \citep{Morpurgo:1956aa}. In 1956, when  Lee and Yang questioned parity conservation in weak interactions  \citep{LeeYang:1956aa}, he quickly reacted with an article on neutrino theory in which he introduced what was later referred to as chiral symmetry \citep{Touschek:1957aa}. It was followed by \textit{The symmetry properties of Fermi Dirac fields} \citep{Touschek:1958aa}.  On the neutrino mass and non-conservation of parity see also \citep{Touschek:1957ab}.}
So it was  for Raoul Gatto and Marcello Cini as well, but the events to follow in the coming months  show that the impression from Panofsky's talk did not leave   their   minds.

\begin{figure}[ht]
\centering
\includegraphics[scale=0.5]{
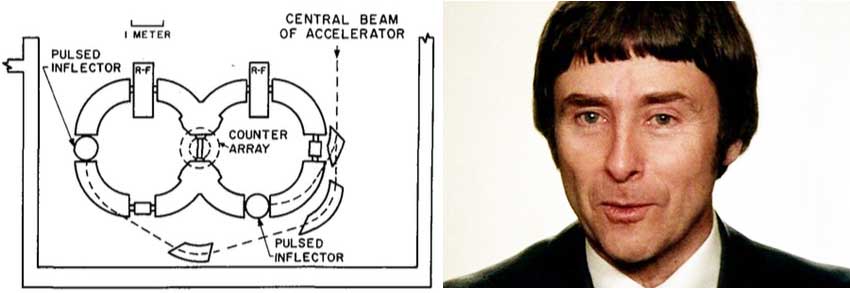}
\caption{Figure 2 from Gerard K. O'Neill contribution to the Proceedings of the 1959 Geneva Conference on accelerators. The figure shows a plan view of the colliding-beam vault at the Stanford Mark III linear accelerator \citep{ONeill:1959aa}. A contemporary photograph of O'Neill is shown in the right panel.}
\label{fig:ONeill1959}
\end{figure}

In September, accelerator and particle physicists met again at CERN, this time to attend the  {\it International Conference on accelerators and instrumentation}, 14-19 September 1959 \citep{Kowarski:1959}. 
{Gerard O'Neill presented the talk  \textit{Storage Rings For Electrons And Protons} focused on the Princeton-Stanford experiment on quantum electrodynamic limits  \citep{ONeill:1959aa,Barber:1959vg}, a scheme of which is presented in Fig.~\ref{fig:ONeill1959}.}

Colliding beams were discussed in the two contributions by O'Neill \citep{Oneill:1959ab} and L. W. Jones \citep{Jones:1959aa}, while Panofsky gave a general talk in which he commented on the experiments on the limit of validity of quantum electrodynamics and mentioned electron-electron scattering, positron-electron scattering and annihilation in flight.\footnote{
{His comment on electrodynamics acknowledges  this to be ``the one area in fundamental particle physics where experiment and theory are in exact quantitative agreement for the full range of energies explored to date'', but he also adds   that ``In principle such experiments could be carried out at lower energy and at high accuracy; however, both the experimental problems and the uncertainty of the higher-order theoretical corrections make this impractical. At the very highest energies, ambiguities might again arise between possible breakdown in quantum electrodynamics and the uncertainty in the calculation of corrections'' \cite[5]{Panofsky:1959aa}.}}

Panofsky was in Europe, at CERN at the time, and a month later, 
came to Italy and gave some seminars, at least one in Rome and certainly one in Frascati. Panofsky's seminar was held in Frascati on  October 26th, 1959, the one in Rome around that date.\footnote{The date of Panofsky's seminar in Frascati is recorded in the list of seminars held at the Laboratory, copy is courtesy of V. Valente to G. P. } According to Nicola Cabibbo, at the end of the seminar in Rome, Touschek asked the question: ``Why not make electrons to circulate against positrons?". Nicola Cabibbo,  who had been Touschek's student,  remembered that:  ``It was after the seminar that Bruno Touschek came up with the remark that an $e^+e^-$ machine could be realized in a single ring, `because of the CPT theorem''' \citep{Cabibbo:1997aa}.\footnote{Raul Gatto, too, well remembered that, ``Bruno kept insisting on CPT invariance, which would grant the same orbit for electrons and positrons inside the ring'' (Raul Gatto to L. B. January 15, 2004). As recalled by Carlo Rubbia: ``I understood that in his mind electron-positron collisions were nothing else than the way of realizing in practice the idea of symmetry between matter and antimatter, in the deep sense of the Dirac equation" \citep[57]{Rubbia:2004aa}.} 

\begin{figure}[htbp]
\centering
\includegraphics[scale=0.7]{
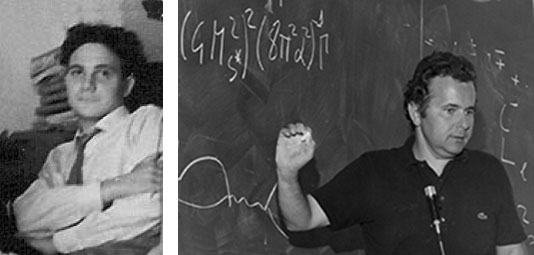}
\caption{Raoul Gatto at left, and Nicola Cabibbo at right.}
\label{fig:Gatto-Cabibbo}
\end{figure}

This is the moment when not only Touschek started thinking about the possibilities which such experiments would bring, but other theorists in Rome began to do some calculations. 
Cabibbo further  recalls that on that occasion Touschek emphasized that instead of an electron-electron collider it would have been much better to build an electron-positron machine.  Soon after, Cabibbo with Raul Gatto  started  to calculate cross-sections:  ``We began to realize that this would have been a perfect machine to study the properties of hadrons, of pions, in particular, and then we began to study the pion form factor.'' As Cabibbo says in  \citep[50]{Cabibbo:2003aa}: ``Ci scatenammo a fare qualche conto sulle sezioni d'urto''.\footnote{
{We rushed into calculating  some possible cross-sections [for electron positron reactions]. }} Images  of these two young physicists are shown   in Fig.~\ref{fig:Gatto-Cabibbo}.

{Together,   Gatto and Cabibbo,  worked on a paper where the possibility of studying  the pion form factor through electron and positrons collision experiments was envisaged. The paper was ready in  February and was then  submitted  to the most prestigious journal for particle physics, {\it The Physical Review Letters}:  received on February 17, 1960,  it was accepted and published within less than a month.}  A similar pattern must also have been followed by Brown and Calogero. Although their paper does not explicitly mention electron positron collisions in its title, the possibility of testing nuclear forces in the annihilation of electrons and positrons is openly mentioned, as seen in \citep[317]{Brown:1960}.\footnote{Francesco Calogero and Nicola Cabibbo had completed their thesis work with Bruno Touschek in 1958  on weak interactions. They had first approached Marcello Cini, but Cini had suggested them to work with Bruno Touschek, as  the most interesting theoretical physicist in the department at the time (from an interview with Cabibbo, taken in May 2003, part of which appears in the  docu-film   \href{https://youtu.be/R2YOjnUGaNY}{\it Bruno Touschek and the art of physics}  by E. Agapito and L. Bonolis, 2003).}

{During this time, namely through fall 1959 and February 1960, Touschek's voice in terms of scientific output is    silent, as there are no publications which he could have been writing during these months.}\footnote{
{The list of Touschek's publications in \citep{Amaldi:1981}, shows  a gap between  neutrino and meson physics papers appearing up to October 1959, and an Internal LNF Report, LNF - 60 / 012,  dated April 1st,  1960,  \href{http://www.lnf.infn.it/sis/preprint/detail-new.php?id=3156}{On the Quantum losses in an electron synchrotron}.}
} 
{He must have been thinking, and perhaps calculating, on  how to realize  the idea he had set in motion. When the two papers by his theoretical colleagues in Rome had been completed and sent for publication,  he had a fair idea of the feasibility of an electron positron machine and was ready to make it public.}  And then, things happened  on  February 17th, 1960. 

The synchrotron had been built and was functioning since the preceding year. 
The existence of this accelerator in the Rome area marked a new era for experimental particle physics, that through the 1950s had continued to be performed using  cosmic rays as a source of high-energy particles, following the  Italian tradition pioneered by Bruno Rossi at the beginning of the 1930s. As a theoretician, Touschek had widely supported these investigations. But now, a need for guidance to future experimental activity with the Frascati machine became of concern for  physicists from University of Rome, who had built equipments and planned experiments. The idea of creating a theoretical group to this purpose was going to be debated in a meeting, which the Director  of the Laboratory, Giorgio Salvini,  called for February 17th, 1960.
 Touschek, who had shown interest in the working of the synchrotron, and was naturally interested in accelerators since his war time
experience with Wider\o e's betatron, was indicated as a possible head of such a group.\footnote{Touschek had worked on the Glasgow synchrotron, and, soon after his arrival in Rome,  had written a paper on the synchrotron theory with Matthew Sands \citep{Sands:1953aa}. He was naturally attracted to follow the activity of the Laboratory also because his maternal aunt had a villa in the  hills near Frascati.} 
The prospect did not appeal to Touschek.  In a later writing, Touschek describes his dislike of the idea of becoming a house theorist, confined to do calculation for the experimentalists. From a 1974 writing prepared for the Accademia dei Lincei, in Italy, we quote:\footnote{B. Touschek, manuscript, B. T. A., Box 11,
Folder 3.92.4, p. 7).}

\begin{quote}
\begin{small}
I did not like the idea. It smelled of what in Germany was known as the `Haustheoretiker', a domesticated animal,  which sells itself and what little brain he has to an experimental institution to which it has to be `useful' [\dots] I  feared --- and this was quite a personal and possibly unjustified fear ---  to end up with the task of proving on theoretical grounds that the  effort and money which went into Frascati were well spent: in short the role of an `Anticassandra', who predicts the past (not the future) and who is optimistic (not pessimistic) about it.
\end{small}
\end{quote}  

\begin{figure}[ht]
\centering
\includegraphics[scale=0.5]{
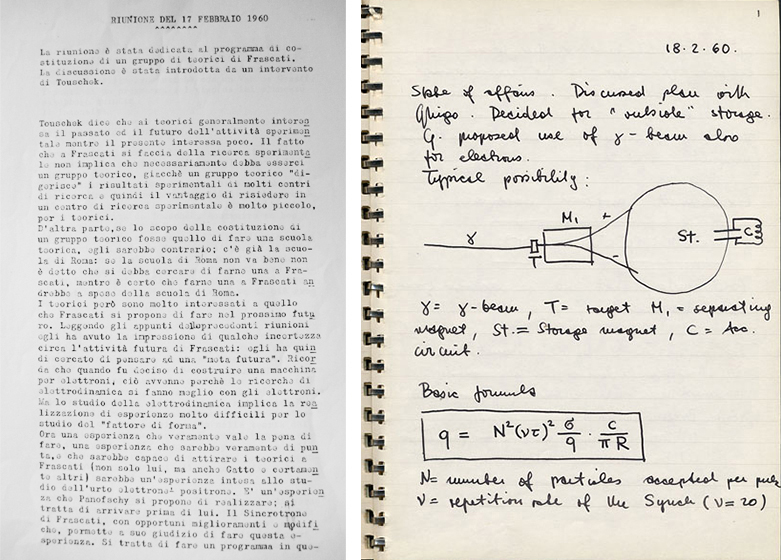}
\caption{At left, we show the first page of the minutes of the Frascati meeting where Touschek proposed, for the first time, that the Laboratories build an electron-positron scattering facility. At right, the first page of Touschek's Storage Ring (SR) notebook, which he started in the day immediately following the Frascati meeting.}
\label{fig:verbale-BT18feb60}
\end{figure}

\noindent
Most  probably he felt the danger of  repeating  the experience of the war years, when he had to do calculations for the betatron or  for  the giant X-ray facility envisaged for war purposes by the Leipzig physicist Ernst Schiebold and which were of interest for the RLM death-ray project.\footnote{A copy of Touschek's calculations preserved in Schiebold's papers at the S\"achsisches Staatsarchiv, Leipzig, was kindly given us by the late Pedro Waloschek.} He wanted to explore new roads in theoretical and particle physics and not to bind his creativity to  calculations for his experimental colleagues. A way out to the pressing of his well wishing Roman colleagues, eager to start experimentation with the synchrotron and receive his advice, had to be found. And thus, at the very beginning of the meeting, Touschek proposes a solution, both to allay his personal fears and to offer a future to the laboratory:\footnote{Minutes of the 17th February 1960,  Report of the meeting  held in Frascati on February 17, 1960 (L.N.F. Report N. 62, December 1960). Copy in Bruno Touschek's papers (Edoardo Amaldi Archives, Physics Department, Sapienza University of Rome, from now on BTA), Box. 12, Folder 95.}
\begin{quote}
\begin{small}
The meeting was called to discuss  the project of the constitution of  a theory group in Frascati. The introduction to the meeting was given by Touschek.\\
Touschek says  that \dots an experiment  really worth doing, a  frontier experiment able to attract theoretical physicists (not just himself, but also Gatto and others for sure) would be an experiment for studying electron-positron collisions.\footnote{In Italian: La riunione \`e stata dedicata al programma di costituzione di un gruppo di teorici di Frascati. La riunione \`e stata introdotta da un intervento di Touschek. Touschek dice 
\dots una esperienza che veramente vale la pena di fare, un'esperienza che sarebbe veramente di punta, e che sarebbe capace di attirare   i teorici a Frascati (non solo lui, ma anche Gatto e certamente  altri) sarebbe un'esperienza intesa allo studio degli urti elettrone-positrone.} 
\end{small}
\end{quote}
\noindent
There was some debate, Touschek kept his point that Frascati did not need a theoretical (physics) group (University of Rome already had one) but, rather,  a program to open up the new energy  frontier in particle physics. He was saving himself and  the Laboratory as well. At the end the idea was approved, Touschek was given the green light to start studies and come back with a workable proposal. In Fig.~\ref{fig:verbale-BT18feb60}, we show the first  page of the minutes of the meeting, at left, and, at the right, the first page of Touschek's SR book, where SR stands for Storage Ring, and is dated February 18, 1960, day immediately following the Frascati meeting.

{Touschek's idea had originally been to use the newly built electron synchrotron to inject electrons and positrons. This was unworkable, in that moment,  both technically and politically.}\footnote{
{Physicists from both Frascati and various Italian universities had been waiting to start experiments with the new machine and could not give up their plans so soon.}} 
{The project was saved by one of the physicists who had built the synchrotron,} Giorgio Ghigo,  who proposed a smaller {dedicated} machine, to be built anew, constituted by a magnet, inside which to place an empty chamber for the simultaneous circulation of electrons and positrons. To the synchrotron it would be  left the challenge to produce enough    electrons to initiate the  beams which would be  orbiting in AdA. 
\BT \ was asked to prepare a workable project and, as Amaldi recalled \citep[32-34]{Amaldi:1981}: ``During the same day, Bernardini, Corazza and Ghigo began to work with Touschek on the first $e^+e^-$ storage ring [\dots]  Touschek had therefore immediately found his first collaborators, but also quickly found the financial resources."  
\\The scientific council of the Laboratories approved AdA's construction  less than three weeks later, on March 7th, 1960. 


\begin{figure}
\centering
\includegraphics[scale=0.53]{
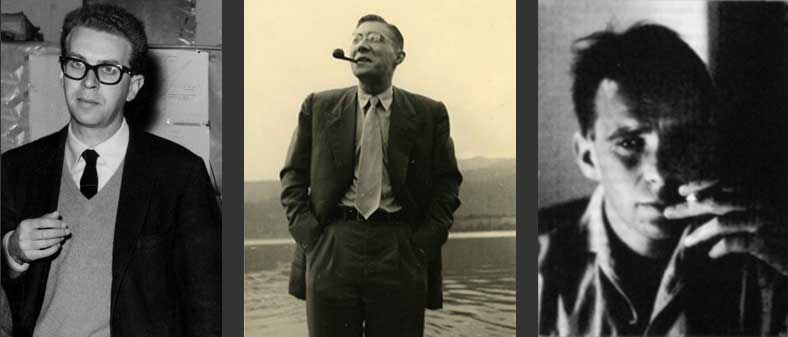}
\caption{At left Carlo Bernardini, who collaborated with \BT \  throughout the entire AdA project and was his closest friend. In center, Giorgio Ghigo, from the Frascati laboratories, who proposed and then  designed AdA's magnet, photo courtesy of Ghigo's family. At right, Bruno Touschek, photo reproduced from \citep{Amaldi:1981}.}
\label{fig:ghigo-bernardini}
\end{figure}

Shortly after,   an exceptional team of scientists and technicians was rapidly assembled  and was able to build a functioning electron positron colliding beam machine within less than a year. In Fig.~\ref{fig:ghigo-bernardini} we show Carlo Bernardini, Giorgio Ghigo and Bruno Touschek. Bernardini, a young theoretical physicist, who joined the AdA team as soon as the machine was approved, was a very close friend of \BT,  and   had  stimulated  his  interest in  what was happening with the synchrotron in Frascati.

\begin{figure}[ht]
\centering
\includegraphics[scale=0.45]{
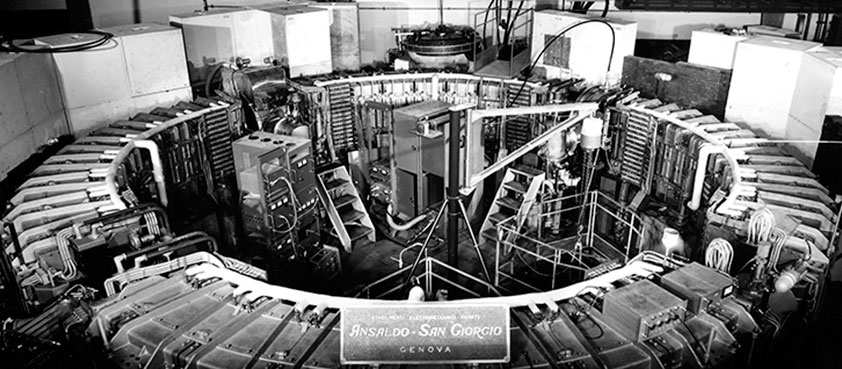}
\caption{AdA, seen in the far back,   near the electron synchrotron in 1961, LNF Archives.}
\label{fig:adasynchro}
\end{figure}

  By February 1961, only  one year from Touschek's proposal, the team had the first electrons circulating in AdA.  During the months  to follow, they  succeeded in  storing beams of electrons and positrons that lived up to 40 hours in the vacuum chamber, a feat which, to their knowledge, nobody had been able to attain.\footnote{When a special, thoroughly cleaned donut chamber was used, ``the pressure went down to less than $10^{-10}$ torr, perhaps $10^{-11}$ torr, thus reaching the limit of sensitivity of the Alpert gauge'' \citep[168]{Bernardini:2004aa}.} The team of scientists who had built AdA  was elated  by (literally) seeing the light emitted by electrons (or positrons) circulating in AdA \citep{Bernardini:1962zza}. 
 But,  
  as it often happens in science,   disappointment  followed the initial excitement.
The Frascati electron synchrotron, {\it Il Sincrotrone} in Italian, which had fed the first electron beam to AdA,   had not been built for this scope --- namely as injector of electrons for an electron positron collider! --- and it soon appeared inadequate for the ambitions and the challenges of a machine such as AdA, which is shown  at the back of the picture, in  Fig.~\ref{fig:adasynchro}, at a small distance from the electron synchrotron, where it was installed hoping to optimize the capture process. Unfortunately things did not improve:    after the first few months of operation, further progress appeared difficult, and the AdA team  started  losing hope to reach their goal. 

\subsection{July 1961:  a visit from Orsay}
\label{ssec:orsay}

 The turning point in AdA's story  came through   the 
 collaboration with the French \LAL \  in Orsay.  
          As we have described in detail in \citep{Bonolis:2018gpn}, AdA's future was saved by the interest aroused by Touschek's announcement of AdA's  working, at  the 1961 CERN accelerator conference  \citep{Bell:1961gi}.\footnote{For the  Proceedings, see also \href{https://cds.cern.ch/record/280184}{International Conference on Theoretical Aspects of Very High-energy Phenomena, 5-9 June 1961, CERN, Geneva.}}
     
     Two French  scientists,  \PM\ and Georges Charpak, 
     decided to go to Frascati and see with their eyes what was happening. 
 Georges Charpak was  at   CERN and his visit to Frascati  was  likely related to    CERN's interest and early discussions about  proton-proton and proton-antiproton colliding beam accelerators.\footnote{Georges Charpak  was  later  awarded the 1992 Nobel Prize in Physics "for his invention and development of particle detectors, in particular the multiwire proportional chamber" .}
 \PM \  came from the \LAL, in Orsay, to  see ``the intriguing things which were happening there", as he recalls in \citep{Marin:2009}.
      Marin, who had been looking for a direction of his research,  was immediately taken in  by AdA's concept and promise.\footnote{He calls it  {\it un  vrai bijou}, a real jewel,  in \citep{Marin:2009}.} 
    {At the same time,  Touschek, who had long been looking for  a solution to AdA's  difficulties, 
 understood that  the \LAL \ could be an invaluable ally, as  the well focused electron beam from the linear accelerator   could in principle  provide the strong currents needed to prove the feasibility of machines such as AdA.\footnote{
 What was most important was the fact that a LINAC  beam is most easily extracted at the end of the accelerator (while the beam extraction from a synchrotron requires a sophisticated procedure): there is no intensity loss and focusing this `naturally' extracted beam is straightforward. Note courtesy of Jacques Ha\"issinski.}

 
  The \LAL\  had been founded in 1957  
 and a linear accelerator had been operational since 1959. The linear accelerator had been designed 
 by Hubert Leboutet from the Thomson-CSF (Compagnie g\'en\'erale de t\'el\'egraphie sans fil) and built by an equipe of scientists and technicians,   directed by  Jean-Loup Delcroix, \citep{Blanc-Lapierre:1963} from the Laboratoire de Physique de l'\'Ecole Normale 
Sup\'erieure directed by Yves Rocard.\footnote{Hubert Leboutet collected his memoirs of the years 1950-60 in  a 1994 publications of the  group Histoire de Thales - aicprat, entitled  {\it des ELECTRONS  et des HOMMES -  Les d\'ebut des acc\'el\'erateurs dans les ann\'ees 50 et 60 \`a  la CSF}, copy  contributed by  private communication from \JH.} 
 

Thus, 
the needs of AdA for a  stronger source of electrons
and those  of the \LAL \ for new ideas and perspectives  definitely met, when it was agreed 
to transport   AdA to Orsay.   The negotiations for the transfer of AdA  between the two laboratories  were  carried out by the laboratory directors first, and then by the researchers themselves. 
On  the Italian side,  there were  Carlo Bernardini and Bruno Touschek, while on the French side the active party had \PM\ and \FL, another  young researcher from  von Halban's linear accelerator experimental team.\footnote{ Hans von Halban had been the first director  of the \LAL. When the decision was taken to build a linear accelerator and a laboratory around it, there came the decision to have a group of physicists to plan for possible experiments together with the necessary technical staff, and a director. The choice fell on Hans von Halban, the Austro-German collaborator of \FJ, who, together with Lew Kowarski, had participated in 1939 to the experiments providing evidence that the number of free neutrons released in the fission of uranium is sufficient to induce fission in other nuclei setting in motion a chain reaction. Halban, who was in Oxford at the time, accepted and led  the  experimental team which  included \PM\ and \FL. In 1961, he resigned from his position,  and \ABL, from University of Algiers, was asked to come and direct the \LAL. \ABL \  was then instrumental in the transfer of AdA from Frascati to Orsay. See also 2013 interview with Maurice L\'evy in  \href{http://www.lnf.infn.it/edu/materiale/video/AdA_in_Orsay.mp4}{Touschek with AdA in Orsay}.}

It took almost one year before everything could be agreed and prepared, but, after letters and visits had been exchanged, finally the date for AdA's travel  into France had been defined to take place around July 4th, 1962 \citep{Bonolis:2018gpn}.

AdA's crossing into France was  not  a simple matter. Late in June, AdA's  arrival had been anticipated by Bruno Touschek to Francis Perrin, the high commissioner for the French Atomic Energy Commission, the CEA, in a letter asking for help in case of difficulties arising  at the Italian French borders. The   Frascati ``convoy", as Touschek had called it in his  letter, consisted of  two trucks, a big one and a smaller one. The smaller one was carrying material and instruments, which would be too heavy to send by plane and which would be needed for installing AdA near the linear accelerator in Orsay. The other  truck, bigger and heavier,   was carrying AdA and   the   indispensable equipment, pumps and batteries, allowing to have  AdA ready for installation as soon as possible, upon arrival in Orsay. This included maintaining   the  extreme vacuum reached  inside AdA's  doughnut,   an extremely rarefied state, with a pressure   that could be as low as $10^{-10}$ Torr.  Such ultra-vacuum had been deemed very difficult  to reach by the American competitors of the AdA team, the Princeton Stanford group. In Stanford a colliding beam accelerator, with two tangential rings for  electrons against electrons,  had been planned since 1957, following Gerard O'Neill's storage ring studies and was now  under construction.\footnote{See \href{https://inspirehep.net/record/919977}{Component Design And Testing For The Princeton-Stanford Colliding-Beam Experiment}.}  The American team had been wary of  experimentation with positrons, not being convinced that the very low pressure  their circulation required  could be within reach of then available  technology. This feat had been accomplished by the AdA team, but it required many months 
to reach such ultra-high vacuum. 
According to Carlo Bernardini, the astrophysicist Lyman Spitzer, pioneer of controlled thermonuclear research in the US, could hardly believe that the Italians had achieved such a ultra-high vacuum, a challenge he had tackled with the different models of his stellarator, the plasma-confinement device he had been building at Princeton University during the 1950s-early 1960s.\footnote{The story was told several times by Bernardini to one of us (L. B.).}
  Maintaining the high-vacuum and carry out the installation in Orsay in short time was essential, if the Franco-Italian team were  to keep the   advantage   on the American team. Thus the team could not run the risk to waste  time, and, once in Orsay,  the machine had to be ready for installation.

 In the sections to follow, the story of AdA in Orsay will now be described in detail.
 
 

\section{AdA's arrival and installation  in Orsay: Summer 1962}
\label{sec:inorsay}

AdA was accompanied in crossing the Alps by the two specialists of the vacuum, 
Gianfranco Corazza and Angelo Vitali, nicknamed {\it Angelino}, who had travelled  with the two trucks and could be ready to take action if problems with the vacuum arose.  When AdA arrived,   it   had been able to cover the 1500 km distance between Rome and Paris in two days and  the vacuum in the doughnut had remained intact, to the great satisfaction  of the Italian scientists who had so carefully prepared Ada for the trip. 
   The scientists from both Italian and the French   team were in a great hurry to start the experimentation, as they knew the competition from the other side of the Atlantic would not be wasting time. Soon after the two trucks left Frascati, the remaining equipment needed for AdA's installation was prepared to be sent by plane. In Fig.~\ref{fig:CB-toFL-letter} we show copy of the letter sent by Carlo Bernardini to \FL, on July 6th, 1962.
   \begin{figure}[htbp]
\centering
\includegraphics[scale=0.075]{
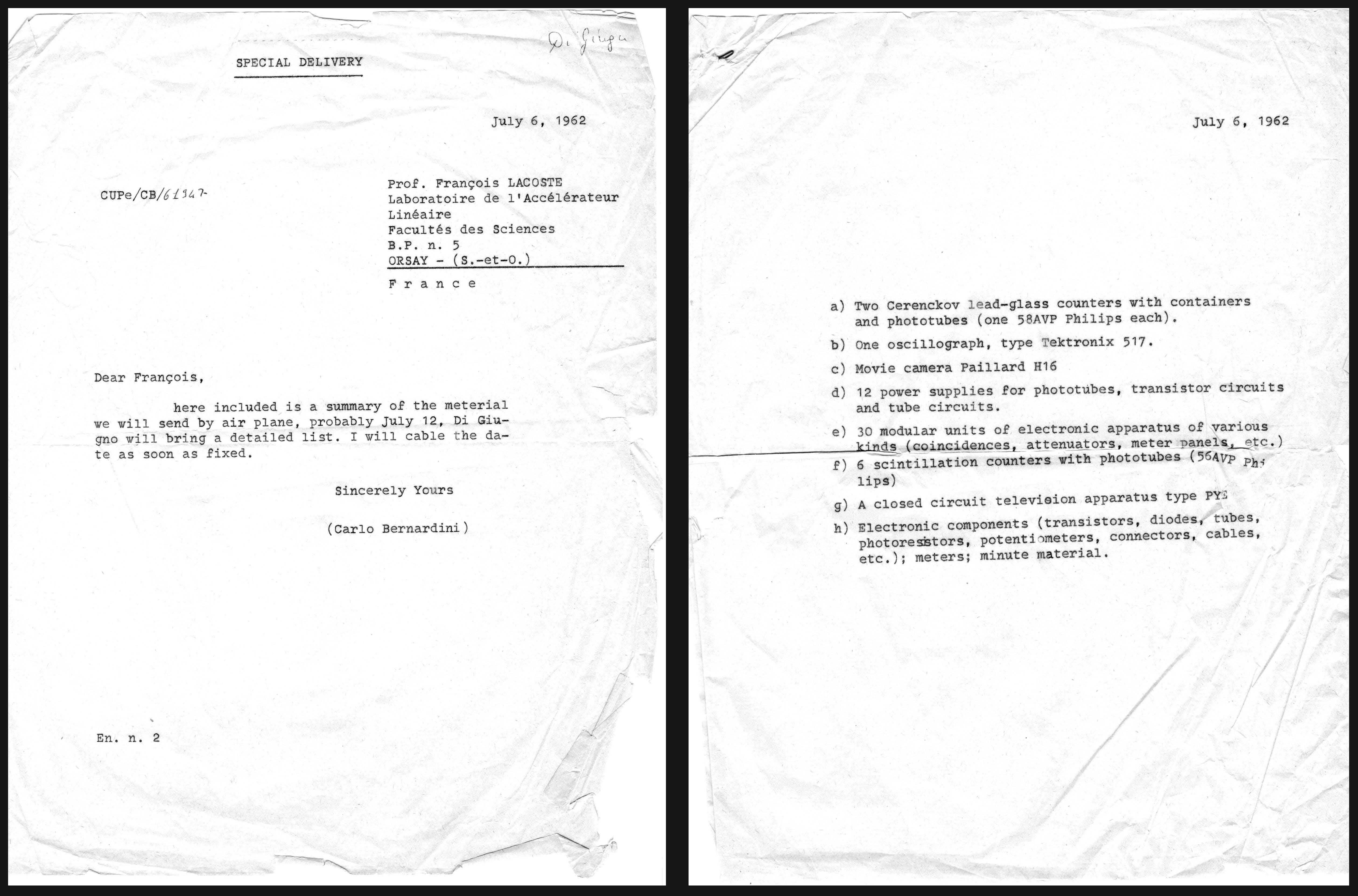}
\caption{Letter sent by \CB \ to \FL\ to anticipate the arrival by plane of some equipment needed for AdA's installation and future experimentation,  courtesy of Giuseppe Di Giugno.}
\label{fig:CB-toFL-letter}
\end{figure}
Once the equipment reached Orsay, and   the Italian technical team led by Giorgio Ghigo   arrived by plane, the installation started in earnest.

 In the left panel of Fig.~\ref{fig:ada-linear-acc} we show a view of the building in Orsay,  which housed the offices and the Linear accelerator, and, in the right panel,   AdA installed in Orsay.
\begin{figure}[htb]
\centering \includegraphics[scale=0.09]{
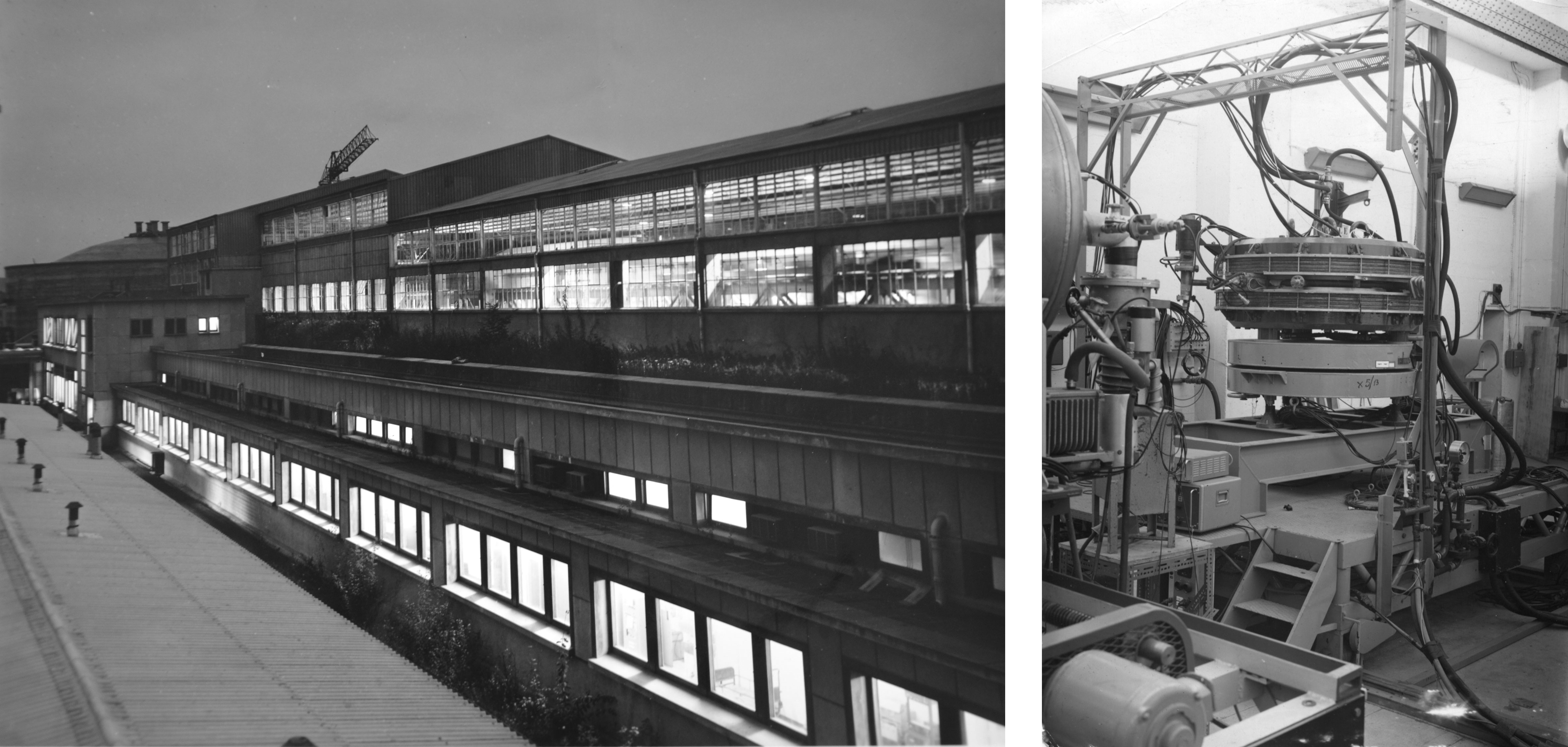} \hspace{.5 cm}
\caption{At left we show  a photograph of the LAL Main building, housing the  Linear Accelerator, and, at right AdA installed in Salle 500 at LAL, both courtesy of \JH. } 
\label{fig:ada-linear-acc}
\end{figure}

On the French side, the scientists  who supervised the installation of AdA near the LINAC, through the month of July,
were Pierre Marin and  Fran\c cois Lacoste. Later on, in November, they would be joined by \JH, a young doctoral student, whose father, Mo\"ise Ha\"issinski, had worked in the Radium Institute, in Paris, directed by Marie Curie. 
The installation proved to be a non trivial affair, and accidents would occur.   In   a 1998 contribution on the occasion of the 1998 Bruno Touschek Memorial Lectures \citep[17]{Haissinski:1998aa}, 
{ \JH\ 
shares  a memory by \PM \ from the installation period:}
\begin{quote}
\begin{small}
AdA was greeted at Orsay by a small team composed of Pierre Marin and
 Fran\c{c}ois Lacoste. Pierre Marin remembers quite vividly an incident which
 marked the installation of AdA  and which could have turned out to be a
 dramatic one. The experimental hall where AdA was to  operate was an 
intermediate energy hall, equipped with a special roof comprising a few water
 tanks having the shape of very large rectangular boxes which provided the
 proper radiation shielding. They could be moved horizontally in order to
 make room for a crane located above them. In the course of the AdA 
installation, it happened that, while the mechanical device used to support 
the ring was hanging on the crane hook, someone pressed a button which
 started the motion of the water tanks. These tanks were so heavy that when
 they reached the crane cable they just pulled the crane together with the
 AdA support. A member of the Italian team saw that the AdA support was 
heading towards a wall and screamed out. Thus alerted, Pierre Marin was 
able to run and stop the water tanks just in time to avoid a catastrophe.
\end{small}
\end{quote}
\noindent
The catastrophe was averted and
by August 1962, the storage ring was installed at Orsay. \JH \  had worked with \FL\ on experimentation at  the Laboratoire de Physique de l'\'Ecole Normale Sup\'erieure, equipped with two accelerators,  and, after a fellowship period  
 in Stanford, at  HEPL (WW Hansen Experimental Physics Laboratory),
  had come back to France for his military service, due to terminate in fall 1962. Thus he was  not present when AdA arrived,  but  joined the team after the summer,  being offered a position as  a doctoral student, as he recalls:\footnote{
From \JH's interview for the movie \href{http://www.lnf.infn.it/edu/materiale/video/AdA_in_Orsay.mp4}{Touschek with AdA in Orsay} by E. Agapito, L. Bonolis, and G. Pancheri. Interviews were held at \LAL,  in Orsay, May 2013.}
\begin{quote}
\begin{small}
I have to say that while I was at Stanford,  a few month before leaving,  I had an  opportunity to see the beginning of  the construction of an $e^-e^-$ ring, and therefore I knew a little bit about the possibilities of such machines, but it took many years in fact for the Stanford machines to take some data. Nevertheless I had some idea what colliding beams were.

At the end [of this period] 
I was offered a position by \ABL, to come as expected and work here. He asked me if I was willing to join the team which had just started to work on AdA, and of course I was very much attracted by this program, again it was frontier physics \dots\\
 I came here,  beginning of November of '62, and AdA was already there, of course. 
 \end{small}
 
\end{quote}
\noindent
\JH \ worked on AdA's measurements until 1964 and his {\it Th\'ese d'\'Etat}, which leads to  the highest education level in France's system,  constitutes the best document of  AdA's achievements, from both the scientific and the historical point of view. In Fig.~\ref{fig:JH-page1-position} we reproduce a 1968 photo of \JH  \ and the  thesis' frontispiece.

\begin{figure}[ht]
\centering
 \includegraphics[scale=0.15]{
 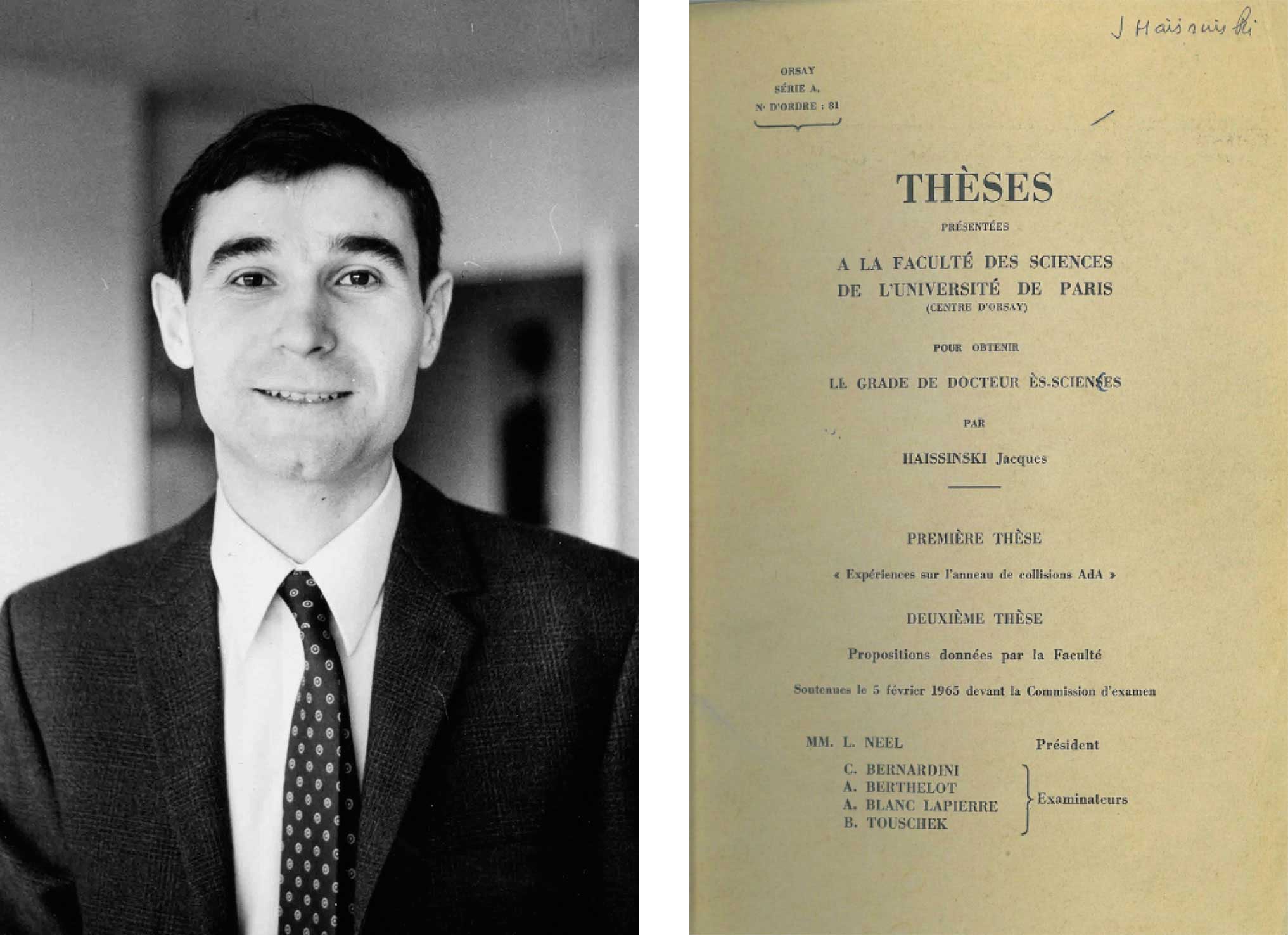}
\caption{\JH \ in 1968 at left, and the frontispiece of his {\it Th\`ese d'\'Etat} at right.}
\label{fig:JH-page1-position}
 \end{figure}

\subsection{First experiments: weekends and long nights or sixty hours in row }
\label{ssec:firstexperiments}
After the July installation,   August 1962 arrived, and with it came the usual summer break, a well deserved pause after the frantic months of planning for AdA's transport and its installation in Orsay.  The French scientists  took their vacation,  sailing or visiting  the countryside, while  the Italian team went  back to Italy, and   escaped the heat  spending a few weeks in  Positano for the Touschek family or in South Tyrol for the Bernardini's.  They all were aware of the challenges ahead of them, and of  the need to store their energies, but also  to  be with their families, before the long  absences to Orsay  expected in the months to come.

In Italy, the  physics community  was generally aware of the Frascati work but there was often a display of scepticism, which Touschek would counter by saying that ``of course electrons and positrons have to meet, it is just a consequence of the CPT theorem". This  deep conviction came from  his  mind frame, as  a theoretical physicist  guided by 
his faith in  the symmetry principles expressed by such mathematical theorem, and was also  supported by his  experience with electrons in Wider\o e's betatron.The Frascati and the Orsay team shared his enthusiasm. In early September, before joining the AdA team in Orsay, where AdA was now ready for the new set-up, \BT,  
Carlo Bernardini and the young Giuseppe di Giugno  
attended the yearly Italian Physical Society meeting in Bologna, where Touschek  presented  an update of AdA's progress. We show  them in Fig.~\ref{fig:bologna1962}, in a photograph taken  during the Conference: Touschek,  in a front seat is seen in the lower right, tanned from his Positano vacation, he is intense and focused, as he knows he will soon  begin the final leg of a journey  into the unknown. 
\begin{figure}
\centering
\includegraphics[scale=0.4]{
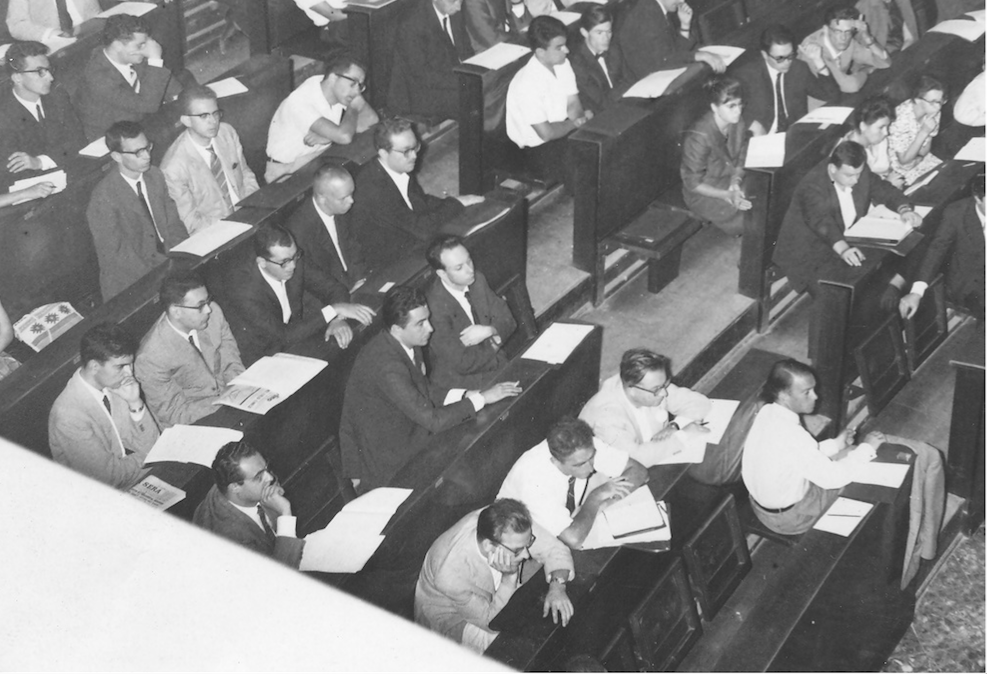}
\caption{Bologna 1962, September 9, 14th, Meeting of the Italian Physical Society, \BT \ is shown on lower right, Giuseppe Di Giugno and Carlo Bernardini, with light color jacket,  on upper left, in a  photo courtesy of Giuseppe Di Giugno.}
\label{fig:bologna1962}
\end{figure}

By  mid-September the team had reassembled in Orsay, ready to begin the new line of experiments. 
\PM \ remembers the AdA period, which stretched from July 1962 until the final runs in '63 and '64, as {\it fievreuse}, feverish. Both the French and the Italians knew they were on the verge of a potentially epochal breakthrough  in accelerator physics. 
As Touschek has said in February  1960, in Frascati, when he   proposed the construction of AdA,   they were not alone in the world to think of electron-positron collisions: 
``It is an experiment which [Wolfgang] Panofsky plans to do: we must arrive there before him".\footnote{Minutes of the Frascati meeting, February 17th, 1960.} They knew   they could not waste any time on the face of the competition from  the other side of the Atlantic, by the American teams of which they were all  aware and probably frightened. The outcome was not assured either, namely to demonstrate that collisions had taken place was not a given, as the discovery of the {\it Touschek effect}  in   winter of 1963 would very soon show.\footnote{When exactly the team discovered the effect, which will be discussed later, is not well established. \CB \ mentions ``a night in 1963" in \citep{Bernardini:2004aa} and confirms it in \citep[70]{bernardini2006fisica}, where he mentions it as {\it una famosa notte del 63}, a famous night in 1963. Since the paper was certainly written by mid-March, it can be expected that the ``night" was some time in January or early February1963. }


Once in Orsay, as soon as the experimentation started, Touschek and Bernardini understood that   a unique opportunity had been dealt to them. The enthusiasm rose, while  the  hopes every run entailed brought  a further step in understanding the working of the accelerator. The Italian team, scientists and  technicians,  would come for a series of runs, which took place during  week-ends,  
when the other French users of the LINAC would let    them   the use of the  beam and they could   experiment injecting  the electrons in   AdA. No personal  testimony by Touschek  for such period is available so far, except   for a well known drawing, the {\it Magnetic Discussion} which may be attributed to this period, according to Carlo Bernardini in {\it Fisica Vissuta} \citep{bernardini2006fisica}
and which we show in Fig.~\ref{fig:mag-discussion}, 
together  with a photograph of \PM\ and \JH, possibly dated  1963.

\begin{figure}[htb]
\centering
\includegraphics[scale=0.8]{
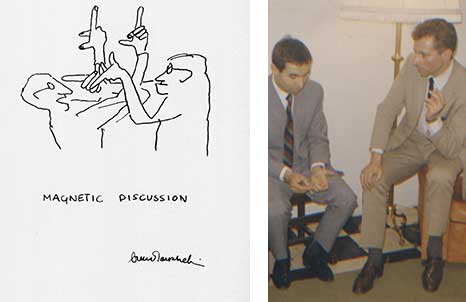}
\caption{It was never fully clear whether the beam first circulating in AdA was of electrons or positrons, infinite discussions would leave no definite answer \citep{bernardini2006fisica}, and Touschek recorded them in the drawing shown in the left panel, reproduced from \citep{Amaldi:1981}. At right an image  of Jacques Ha\"issinski and \PM \ from a photograph taken by Yvette Ha\"issinki, courtesy of \JH.}\label{fig:mag-discussion}
\end{figure} 

 A detailed  record  of the results was kept by Touschek, in two large notebooks, where he reported all the calculations he was doing to understand the behaviour of the particles in the beam, and the many hypotheses and checks following the measures. A copy of the first AdA logbook exists, in  Bruno Touschek's papers (BTA), together with the original of the second logbook, a further notebook was unfortunately lost.\footnote{The original  of the first AdA logbook, with starting date 4 April 1960, is in possession of Touschek's son Francis Touschek, with the  full copy to  be consulted in  BTA (Box 11, Folder 89), together with the Notebook ``Ada II'' (Box 11, Folder 91) and a smaller Notebook (``Quaderno di AdA'', Box 11, Folder 90).}  To these notebooks,  we shall often refer to pinpoint some dates through this chapter.

 \noindent From { \FL}'s interview in Orsay, May 2013:\footnote{
 Extracted from recorded interviews for the  movie \href{http://www.lnf.infn.it/edu/materiale/video/AdA_in_Orsay.mp4}{Touschek with AdA in Orsay} by E. Agapito, L. Bonolis, and G. Pancheri. Interviews were recorded at \LAL,  in Orsay, May 2013. In particular, see video-cassette  ``Lacoste ACO desktop sequences".}
\begin{quote}
\begin{small}
I participated in   the start of AdA in Orsay, we were in a small room which doesn't exist anymore. 
We were using the electron beam of 500 MeV station  to produce gamma rays which were targeted inside  AdA to produce electrons and positrons. The storage rate was quite small so  we needed about 48 hours to have sufficient number of electrons. You could count electrons one by one  by the differences  in the light {[emitted by the accelerated electrons]}. 
\end{small}
\end{quote}

\noindent
Another testimony comes from {\JH}'s interview in May 2013. The interview took place in the so-called {\it Igloo}, a round building where the original ring ACO is maintained as part of the museum Sciences ACO. 
 ACO stands for Anneau des Collisions d'Orsay. It was a circular electron positron storage accelerator, which was designed and built in the years '63-'65 under the supervision of Pierre Marin. Its design benefited particularly  from the experience that the AdA team  got with AdA. \JH \ remembers:
 \begin{quote}
 \begin{small}
 When I joined the AdA collaboration, AdA was already installed in the experimental hall,  a few  tests  had been made. First thing I was in charge was to try and upgrade the line which transfers the electrons between the linear accelerator and the ring in order to increase the injection rate. So that was my first work. The injection rate was much higher than what it  was in Italy.  But it would  still  take many hours, between ten [to] fifteen hours to fill the ring  with electron or positrons. One  has to remember that the goal of bringing AdA to Orsay was to try and to  study   collective effects, effects that  take place when there are many particles which are in the beam. In particular the main goal was to check that the  two beams, the electron particles and the positron bunch,   were crossing and overlapping precisely in order to get the maximum  rate of collisions. And that took two years, two and a half of experimentation here at Orsay. The so called  data taking runs  used to take place during the week-ends, they would start on Friday evening and end up  next Monday morning, which means that we were working sixty hours in a row, non stop. And of course this was a bit stressing and tiring, but I think we were all very enthusiastic and most of the members of the team would stay in the control room most of the time. Sometimes of course there was some possibility of sleep far away, but most people were still there most of the time. 
  \end{small}
 \end{quote}
\noindent
Two photographs from those times are shown in Fig.~\ref{fig:JH-Corazza}.}
\begin{figure}[htb]
\centering
\includegraphics[scale=0.21]{
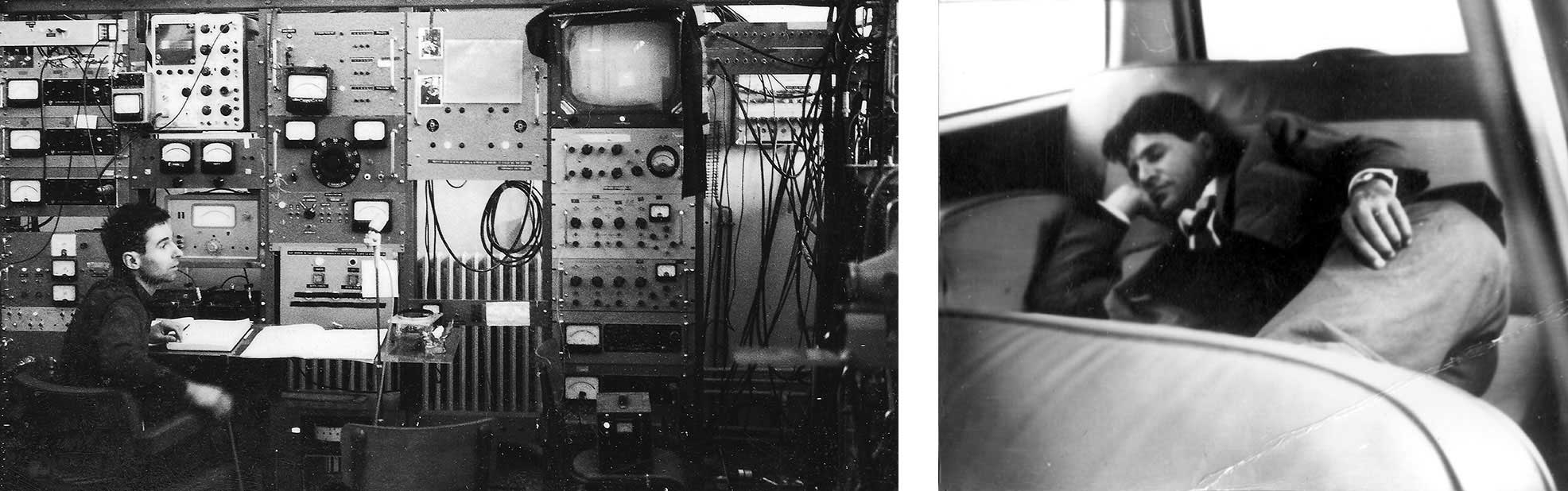}
\caption{
\JH\  working at AdA's control station in Orsay, at left (photo courtesy Jacques Ha\"issinski),  and, at right, Gianfranco Corazza, taking a temporary rest,  sleeping in the SIMCA car the Italian team used to rent, in Orsay, courtesy of G. Corazza.}
\label{fig:JH-Corazza}
\end{figure}
As soon as experimentation started, the injections of electrons and positrons in the AdA ring with  the LINAC, which was a much  stronger source than the Frascati synchrotron,  had showed a marked improvement,    and   raised high hopes  for the next step in experimentation. The technique used was the same as in Frascati, namely, electrons and positrons were first generated inside the doughnut, by a photon hitting an internal target. Then one type of charges, say the electrons, would be bent by AdA's magnetic field  into  the expected circular orbit, while the other, oppositely charged, particle would just be wasted away. Once a sufficient number of, say,  electrons had accumulated, AdA would be  flipped over like a pancake \citep[21]{Haissinski:1998aa}, namely rotated around  a horizontal axis and the magnetic field would change direction. It would then be  the  turn of the electrons to waste away, while the positrons  would now be directed to move   into the same circular orbit as the electrons, but circulating in the opposite direction.

This effect is the one represented graphically by Touschek's  drawing of Fig.~\ref{fig:mag-discussion}, which  represents the classical law of ``the left hand", which established how a magnetic field acts on a charged particle moving perpendicular to it. The mechanism  flipping  AdA, turning it upside down,  was called  {\it il girarrosto}, in Italian, the  roasting spit in English.   {\JH} \ comments:\footnote{Orsay 2013 interview, 
as mentioned.} 
\begin{quote} 
\begin{small}
As I said, it used to take a long time to store the particles, and of course during that time Bruno Touschek was monitoring very precisely the rate at which particles were stored and when he felt that the beam was not up to the best possibility, I remember that he would use a watch and put numbers, checking the rate. 
\end{small}  
 \end{quote}
 \noindent
Was Bruno repeating a routine followed twenty years before, in the M\"uller factory, in Germany, during the dark years of the war, when  trying to get Wider\o e's betatron work? 
{ \JH}\ continues:
\begin{quote}
\begin{small}
After a while we had to understand that something had to be done because he [Touschek] was getting more and more unhappy because probably the people operating the accelerator were falling  asleep or at least dozing and then I would run to the other control room to the linear accelerator, which was 100 meters away, and spend, sometimes, hours trying to get the engineers or the technicians motivated to get the   best  possible beam  that   AdA  could  benefit from. 

During the year and a half in which the data was taking, it was not always easy, there has been a number of surprises, and in physics surprises are difficulties to  overcome \dots When the ring was flipped over to change the injection from electrons to positrons, sometimes, the beam which was stored was lost, it took us a while to understand what was going on. By sheer  luck Bruno Touschek and also Pierre Marin, I remember,  both checked what was going on and they noticed that when they were flipping the ring, there were some magnetic dust  particles within the doughnut which were   falling across the beam  and  which would eliminate all the particles. 
\end{small}
\end{quote}
\noindent
This difficulty with losing the beam when changing electrons with positrons during the injection, is echoed in {\FL}'s words during the   already mentioned 2013 interview in Orsay:
\begin{quote}
\begin{small}
The long storage time was a problem  because we also would lose electrons now and then,  either through short stops of the electrical supply by the EDF [Elecricit\'e de France] which was not  stable or also when we flipped from the electron to the positron side  and the small powders could fall and would destroy the beam. You had to start all over again and I  remember a technician of the Frascati [team] who would tell  \dots  every time we lost the beam: ``elettroni tutti morti". And this I still remember because it was very frustrating. 
\end{small}
\end{quote}
\noindent
Then as Bernardini  \citep{Bernardini:2004aa} also mentions, the puzzle could be solved when 
\begin{quote}
\begin{small}
\dots  by sheer chance, Bruno
and Pierre Marin  were looking through the porthole in the donut searching
for the malefic elfs who were destroying the beam  and found them in the form of
fluttering diamagnetic dust left over from the welding of the donut and moving under
gravity along the magnetic field lines and passing through the beam.
\end{small}
\end{quote}
\noindent
Giorgio Ghigo is quoted by \PM \ for having suggested  the  solution which was finally adopted: Marin  calls   it  {\it la parade}, the perry, in  a sport-like term which clearly reflects the spirit of the team who was fighting a battle against time, to  be the first to see the particles collide.  

 Giorgio Ghigo had   the  brilliant intuition that the same effect  of inverting  the direction of magnetic field (by flipping AdA upside down) could be obtained   by  changing the direction by which the positrons would enter it! To this effect, after the electrons had already been stored in the ring, AdA was first translated away, then rotated by 180 degrees around a vertical axis, and finally translated back to the point of arrival of the LINAC beam, so that the relative orientation of the positron velocity and the magnetic field would now send the positrons to be bent so as to enter the same orbit  as the electrons, but in opposite direction. 
 {``For obvious group-theoretical reasons\dots'', commented Touschek \citep[5]{Bernardini:2003aa}.\footnote{As he wrote to his parents in June 1943, he had studied group theory because  he would have liked to definitely dominate this area as completely as possible after the war.} Gianfranco Corazza found a clever solution to realize the new scheme proposed by Ghigo. With the help of Antonio Marra he had a translating platform built allowing the two faces of the target to be exposed to the LINAC beam, so that  the magnetic dust, the ``small powder" remained stuck to the bottom of the donut. } It was, as \PM \ says, a very elegant solution. Later on, when discussing storage rings, and AdA's injection mechanism  in particular,   in his presentation to the 1963 Spring Meeting of the American Physical Society, Gerard O'Neill would admiringly call  it simple, but very efficient.
 
\begin{figure}[htb]
\centering 
\includegraphics[scale=0.45]{
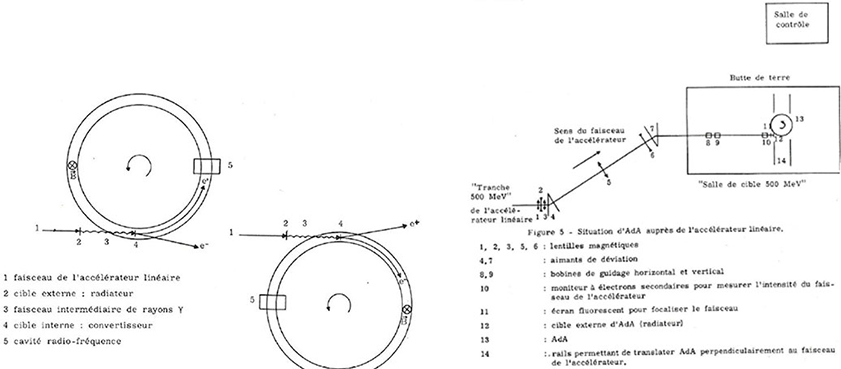}
\caption{Figures 4 and 5 from from \JH's thesis show  the positioning of AdA with respect to the linear accelerator and the translating platform at right, with mechanism of injection of electrons and positrons,  at left. }
\label{fig:JHinjection}
\end{figure}

A description of the injection scheme adopted in Orsay is shown in Figs.~\ref{fig:JHinjection} from \JHT. 

\begin{figure}[htb]
\centering
\includegraphics[scale=0.35]{
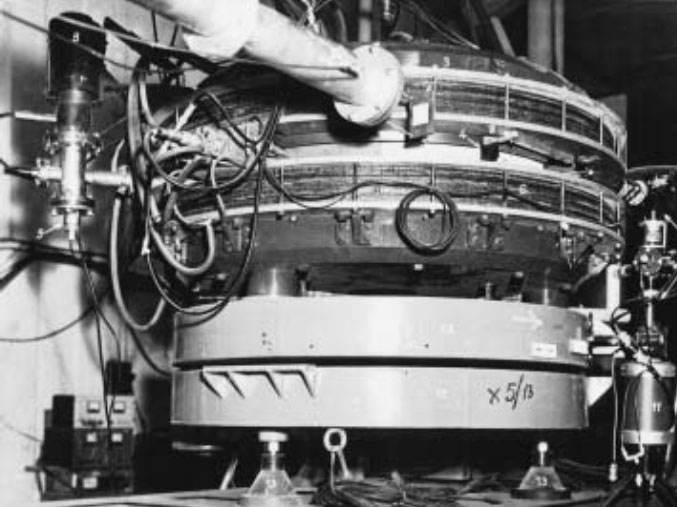}
\caption{AdA on the rotating and translating platform at Orsay. On the left the injector beam channel. Courtesy of Jacques Ha\"issinski.}
\label{fig:AdA_platform}       f
\end{figure}

The new arrangement for AdA is visible in Fig.~\ref{fig:AdA_platform}.
In this interview, {\JH}  further remembers:
 \begin{quote}
 \begin{small}
Another difficulty that we ran across, not a major one, nevertheless it took some time to  repair it, was that the detectors which were used to look at the annihilation or the collision at least between electrons and positrons, they were big blocks  of lead glass which used to detect the \v Cerenkov light produced by the photons coming from the beam collisions. Because these detectors were kept in the experimental hall where  the primary  beam from the LINAC was coming in, there was very high radioactivity  and this radioactivity progressively blackened totally the two \v Cerenkov, so we had to restore the transparency of these blocks and this took some time. That was not a major problem, but nevertheless it was one of the hurdles we had to overcome.\footnote{The solution was to heat up the \v Cerenkov detector, a heavy glass cylinder, just below the softening temperature, and then gradually cool it to avoid internal tensions that might create fractures. The \v Cerenkov was brought to Frascati and Corazza took care of this delicate procedure which took about 48 hours \citep{Corazza:2008aa}.}

The data taking runs, all started on Friday evening. The very first measurement which had to be done was to calibrate the instruments,  that is to measure carefully the  intensity of the particles which were circulating in the beams. This was essential, since what we were looking for were collective effects,  which  depend on  the `number' of particles stored;  there were perhaps hundred times higher than what could be achieved in Frascati. And during this period I remember Bruno Touschek and Carlo Bernardini were watching us very carefully  all the time and taking notes on the log book. 

Bruno Touschek would rather stay away from the knobs, the hardware, perhaps he was dissuaded from doing that,  because he was so quick in changing things that people did not want him to interfere too much with the program. But he  never missed any one  of these runs in Orsay, he was always there from the beginning to the end, always had  input making plots and \dots  always putting things forward.   It was essentially Ruggero Querzoli the senior member of the collaboration  who was telling him to leave us continue the program, which was planned. Sometimes of course Bruno Touschek would get a bit impatient, but most of the time he had some smile on his face, because he had a very good sense of humour, he always saw things which were very special in his own way and he always had some  interesting remarks to make  about what was going on. 
\end{small}
\end{quote}   
 \subsection{Italians in Paris}
 \label{spec:italians}
 The arrival by plane of the   technical core of the AdA team  is humorously recalled by one of the Frascati technicians, Mario Fascetti, in 
  the docu-film  \href{http://www.lnf.infn.it/edu/materiale/video/AdA_in_Orsay.mp4}{Touschek with AdA in Orsay}.\footnote{For the Italian  version see \href{http://w3.lnf.infn.it/ada-anello-di-accumulazione/}{Touschek con  AdA a  Orsay}.} 
 Fascetti   recalls their arrival at  Orly,  with many   baggages, and reminisced that Giorgio Ghigo, upon arrival,  rented a Renault for the drive to Orsay. But, as  Ghigo started the engine of the Renault,  emotion and inexperience gave  way: he  engaged  the car in the third gear, the car started and  immediately jumped to a violent stop.     The   luggage, not so well secured on the top of the car, fell off  rolling    on the highway. Luckily, as Fascetti says, nothing happened. 
 
As for the other members of the team,  
 Carlo Bernardini in a 2013 interview in Frascati remembers:
 \begin{quote}
 \begin{small}
 We used to fly to Orsay from Rome all the  week-ends. The colleagues, our friends in Orsay, dedicated all their week-ends to us,   the families were not so happy about these absences during the week-ends. 
 \end{small}
 \end{quote}
 \noindent
 They would leave Rome on Friday evening and return either on Sunday or Monday night. Bernardini, who had teaching obligations, would take the evening flight  back  to Rome, a Viscount aircraft,  who would then continue towards Bangui and Brazzaville. Touschek preferred  the train, and take the  overnight trip between  Rome Termini Station and  the Gare de Lyon, in Paris. 

\begin{figure}[htb]
\centering
\includegraphics[scale=0.49]{
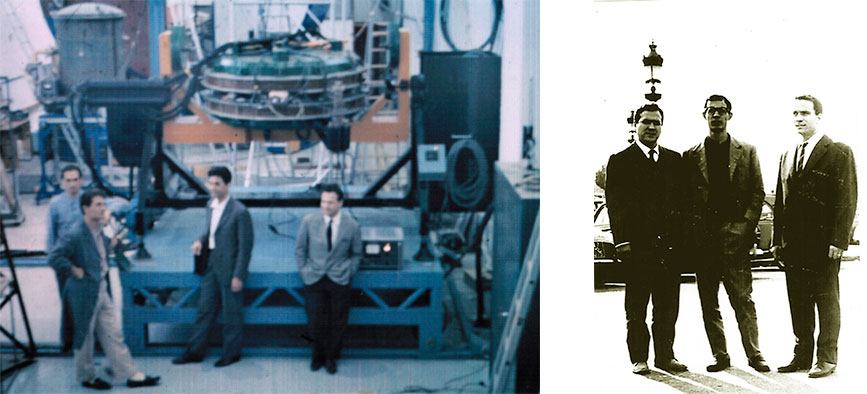}
\caption{Members of the Italian team  in Orsay in 1962, with  the left panel showing Giorgio Cocco, Bruno Ilio, Gianfranco Corazza and  Angelino Vitale  with AdA in the background, while the right panel shows Angelino Vitale, Peppino Di Giugno and Mario Fascetti, courtesy of Mario Fascetti, appearing at the extreme right. }
\label{fig:techs-in-orsay} 
\end{figure}

In those days, the cultural differences between the two teams were much stronger  than those we can see now. \FL\    reminisces:
 \begin{quote}
 \begin{small}
  The organisation of the Frascati team was much more military than for us: the technicians would call Touschek and Bernardini ``Dottore", which was not at all our custom,  as we would, all the physicists between them, and the technicians,  all would call also us by our last names.
  \end{small}
 \end{quote}
\label{ssec:italians}
The Italian team  included senior scientists  --- Bruno Touschek, Carlo Bernardini, Giorgio Ghigo, Ruggero Querzoli and Gianfranco Corazza --- as well as a young post-doc, Giuseppe (Peppino)  Di Giugno, and four  technicians: Bruno  Ilio, Angelo (Angelino) Vitali, Mario Fascetti and Giorgio Cocco.  While most of the scientists had travelled  abroad to conferences and the like, the Paris transfer   was  a wholly new  adventure for the young Peppino Di Giugno and the technicians.\footnote{Di Giugno would later leave physics for electroacoustics and digital sound,  and  work with 
{Pierre Boulez,  Luciano Berio,  Robert Moog.}} One of them, Mario Fascetti, remembers that, before leaving Frascati, the Laboratory director, Italo Federico Quercia, bought for them tickets for a show  in one of the famous Parisian strip-tease theatres, the {\it Caf\'e Mayol}: not quite the Moulin Rouge, but still enough for the young technicians to enjoy something not existing in Italy, at the time. Fascetti kept the (cancelled) ticket to this day, as a memento of the greatest adventure of his life, together with photos taken in Orsay and Paris, two of which, dated as of July 1962, are  shown in Fig.~\ref{fig:techs-in-orsay}.
Among his memories, one example of the enthusiasm and drive of those days is worth remembering. In an  interview  recorded in June 2013,  Fascetti recalls a week-end when the whole experiment  almost failed:
\begin{quote}
\begin{small}
I was always available to make sure that the high frequency apparatus would be working in all its parts. But, once, an accident happened. It was Friday. We were allowed to use the linear accelerator only on Saturdays and Sundays. Next to the radio frequency station, there was a tall stack of heavy co-axial cables, two or three meter high.   I was in the laboratory, but not there, and  someone passed by, brushed against the stack, which broke,  and all the cables, kilos and kilos of it, fell from high    on top of 
the radio frequency apparatus,  damaging  it heavily. It was panicky. That meant to stop  experimentation with Ada and fix it. But how? We could only work on Saturdays and Sundays [with the beam], and I was the only one who could fix it, as I had worked to  build it [in Frascati].  I tried a call to Rome, and speak with my boss, ingegner Puglisi, but could not find him. Then, it came to my mind that I should be able to do it, by myself: `I have built it after all', I told myself. So I asked one of the French scientists, either Marin or perhaps  Lacoste, and asked for two of their technicians to help me and repair the damage. This was a very difficult  thing to do, since, as I remember well,  the French personnel used to disappear as as soon as their work hours were off, at 4 or 5 in the afternoon. But somehow they were able to convince two people \dots That night we worked and the morning after the accelerator was working and available to all the experimenters.
\end{small}
\end{quote}   
\noindent
The beauty of the {\it Ville Lumi\`ere} did not escape the attention of the young technicians of the team, but they saw its marvels with a special twist. When asked by Giorgio Ghigo what they thought of the Eiffel tower they had just visited, they rapsodied about the iron bolts which kept it together.\footnote{Andrea Ghigo's memory in \href{http://www.lnf.infn.it/edu/materiale/video/AdA_in_Orsay.mp4}{\TwAO}.}

\begin{figure}
\centering
\includegraphics[scale=0.17]{
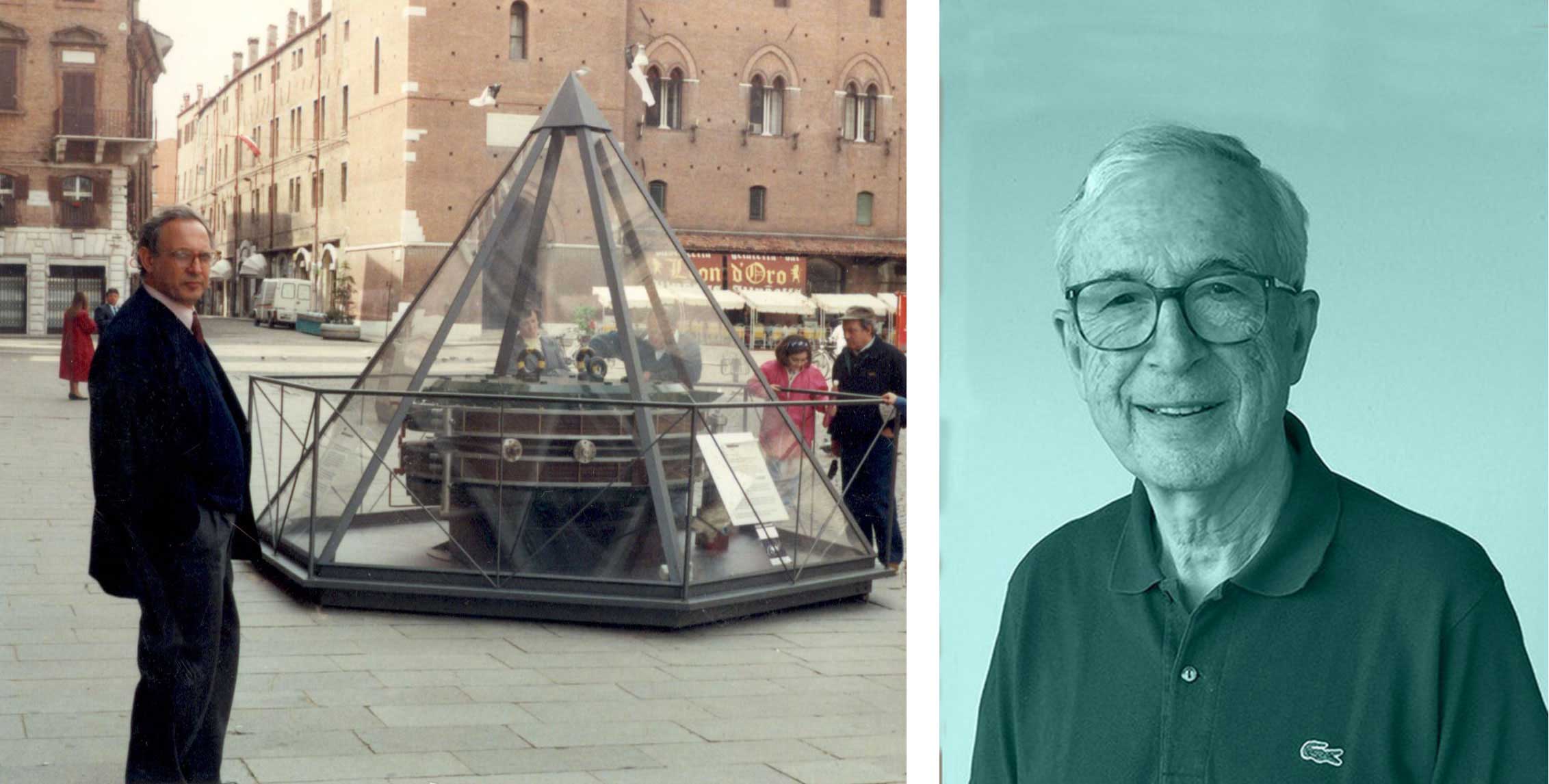}
\caption{ Giuseppe, called Peppino, Di Giugno with AdA  in Ferrara in 1991,  and a recent photo of \FL. Di Giugno's photo is courtesy of the author, \FL's is from  \url{https://www.deasyl.com/english/.}}
\label{fig:Digiugno-lacoste}
\end{figure}

  In a 2013 interview, in Frascati, Di Giugno also reflects on his experience and compares it with present day high energy physics research, striking a note which had already resonated in \FJ's last thoughts, and would be present in Touschek's, as well, at the end of his life, when he was at CERN in 1978:
\begin{quote}
\begin{small}
I always say that I consider myself to have been extremely lucky to have been part of the AdA group. To be part of such group  is one thing which can happen only once in life \dots Around 1975 I left high energy physics, mostly because it had become like a factory, and today I could not work at CERN with two or three thousand people. 
\end{small}
\end{quote}
\noindent
Paris reserved many surprises to the young Italian, as when he recollects the time he visited  his  friend  Fran\c cois Lacoste at home:

\begin{quote}
\begin{small}
One evening, one of the French collaborators, Fran\c cois, invited me for dinner at his home. He arrived with an enormous car, I forgot what it was, a Cadillac or a Bentley, perhaps a Rolls Royce. He lived in a palace on the Champs \'Elis\'ees. I was stupefied: more than a palace, it was a museum, with columns, waiters, paintings right and left. While we were eating, I was looking around, and saw an embalmed crocodile on the floor, and all around us paintings of crocodiles, and also a marble crocodile. I asked him ``how come you have so many crocodiles in this house?" and he answered that this was the house symbol.

I had clearly not known  who  was our friend Lacoste!
\end{small}
\end{quote}
In Fig.~\ref{fig:Digiugno-lacoste} we show photos of the the two friends in later years.\footnote{Di Giugno's photograph was taken during the INFN public exhibition {\it Dai quark alle Galassie}, held
in Ferrara, April 13- May 4, 1991.}


\section{The winter of 1962-63:}
\label{sec:touschekeffect}
 \FL \ helped in installing  AdA and collaborated during the first runs of the machine, but soon he left particle physics to work in aerospace industries. 
 The team missed him, but in the meantime   \JH\ had joined the AdA group.  As  Bernardini says, ``luckily   we  could count on Jacques'', 
 who  took over with great dedication and enthusiasm. His {\it Th\`ese d'\'Etat} at the end of three years of experimentation with  the only existing electron-positron ring in the world  at the time, represents the best document of the great AdA adventure.  

By December 1962, the new arrangement for injection was ready for operation \citep[117]{Touschek:1963zz}. 
Bernardini was very proud of having invented a novel injection procedure including an amplitude modulation of the radio frequency during the now very short LINAC pulse, which increased the accumulation rate by no less than two orders of magnitude  \citep[70]{bernardini2006fisica}. At this point, the puzzle of the magnetic dust, which killed the beam --- {\it elettroni tutti morti} ---  when AdA was flipped over, had been understood and solved,  and the injection rate 
should have been  good enough to allow the team to address the measurement of the interactions between the beams.  Good enough, but not sufficient to see production of new particles, so Touschek proposed to look for a process, which had a higher probability to occur and hence needed a lesser number of stored particles. It was an old friend of Touschek, namely annihilation into two $\gamma$ rays, which he had considered from the very beginning, on the day he had first sketched his ideas about AdA.\footnote{Touschek had proposed that the process of annihilation into two $\gamma$ rays  be used a standard process against which to measure all the others, and it is conceivable that term luminosity --- because the two gammas would play the role of a unit such as the {\it lumen} ---  was invented by him. The concept of luminosity, namely how to calculate the rate of interaction in colliding beam experiments, was of course known, having been discussed early by G. O'Neill in his first works on storage rings \citep{ONeill:1956aa}.}

\begin{figure}
\centering
\includegraphics[scale=0.18]{
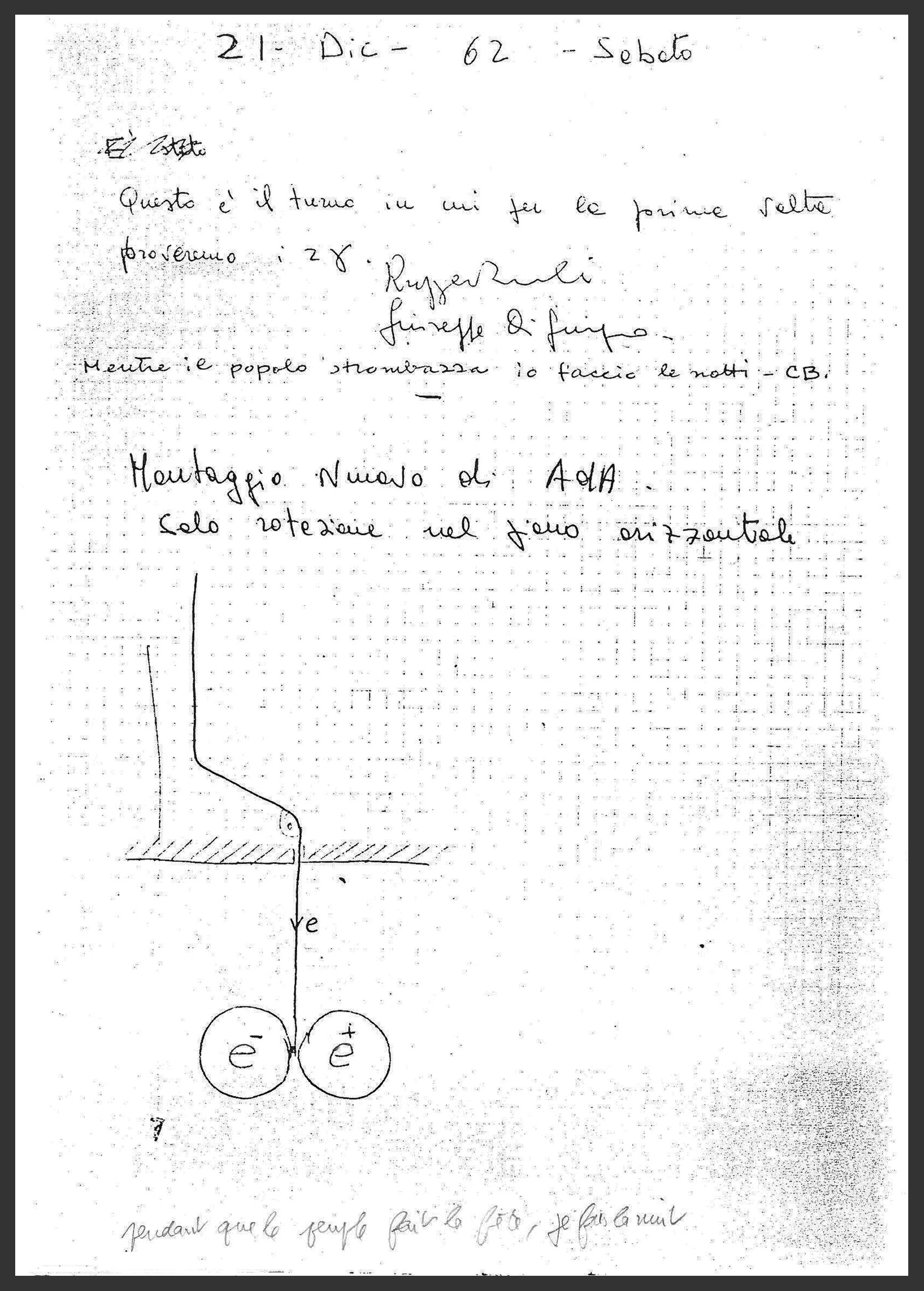}
\caption{
Copy from  AdA's log book,  kept in Touschek's papers at Sapienza University (BTA), saying
 that on December 21st, 1961 ``This is the run in which for the first time we shall try [to look]  for  $2\gamma$", signed by Ruggero Querzoli and Giuseppe di Giugno. This is  followed by a few lines by Carlo Bernardini (CB), 
 ``{\it While the people are trumpeting, I work through the nights}", and  a sketch of the new AdA set up with rotation only on the horizontal plane. At the very bottom, one can see the French translation of Carlo's lines, i.e. ``{\it pendant que le peuple fait la f\^ete, je fais la nuit}'', most likely by \JH.}
\label{fig:21dec1962}
\end{figure}

\subsection{Looking for two $\gamma$ rays}
On December 21st, 1962, the first run to look for $e^+e^-\rightarrow 2 \gamma$ started. We know the exact  date when experimentation in  search for the two photon annihilation process, started, from a page from Touschek's log book 
which is reproduced in Fig.~\ref{fig:21dec1962}.

\begin{figure}
\centering
\includegraphics[scale=0.45]{
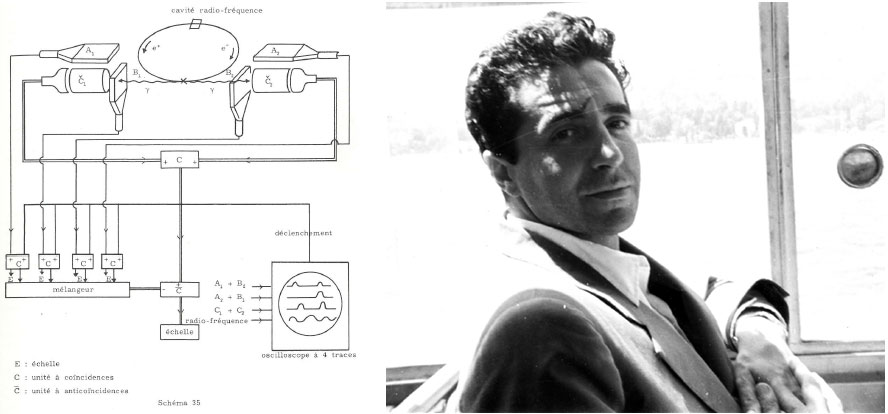}
\caption{The logical scheme for measuring the process $electron + positron\rightarrow 2 \gamma$, from \JHT. The $\gamma$ rays exiting AdA (center top) are depicted as two wavy lines labelled $\gamma$, to be detected by an assembly of electronics and detectors. The apparatus,  was constructed by Giuseppe Di Giugno and Ruggero Querzoli, who is shown here  in the right panel.}
\label{fig:2gamma}
\end{figure}

Experiments    in particle physics are a complex ensemble of hardware and software: there is a logical construct to identify the events one looks for, then the actual building of the electronic instruments which will collect the particle's signals, and the digital realization of the idea, fed into a computer program. It was 1962, nobody had yet built an apparatus for a colliding beam experiment, which could detect, two $\gamma$ rays, exiting the accelerator in opposite directions, in coincidence with each other and with the beams  circulating in the storage ring. The AdA team did it. They  could count on the  exceptional digital vision of Giuseppe Di Giugno, and on  Ruggero Querzoli's capacities as an experimentalist who had built one of the particle detectors for the  Frascati  synchrotron. The 2--$\gamma$ experimental apparatus for AdA was built in Frascati and assembled in Orsay, and was ready by December.  In the left hand side of Fig.~\ref{fig:2gamma}, we show the logical circuit devised by Peppino Di Giugno for the experiment.\footnote{Di Giugno personal communication to G. P., July 2018.} In the right panel, a photograph of Ruggero Querzoli.

 As we shall see,  no firm conclusions could be drawn from the measurements, as there were inconsistencies and contradictions between expectations and observations, although the number of stored particles appeared now to be sufficient to observe annihilation into two $\gamma$ rays. Touschek was concerned that this could be an effect of {\it radiation damping}, a phenomenon he was very familiar with since his betatron times.\footnote{ A trace of Touschek's work on radiation damping in a betatron done during the war, can be found in a letter to Arnold Sommerfeld from Kellinghusen, dated September 28, 1945 (Deutsches Museum Archive, Munich, NL080,013). A German version, ``Zur Frage der Strahlungsd\"ampfung im Betatron,'' is among Rolf Wider\o e's papers at the Library of the Wissenschaftshistorische Sammlungen of Eidgen\"ossische Technische Hochschule. An undated English version, written in G\"ottingen (The Effect of Radiation-Damping in the Betatron), apparently prepared for submission to the \textit{Physical Review}, is in BTA, Box 4, Folder 15. Interestingly, in the last two lines Touschek added: ``Thanks are due to Dr. Wider\"oe for discussions on this problem, which has been treated by the author during March, April 1945'', \textit{that is when he was Gestapo's prisoner}. A further, but slightly different version (Radiation Damping and the Betatron) can be retrieved at the Archive of the Defense Technical Information Center, \url{https://apps.dtic.mil/docs/citations/ADA801166}. This one is dated G\"ottingen, September 28, 1945, the same date of the letter written to Sommerfeld.} 
   However,  as it often happens,  it was not the spectre of well known phenomena  which bothered AdA, but  was a combination of totally new effects, which in a few months would  lead to Touschek's greatest insight and to the main legacy of the AdA group, the understanding of how to plan and build higher energies particle colliders. 
\subsection{The {\it AdA effect}}
The turning point of AdA's path towards becoming  a milestone in particle physics, took place in early 1963.
The process  of electron positron annihilation into two gammas had been searched  for, but the results of the experiment did not correspond to the expectations. There arose the suspicion that the volume of interaction was not correctly estimated and that the volume occupied by the electrons and positrons in their respective bunches could be larger than originally thought.\footnote{
For the same number of particles in each beam, the number of collisions  which can be experimentally observed decreases if the volume occupied by the particles increases.  }

As  the year 1963 came by and rolled into its winter months,  the team  started focusing on the beam life time and its size, and measurements of the correlation between the number of injected particles and the beam life time were  studied. This is when the new surprises arrived.  Let us hear the voice of some of the protagonists, through their interviews or written memories. 

\begin{figure}[htb]
\centering
\includegraphics[scale=0.21]{
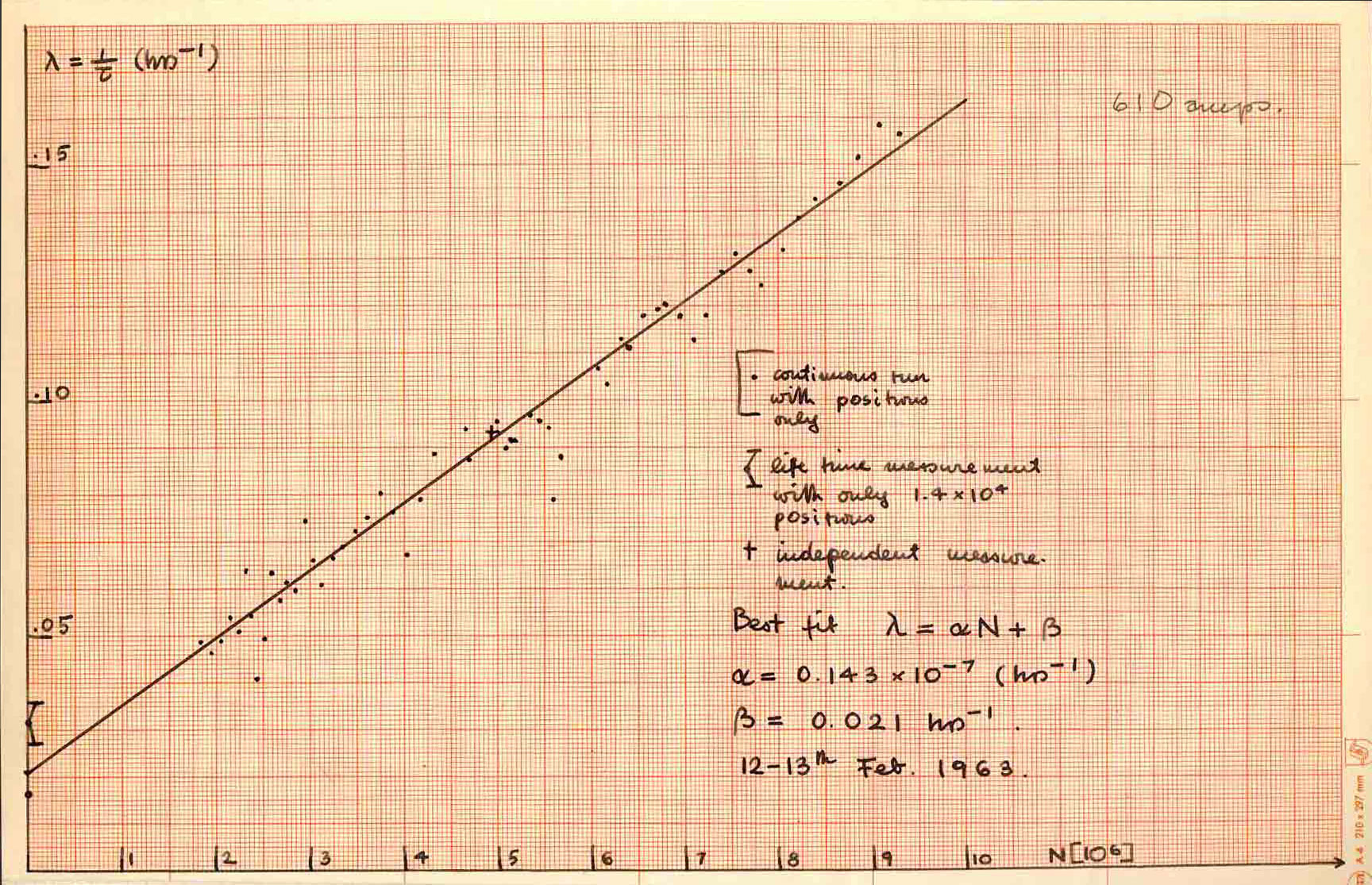}
\caption{The linear correlation between the number of particles in one bunch and the inverse of the lifetime, at the nominal AdA energy, in a plot by Touschek's own handwriting (Original document in Carlo Bernardini, personal papers). The data were reproduced  in Fig. 1 of \citep{Bernardini:1964lqa}.}
\label{fig:positroni-run}
\end{figure}

\begin{figure}[htb]
\centering
\includegraphics[scale=0.18]{
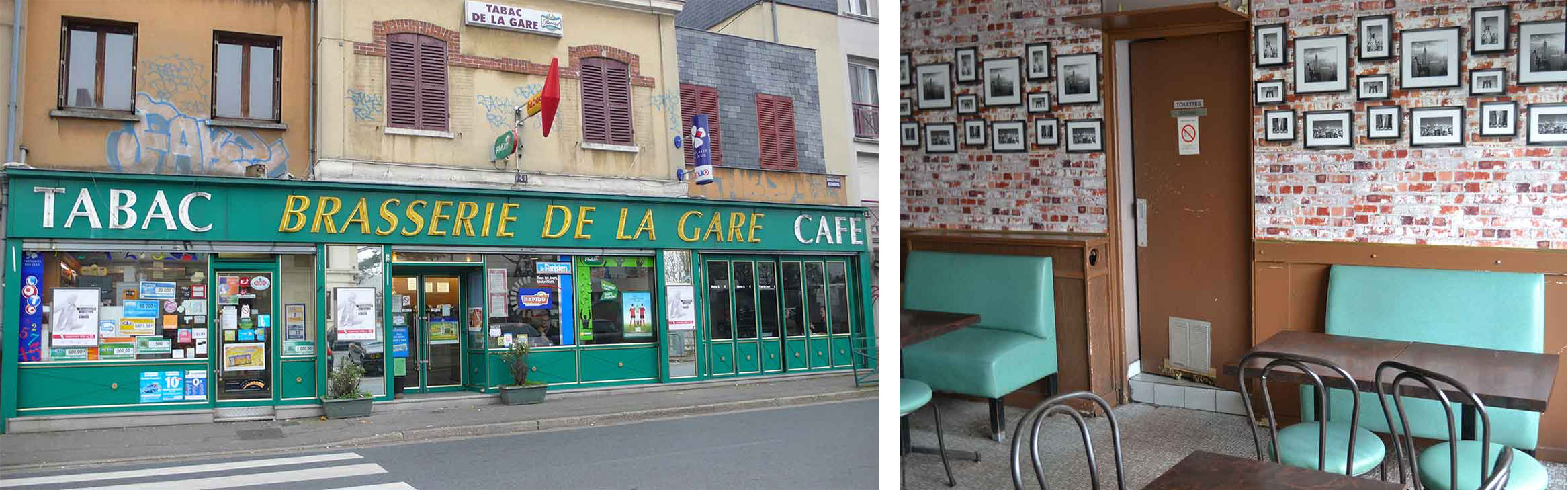}
\caption{The Cafe de la Gare, facing  the Orsay railway station, in 2013. }
\label{fig:cafe}
\end{figure}

\begin{figure}[htb]
\centering
\includegraphics[scale=0.45]{
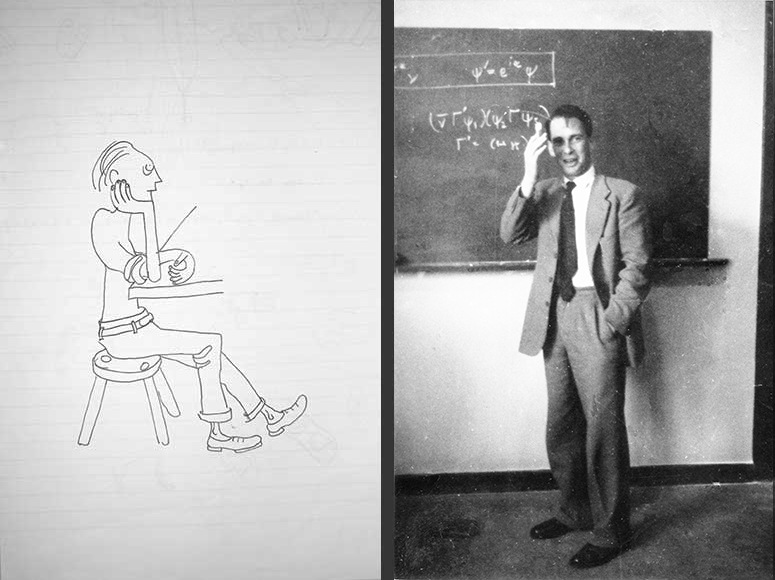}
\caption{Drawing by Bruno Touschek, date unknown, and his  photograph, dated  around 1960, both courtesy of Elspeth Ionge Touschek.}
\label{fig:pensatore}
\end{figure}

 They were in Orsay, like  every week-end, ready for an exceptional charge in AdA, a luminosity test in the best possible conditions.   
  The charge of the first beam began at dinner time. Everything was working. The injection speed kept constant during three-four hours. They had three monitors: a monitor showing the injection speed from the synchrotron light detected by the phototube, a monitor showing the LINAC intensity from the control room, a monitor for the gas pressure in the vacuum chamber measured by the Alpert vacuometer. Shortly after midnight they had $10^7$ electrons circulating in the vacuum chamber, a number compatible with the life time at the measured pressure, which was slightly over 10$^{-9}$ Torr. They were already dreaming to reach the fateful number of 10$^9$ circulating electrons which would allow them to observe some meaningful process. But as soon as they reached 10$^7$ stored electrons, the injection speed appeared to be conspicuously slowing down, as recalled by Carlo Bernardini:\footnote{Carlo Bernardini, personal communication to L. B., December 9, 2009.}
\begin{quote}
\begin{small}
{After four or five hours of steady charging, we saw, plotting the data, that some kind of saturation was beginning, ``as if the beam lifetime depended on the stored current''. This was the
immediate perception, confirmed by a simple plot.}
\end{small}
\end{quote}
{In Fig. \ref{fig:positroni-run} we show the plot of one such measurement, corresponding to a positron  run for 12 and 13 February.  }
Continuing Bernardini's quote:
\begin{quote}
\begin{small}
Bruno went crazy. 
\\Corazza and myself had immediately offered  what appeared to be the most trivial  explanation: such an intense synchrotron radiation is extracting residual gas from the walls of the donut. The intensity
of radiation depends on particle's number, therefore
the scattering lifetime depends too. Looked plausible;
but no change of pressure  was registered from the Alpert vacuometer, even if the
sensitivity was adequate. 
\\Bruno  became very pensive, but then we saw he had a flash of thoughts and said: ``I will think about it; try to make some measurements at lower energies of
AdA." We were working at 220 MeV.  He left and went to the Caf\'e de la Gare,  as usual when he was preoccupied. 
\end{small}
\end{quote}
Fig.~\ref{fig:cafe} shows two  recent pictures of the Caf\'e de la Gare, and  in Fig.~\ref{fig:pensatore}  a drawing by Touschek, perhaps reminiscent of that night, together with a photograph of Touschek, dated around 1960.

\noindent 
Returning to Bernardini's quote:
\begin{quote}
\begin{small}
He came back a couple of hours later, very excited, with a paper tablecloth full of calculations. It was nearly 4 a.m. in a cold morning. He cried: ``\textit{The beams are in a bath!}''
\\``A bath? Which bath?'', we asked. Touschek answered: ``Consider a long bath with three
orthogonal kinds of lateral walls: two are transversal
and infinite, one is longitudinal and finite. If a
swimmer is swimming in this pool, where can the water
 splash off the bath? From the \textit{bordelli}!\dots'' he said in
Italian, to mean the lower borders.
 \\He had calculated M\o ller scattering at different energies: ``Well, mister M\o ller is swimming in AdA!''
``What?'' We asked with one voice.
``Look'', he answered, ``here is the whole calculation \dots''.
\end{small}
\end{quote}
 \noindent
 What he meant, in his colorful way of representing the physical processes taking place within AdA, was that there was M\o ller scattering in the bunch, namely electron-electron scattering. He had understood that at the beam intensities they had reached, saturation would occur because electron-electron scattering  in the beam's bunches was transferring energy from the betatron oscillations in the radial directions into the synchrotron longitudinal oscillations whose amplitude is strictly limited by some stability criterion. Many particles were lost when the density in the beam had reached a relevant value. He had also calculated the energy dependence, showing that it was luckily decreasing with energy.
 \\Actually, as Touschek had suggested before leaving for the Caf\'e de la Gare, they had
checked the very unusual energy dependence of the effect:

\begin{quote}
\begin{small}
{The data were already there; we got more in the next few
days. There was an astonishing agreement between
measurements and Touschek's plot on the tablecloth of the 
Caf\'e de la Gare. Bruno had computed by hand all details but
asked me to check the scale factor, which I did immediately
in the late morning. Scale, exponents etc. were perfectly
working. The reason why Bruno was  calmer was that the
energy dependence did not anticipate a disaster for higher
energy rings like ADONE: this was a low energy effect.
But we had the opportunity to see it with AdA just because
the vertical size of the beam was much smaller than
predicted with multiple scattering.}\footnote{Text from slides used by Carlo Bernardini during a seminar on the Touschek effect to PhD students in Frascati Laboratories, December 2010 (L. B. personal papers).}
\end{small}
 \end{quote}
 \noindent
 { \JH} remembers well what happened that night:\footnote{May 2013 interviews in Orsay for the docu-film \href{http://www.lnf.infn.it/edu/materiale/video/AdA_in_Orsay.mp4}{Touschek with adA in Orsay}.}
\begin{quote}
\begin{small}
And then of course one surprise was the fact that the more particles were stored in the AdA ring, and the shorter was the lifetime. That is to say, we were losing the particles at a faster rate when \dots and this was totally unexpected and it's  Bruno Touschek who understood what was going on. It was a typical collective process which was taking place within the bunches, particles belonging   to the same bunches  were making oscillations, they are moving one with respect to the  other,  and some of these collisions are accompanied by a transfer of energy, there is some relativistic effect which  plays into this process  and at the end the two particles which have collided are just lost. 

And so, first, what Touschek understood is that it was an internal process like that and then,  quite rapidly during the night, after having understood that this was the basis of the process, he made the calculations and showed that this was exactly what was observed. So, later on, this took the name of the \textit{Touschek effect} and it was one of the major features which had to be taken into account in the design of the storage rings  which were built after AdA. 
\end{small}
\end{quote}
\noindent
Di Giugno, as well, remembers those fateful hours:\footnote{June 2013 interviews in Frascati  for the docu-film \href{http://www.lnf.infn.it/edu/materiale/video/AdA_in_Orsay.mp4}{Touschek with adA in Orsay}.}
\begin{quote}
\begin{small}
When Touschek then explained how it worked, it was amazing how simple the explanation was. With pen and pencil, he showed how this fact depended on energy, and there was one factor here and another there, and it was amazingly simple\dots
\end{small}
\end{quote}
\noindent
As {Bernardini} recalls, this was an example of what they used to call Bruno's serendipity:
\begin{quote}
\begin{small}
He had such a precise idea of what was happening in the beam, as if he could see the electrons with his eyes.
\end{small}
\end{quote}
\noindent
{The  {\it AdA effect}, as it was initially called, could have been devastating for the operation of the larger machine, ADONE. Luckily, what Touschek's calculations showed --- and measurements  immediately confirmed ---  the  effect diminishes rapidly with energy:
both ADONE and ACO appeared  to be on the safe side! ``Bruno was walking frantically from a wall to another in the Lab, like a billiard ball. He was terribly excited: it doesn't happen very often to be happy because one has understood the why of a misfortune\dots'' \citep[71]{bernardini2006fisica}.


{The results of the measurements and observations by the AdA  team during the months up to February 1963, can be summarized following   \PM. In \citep[49]{Marin:2009}, he writes:
\begin{quote}
\begin{small}
AdA's contribution  was of capital importance for Orsay, which did not possess any culture of circular accelerators whatsoever, but it was  also fundamental to the whole line of lepton storage rings which  followed.\\
Firstly, there is the effect, which reduces the life of the beam in accordance with the number of particles present inside the pack. Experimentally speaking, it is necessary to separate this effect from the one of the life-span due to the vacuum, because the pressure inside the vacuum chamber grows also with the synchrotron radiation which makes the molecules of the walls desorb from the walls. 
Also, one  needs to grasp the importance of the phenomenon of coupling between the horizontal and the vertical oscillations of the particles in the beam. It  turned  out that the beam was  20 times higher than anticipated, when calculating only the simple effect of the vertical emission of the synchrotron radiation. The importance of the 3 terms was finally understood and, in 1963, gave cause of the publication, in {\it The Physical Review Letters}, of the first collective effect ever observed in a lepton storage ring. It will then be known, by the world, as the  {\it Touschek effect}, an exchange of energy between two relativistic particles, animated by small radial oscillations and who are lost by pairs.
\end{small}
\end{quote}}

\subsection{After the discovery of the {\it Touschek effect}}
Soon after the discovery and the understanding of the  effect, the team prepared  the third article on AdA's experimentation which was submitted to {\it The Physical Review Letters}, on April 1st, 1963, where it was published just a month later: ``Eventually, we believe that the consistency of the data we dispose of at present is fairly good and encourages in proceeding further with storage ring machines since these data do not show any failure `in principle' of the main idea'' \citep[13]{Bernardini:1963sc}.

{At around  the same time,} Touschek wrote to Burton Richter in Stanford  to inform the Princeton-Stanford group of what they had found.\footnote{BTA, Sapienza University, Box 1.}

The Frascati and the American group had been in very friendly and collaborative terms at least since January 1962, when O'Neill had congratulated Ghigo and Bernardini on their initial successes with AdA, as we can see from the left panel of Fig.~\ref{fig:oneill62-3}.
At the time,
O'Neill had written: ``\dots many congratulations on your success in storing a beam with so long a life time. I look forward with much interest in hearing the details", concluding  with ``Very best wishes to all of you and we hope to hear more good news".\footnote{In 1962 O'Neill came to Europe, visiting both Frascati and Orsay, as from the unpublished report entitled {\it Summary of visits to Frascati and Orsay} \citep{ONeill:1962aa}.} The tone of this letter is very different from the one received by  Touschek after he informed Richter  in  March 1963 of the discovery of the {\it AdA effect}.  By  that time, the AdA team was  submitting  the article, which would be rapidly accepted and published in just a month. 

The answer Touschek  received  from Gerard O'Neill, on March 28th 1963,   was not encouraging.  It was rather cold, even  dismissive both of what the Franco-Italian team had discovered, namely the Touschek effect, and pessimistic of the value of AdA's future prospects, given the limits imposed by the new effect on the  luminosity at AdA's energy. The 1963 letter is   shown in the right panel of Fig.~\ref{fig:oneill62-3}.
\begin{figure}[ht]
\centering
\includegraphics[scale=0.136]{
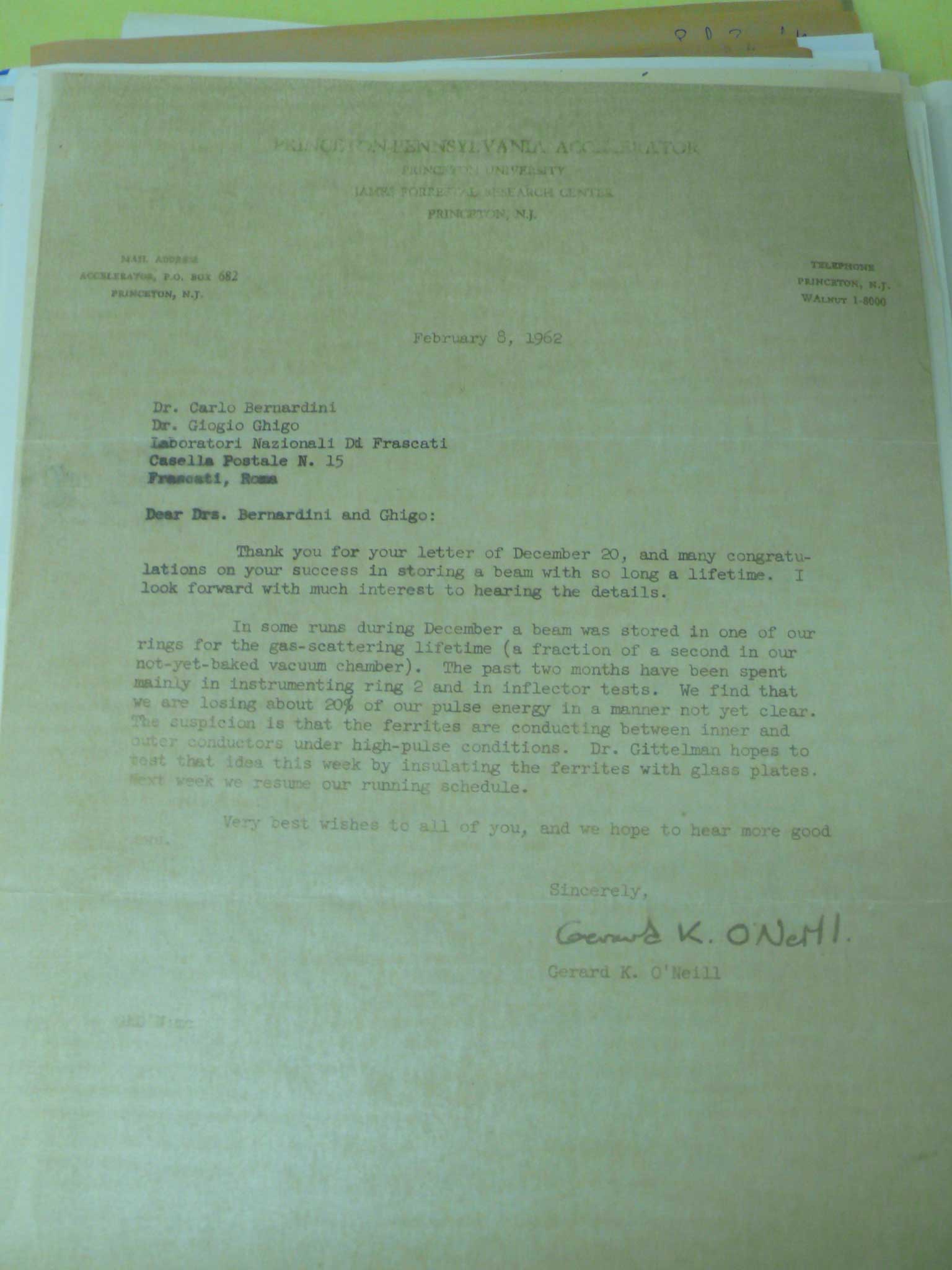}
\includegraphics[scale=0.136]{
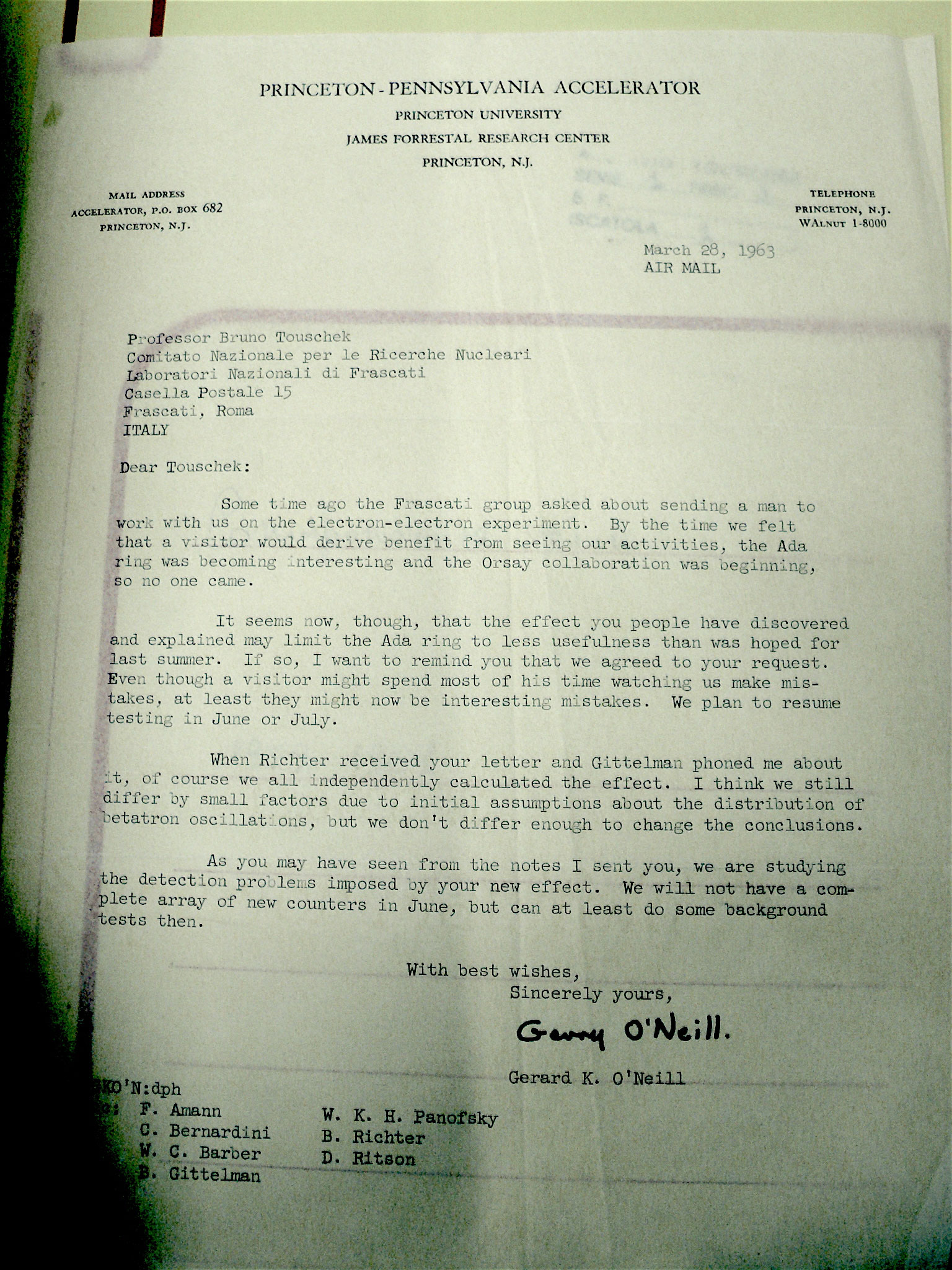}
\caption{Two letters from Gerard O'Neill to the AdA team. BTA, Sapienza University, Rome.}
\label{fig:oneill62-3}
\end{figure}

About the Touschek effect, O'Neill writes that ``It seems now that the effect you people have discovered and explained may limit the AdA ring to less usefulness than was hoped for last summer." And then he adds that they all calculated the effect, after receiving Touschek's letter, and basically agree on the conclusions. 

{While it is true that at this point AdA could not aspire to observe electron-positron annihilation into new particles, the importance of AdA's studies on beam dynamics and its impact on planning  future machines did not escape the American team. A reflection on both the above points can  be found in the talk which O'Neill gave at the end of April,  in  Washington D.C., at the Spring Meeting of the American Physical Society, 22-25 of April \citep{ONeill:1963ad}.\footnote{See \url{https://www.osti.gov/biblio/4875549-storage-rings}.} The importance of what the team had discovered is highlighted by the fact that, in this talk, O'Neill already calls it the {\it Touschek effect}, thus extrapolating it from the limitations of a small, low energy collider such as AdA. After describing how the Princeton-Stanford project was born and its present stage of operation, O'Neill opens a new paragraph by saying that ``In 1960, a group at Frascati, the Italian National Laboratory, became interested in colliding beams. They built an electron-positron simple storage ring called AdA [\dots] This group [\dots] has made  rapid progress and has passed some of the important milestones in colliding beam technique   sooner that we have. The differences are mainly in the areas of vacuum technique and the method of injection." About the method of injection, O'Neill is clearly impressed by the way injection is accomplished, he calls it an ``inefficient, but simple method" and admits that   the sacrifice in efficiency has the great advantage that nothing can go wrong and that its simplicity makes it possible for the vacuum in AdA to be good.  He acknowledges the  importance of the effect discovered by the Orsay-Frascati team,  and in his talk  the name {\it Touschek effect} appears, to our knowledge, for the first time. {Interestingly, O'Neill called it as the ``Frascati effect'' in an article published on \textit{Science} in summer 1963. In this article, O'Neill mentioned the Touschek effect as the third of ``three problems, not foreseen earlier, which will limit but (we now think) not prevent the carrying out of electron-electron colliding-beam experiments'' \citep[683-684]{ONeill:1963aa}.} 

\subsection{
{Single bremsstrahlung:  the calculation  that helped to  confirm that  collisions had taken place}}
\label{ssec:siglebremss}

O'Neill's March  letter, with its put-down of AdA's prospects,  had been  sent in copy to the Stanford  scientists, Pief Panofsky, Burt Richter and David Ritson, as well as to Bernard Gittelman, O'Neill's   Princeton colleague. It  had  also been sent to Fernando Amman and Carlo Bernardini, the Italian members of the ADONE and AdA team.
The large number of recipients, who would be reading the letter in copy, suggests   the attempt, by the US colleagues, to establish some official status on the  experimentation with the date  of March 1963, which could eventually   give credibility to Richter's later claim that AdA had been no more than a ``scientific curiosity" \citep[169]{Richter:1997aa}. 

Anyone would have been discouraged reading such letter, which  reflected  the opinion of so powerful  competitors:  not so Bruno Touschek and the intrepid Franco-Italian team. If AdA's luminosity was still too low to observe annihilation into new particles and even  annihilation into two $\gamma$ rays,  there was another process which had a higher cross-section and which could demonstrate that the beams would collide, namely elastic electron-positron scattering.\footnote{First attempts to observe the process $e^+e^-\rightarrow 2 \gamma$ had started  on Saturday, December 21st, 1962, but no conclusive evidence had been obtained \citep[17]{Haissinski:1998aa}.} But electron-positron scattering, namely $e^+e^-\rightarrow e^+e^-$ would be very difficult to distinguish from various  background processes. On the other hand, according to Quantum Electrodynamics, such a reaction is always accompanied by photon emission, the so called {\it bremsstrahlung } process which corresponds to the reaction $e^+e^-\rightarrow e^+e^- +\gamma$. The calculation of a similar  process --- bremsstrahlung from electron-electron scattering, namely for   $e^-e^-\rightarrow e^-e^- \gamma$ ---  was present in the literature, but 
 no calculation of the cross-section of the case for electron-positron  scattering was known to have  been done as yet: it had not been included in the {\it Physical Review} paper by Cabibbo and Gatto about electron-positron processes \citep{Cabibbo:1961sz}, nor in the  more recent    Ba\u ier's article, March-June 1963  \citep{Baier:1963}, which had followed it.\footnote{\url{https://doi.org/10.1070/PU1963v005n06ABEH003471}.} To prove that collisions had taken place, one needed not only an in-depth vision of the geometry and dynamics of the particle beams, 
 which the team had now conquered, like  no one else in the world.  They 
 also needed a check against the  theoretical expectations for the process.  It became clear that  one could record the correlations between the photons emitted by electrons or positrons and the intensity of the particle beams, and  the team prepared the first of the four runs which would finally establish the first observation of electron-positron collisions. These measurements would allow calculation of how many particles were produced when  
 the photon was recorded. But was this number in agreement with the expectations from theory? 
 This was a calculation which needed to be done, fast. And indeed it was done, between Rome and Frascati, as we shall see next. 


{It had in fact  happened that,  around that time, early 1963,  two  top physics students at University of Rome,  Guido Altarelli and Franco Buccella,  were close to finishing their course of studies and   had been   looking for a thesis in theoretical physics.\footnote{The course of 
{physics studies in Italian universities lasted four years} and led to a {\it Laurea in Fisica}, with the title of Doctor in Physics, 
{upon completion  of  an original  research article or of a  review paper, such that it could be published in peer reviewed international journals.} }  Early in 1963, Franco Buccella, under the advice of his cousin Luciano Paoluzi,  a young researcher at University of Rome,    approached Raoul Gatto, already highly regarded and soon to become Professor at University of Florence.  Gatto 
suggested Buccella  start looking, rather generally, 
 at the calculation of ``radiative corrections"  to electron positron annihilation, and directed him to study the first four chapters of the book by Jauch and Rohrlich, on Quantum Electrodynamics   \citep{Jauch:1955aa}.\footnote{Gatto may have been influenced by Touschek, who had studied    the  emission of radiation, during his betatron days. Touschek   knew that  it would influence the 
 extraction of physics from electron  machine performances, and was probably thinking of collective radiative effects, such as those he  had studied in Glasgow with Walther Thirring \citep{Thirring:1951cz}. He would take up again this problem in 1966, as it became crucial for ADONE's performance \citep{Etim:1967}.} As Buccella puts it:\footnote{Private communication by Buccella to G. P., May 9th, 2018.} 
\begin{quote}
\begin{small}
I started reading the book, but, after going through the first four chapters,  it was for me totally incomprehensible. I was in my fourth  year of physics studies, still had to finish a number of exams, and  I did not have any clue as to what I was supposed to calculate nor how  to start doing the calculation, so I postponed the thesis to concentrate on the exams.
\end{small}
\end{quote}
\noindent
 Around the same time, the other student,  Guido Altarelli,  had gone to see Bruno Touschek and asked for a thesis, in theoretical physics, 
as  his friend Buccella had done with Gatto. Altarelli either did not understand or was not pleased with what Touschek proposed, and began arguing with him. During the discussion, Touschek suddenly jumped up and rushed to a chair, where his jacket, which had been hanging next to a small heater, had started going on fire. Altarelli then decided this professor was not his favourite and turned to Gatto for a less exciting thesis adventure.\footnote{This episode, by private communication from Altarelli to G. P., in 2008, was published in the Italian journal {\it Analysis}, issue 2/3 - June/~September 2008,  in an article about \href{http://www.analysis-online.net/wp-content/uploads/2013/03/greco_pancheri.pdf}{Frascati e la fisica teorica: da AdA a DA$\Phi$NE}. Guido Altarelli later became  Professor of Theoretical Physics at University of Rome, and  Head of  CERN Theory group. He passed  away on 30 September 2015.}  }

 {The presence of a  heater in Touschek's office,  places the episode  in the winter. It may thus have been  February or early March 1963 when Altarelli went to see Gatto.\footnote{ In the {\it Analysis} article, which was written following a private communication with G. Altarelli,  there is mention of  the month of November (1963), but, following a conversation with Buccella in May 2018,  this date must be wrong. In fact,  the thesis was finished through  the summer of 1963, and the two students graduated in November 1963, as Buccella has recently confirmed.}
 
  In March, after the discovery of the Touschek effect,  the AdA team already knew   that annihilation processes could not be observed with AdA (not enough luminosity and too small cross-sections), 
  and Touschek, as we have seen,    turned  to the bremsstrahlung process to prove collisions.  
  {In the winter, 
 Touschek had  observed that while  two-$\gamma$ annihilation was out of question, events with photons had been observed and thought to be due to emissions of single or double photons accompanying the scattering of positrons against the electrons. If one could prove this, it would have been proof that collisions, if not annihilation, had taken place. 
 Touschek had made an order of magnitude estimate to prove that the cross-section for single bremsstrahlung was consistent with AdA's observation, but a more precise proof had to be done.\footnote{This was discussed by Touschek in his presentation at The Brookhaven Summer Study Meeting on accelerators, which took place in June 1963 \citep{Touschek:1963zz}.} In fact, it was being done, in Rome,  by the two  students of Raoul Gatto.

 And indeed, in early Spring,  after discussing the question with Touschek,   Gatto had   better focused  
  on what was needed for AdA. Not  having heard  from Buccella, Gatto  thought Franco had given up and, when Altarelli asked for a thesis, proposed him to do the calculation of the single bremsstrahlung.\footnote{Gatto was moving between Rome and Florence, and  was regularly in contact with Touschek, when  being at University of Rome, and may have been told by him  which process could be of use to help  in AdA's current plight to demonstrate collisions.}  Buccella and  Altarelli, 
 {whom we show in Fig.~\ref{fig:altarelli-buccella},} were friends since childhood, Altarelli    very tall and  lean,  with a sarcastic vein, Buccella  shorter and handsome,  more naive, and  gifted with a great sense of humour.  
 \begin{figure}
 \centering
\includegraphics[scale=0.30]{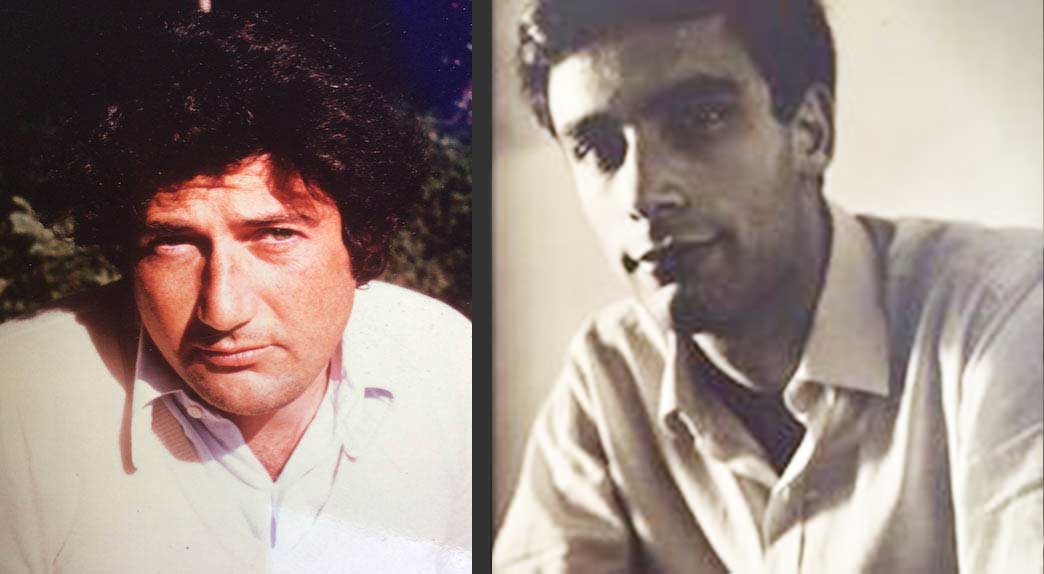}
 \caption{Left panel,  Franco Buccella at the time of his thesis, 1963, personal contribution, and Guido Altarelli, photo courtesy of Monica Pepe-Altarelli.}
 \label{fig:altarelli-buccella}
 \end{figure}
 
 Gatto told Altarelli that  a certain ``Buchella" had been in contact for this thesis, but had since disappeared. When Guido Altarelli was told by Gatto that a similar  calculation had already been ``offered"  to his friend, although Franco had not formally accepted it,  he did two things: firstly he inquired from other theorists how the proposed calculation could be done, then, loyal to his fried,  asked Franco to join him in the work.\footnote{Private communication by Franco Buccella, May 2018.} It was not exceptional  in those days for two  students to  work together on  a theoretical thesis, in particular  when the topic   implied original work, with long complicate calculations, in need of cross-checks. 
 
 Altarelli found out that  the  technique which would allow them to tackle the problem was  that of Feynman diagrams, and the two friends got down to work.\footnote{At the time, at University of Rome, it was customary to assign a theoretical thesis  without initially indicate to the student how to attack it. The students had to study the textbooks by themselves, i.e. the one by Jauch and Rohrlich as in Buccella's  case, and then basically start on a specific calculation, without initial guidance. If they survived this first  obstacle, they could go on with the thesis, and the professor's guidance. Very often, the student would ask the advice of young post-graduate theoretical physics researchers, and get some idea of where to start from.} 

\section{The summer of 1963}
\label{sec:summer1963}
{Summer arrived, and with it the summer meetings, following  the  well established tradition   to present the year's work in  various world wide venues. In just two years' time, storage rings had now become an important topic in accelerator physics, and two were the main events ahead of   the community of accelerator physicists,  the   first was a 5 weeks long Summer Study  in Brookhaven, in the USA from 10th June to  19th July,  and the second  the traditional accelerator conference to be held in Dubna, in the USSR, in August. The first near New York, the second  
not far from Moscow.

\subsection{Summer studies in Brookhaven}
Touschek was invited to lecture at  Brookhaven National Laboratory, in Upton, near New York, where a Summer Study on Storage Rings, Accelerators and Experimentation at Super High-Energies had been organized. In his talk, entitled {\it The Italian Storage Rings} he described the status of experimentation with AdA and then the situation of ADONE, the bigger project \citep{Touschek:1963zz}. 
{Touschek relates that although they had observed about 10 genuine beam-beam events, they did not observe annihilation into two gammas,  
 as the two \v Cerenkov counters did not signal the expected coincidences. Given the expectations, both for the theoretical cross-section and the registered number of particles in the beam, they had to conclude that the cross-sectional area  of the beams, the other factor  from which depends the luminosity,  was larger  than expected, being at least 20 times higher. He then put forward the hypothesis that the few observed events, the genuine 10 beam-beam events, were probably due to radiative electron-positron scattering. To substantiate his hypothesis, he gave an order of magnitude estimate of both single and double bremsstrahlung processes. He then stressed  the fact that ``Life time measurements are of paramount importance when considering the feasibility of coincidence experiments."}

{The question of the volume  of the beam is amply discussed, but it would seem only from a theoretical point of view, namely that if the dimensions had been of the {\it natural} type, one should have seen annihilation into two gammas. But since they did not see that, the observed gammas had to be radiative events, which had a bigger cross-section, and the culprit were the dimensions of the beam. He then proceeds to relate the life time measurements which had taken place that year. He says that  ``the first accurate measurements were made in February 1963". Various discrepancies about the life time are discussed and finally he says that such discrepancies can be eliminated if one assumes  the beam is considerably higher than ``natural" and that this increase in height is  due to  some unwanted coupling between radial and vertical betatron oscillations \dots} ``Accurate photometric measurements carried out by Marin and Ha\"issinski show a beam height of less that 150 microns. Measurements of beam height now in preparation involve the observation of  forward-scattered electrons instead of the annihilation reaction.'' \citep[197]{Touschek:1963zz}}

Gerald O'Neill was also present and made two different presentations  \citep{ONeill:1963ac} \citep{ONeill:1963ab}, where he recalled that the Princeton-Stanford storage rings 
of electrons against electrons
 ``was designed to provide a sensitive quantum-electrodynamic test'' for which the energy was chosen to be 500 MeV. He also remarked that a comparison with the AdA storage ring showed ``a rather fortunate difference in parameters which has made these rings complementary.'' 
Surprisingly, Lawrence W. Jones, 
who gave the summary talk, 
 did not mention AdA in the twenty-pages overview on the experimental utilization of colliding beams, not even in the brief historical introduction outlining US proposals and attempts before 1957 and ``main advances'' since then  \citep{Jones:1963aa}.

\subsection{Dubna and the new institute in Novosibirsk: {\it Une grande premi\`ere}}
\label{ssec:dubna63}
 
{The summer of 1963 is also} when the AdA team  learnt that beyond the Iron Curtain, and further away, beyond the Urals, in the Novosibirsk Laboratory of the Siberian branch of the USSR Academy of Sciences,  a team of  Russian scientists, led by 
Gersh Itskovich
 Budker,  unbeknownst to everyone in the West, had  started the construction of an electron-positron colliding beam accelerator.

This may have increased their sense of urgency, but there is no trace of this  in Bernardini's recollections of the occasion, until the year 2000, when Bernardini, who was writing a book,  asked \PM\ about the Russian activities.\footnote{Fernando Amman and Carlo Bernardini were among the 25 scientists invited to travel to the new Laboratory in Novosibirsk, but Bernardini's only mention of this trip is about a forced   stop-over of the plane in Tomsk  \citep[72]{bernardini2006fisica}}}  A posteriori, the Russian work did not directly challenge  AdA's primacy in proving that collisions had taken place, but later on it was  the Russian electron-positron storage rings, VEPP-2, and the French ACO which produced the first important particle physics results, when  they  were first to study the properties of  the so-called vector mesons, $\rho$, $\omega$ and $\phi$,  new particles which   added important mile-stones to  the understanding of the building blocks of matters, the quarks. 

But, let us proceed in order. After the publication of the paper on the Touschek effect, the team was invited to present their results at the summer conferences, which were now, and still are, the stage where new discoveries or important scientific developments are presented after the year's work. In 1963, the highlight of the conference season was the International Conference on High Energy Accelerators, which was held in Dubna, in late August, in the week from  21st to the 27th \citep{Kolomenskij:1965bnh}.  At this conference, work on storage rings, including electrons and positrons, was not an exotic, out of view item any more. Here is a list of some of the talks: 

 \begin{itemize}
  \setlength\itemsep{0.1 em}
  \begin{small}
\item [] \textit{On Particle Injection Into An Electron-Positron Storage Ring}, Yu.N. Metal'nikov, V.A. Petukhov
 \item [] \textit{Stability Of The Motion Of Accelerated Electrons In Storage Rings And Related Experiments},
F. Fer (Inst. Curie, Paris)
\item [] \textit{Transverse Space-Charge Effects In Synchrotron And Storage Rings}, F. Fer, C. Delcroix
\item [] \textit{Choice Of An Operating Point For A Positron Electron Storage Ring}, B. Richter, D. Ritson
\item [] \dots 
 \item [] \textit{Particle Dynamics In Storage Rings}, E.D. Courant (Brookhaven)
\item [] \textit{Effect Of Synchrotron Radiation On Colliding Beams In Storage Systems}, Yu. M. Ado (Lebedev Inst.)
\item [] \textit{The Cern Electron Storage Ring Model}, F.A. Ferger, E. Fischer, E. Jones, P.T. Kirstein, H. Kozlol, M.J. Pentz (CERN)
\item [] \textit{Lifetime And Beam Size In Electron Storage Rings}, C. Bernardini, G.F. Corazza, G. Di Giugno, G. Ghigo, R. Querzoli, J. Haissinski, P. Marin, B. Touschek \citep{Bernardini:1963xza}
\item [] \dots
 \item [] \textit{Recent U.S. Work On Colliding Beams}, L.W. Jones (Michigan U.)
\item [] \dots 
\item [] \textit{The Orsay Project Of A Storage Ring For Electrons And Positrons Of 450 Mev Maximum Energy},
Storage-Ring Group (H. Bruck (Orsay, LAL) for the collaboration)
\item [] \textit{Studies Of Colliding Electron-Electron, Positron-Electron And Proton-Proton Beams}, G.I. Budker, A.A. Naumov. \citep{Budker:1963aa}
\item [] \textit{Status Report On The 1.5 Gev Electron Positron Storage Ring Adone}, F. Amman et al. \citep{Amman:1963aa}
\end{small}
\end{itemize}

The Proceedings of this conference \citep{Kolomenskij:1965bnh} can now be found at the site \url{http://inspirehep.net/record/19356}, but for quite a long time, they were practically unavailable to the Western world, having been written in Russian. From the AdA team, both Pierre Marin and Carlo Bernardini attended the conference, but,
interestingly enough, their published  memories about  the conference  are rather different. 
Bernardini hardly mentions the Russian progress  in his published accounts {of the Conference} 
whereas  
 the atmosphere of the Conference is vividly recalled and comes  alive  in \PM's words \citep[52]{Marin:2009}:
\begin{quote}
\begin{small}
In 1963, the Soviets had organized a conference in Dubna on high energy accelerators, where a number of colleagues had been invited from all over the world. From Orsay, there came R. Belb\'eoch, L. Burnod and myself and Henri Bruck, who presented the ACO project. In this occasion, Budker gave an important  talk on their activity  in  progress on colliders, which had been talking place since 1956 at the Kurchatov Institute, and later  continued at the IPN (Institute  of Nuclear Physics) in Novosibirsk where he had moved. No one in our group  understood any Russian and we only had, to guide us, a list of the abstract of the papers presented at the Conference. With the result, that we did not understand much of what the talk was about. It is however extraordinary that, at the end of the conference, Budker invited some twenty participants to visit Novosibirsk.
\end{small}
\end{quote}
\noindent
Bernardini presented the team's work on the beam  life-time in storage rings \citep{Bernardini:1963xza} but his recollection of the trip to Novosibirsk is rather vague: he recalls that the airplane from Moscow had to have an unexpected stop in Tomsk, but he says nothing about what they saw in Novosibirsk \citep[72]{bernardini2006fisica}. And still, as \PM\ puts it in his memoirs,
\begin{quote}
\begin{small}
C'\'etait une grande premi\`ere.
\end{small}
\end{quote}
\noindent
This was indeed a grandiose debut, a first night to which they had been invited from all over the world.
It was an amazing visit to a totally new laboratory, which had been built in just a few years, moving people and equipment from Moscow to the new location of Novosibirsk, beyond the Urals, where the Siberian Branch of the Soviet Academy had  established  a great laboratory under the direction of Budker. The astonished visitors, who had not really understood what Budker had said in Russian during his presentation in Dubna, saw  two colliders in advanced stage of construction,  an eye-opener of what the Russian scientists had been doing, without the West even knowing anything of it.
For the French scientists, who had just began the construction of ACO, whose first elements were still being finished in the machine shop, it was a shock  to be facing the electron-positron collider VEPP-2, with beam energy of 700 MeV (higher than ACO's), in an advanced stage of construction. We show in Fig.~\ref{fig:vepp} copy of the two photographs of VEP-1, and VEPP-2, which accompanied later the contribution to the proceedings by Budker  and his collaborators.

\begin{figure}[ht]
\centering
\includegraphics[scale=0.33]{
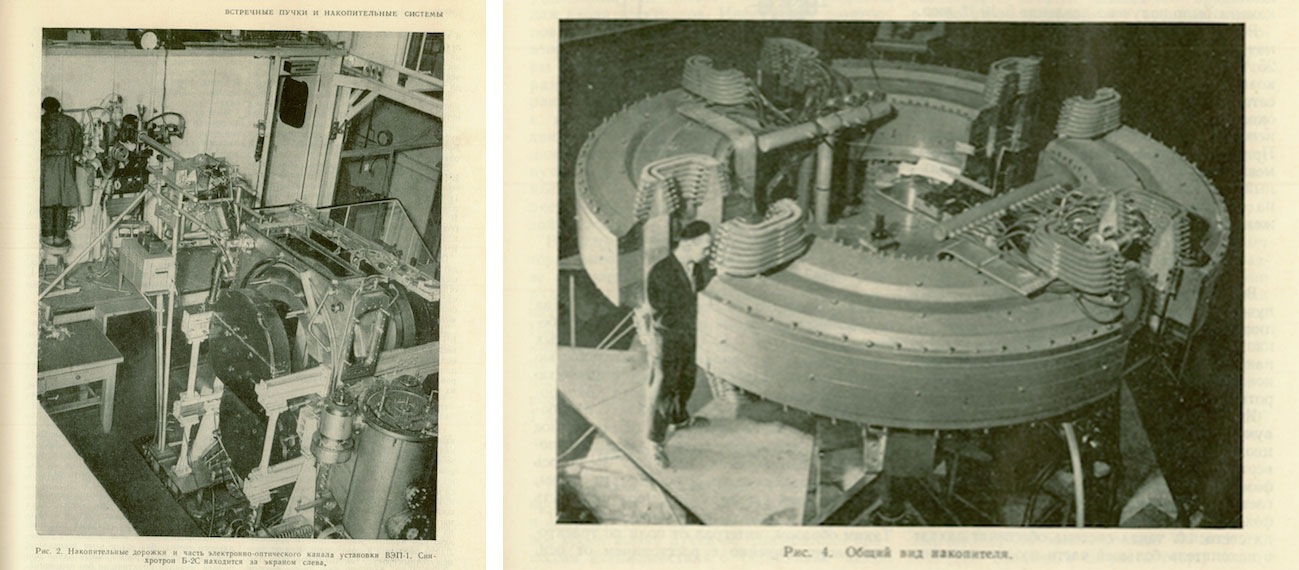}
\caption{From the 1963 Dubna Conference Proceedings, VEP-1 and VEPP-2, left and right photographs respectively \citep{Kolomenskij:1965bnh}.}
\label{fig:vepp}
\end{figure}

The Conference Proceedings were in Russian, and were sent abroad, as it was customary, to the participants's libraries. The Americans  
translated it in English a couple of years later, but the volume in English was not widely circulated and the individual contributions were posted on \href{http://inspirehep.net/record/19356}{inspire.net} only recently. The whereabouts of the Proceedings in France  are related by Pierre Marin in his book, but it is worth recalling them as from a letter Marin wrote to Bernardini on December 26th, in   the year 2000. We show  the letter in Fig.~\ref{fig:PM-toCB-2000}.\footnote{Carlo Bernardini personal papers.}

 The Pierre Marin's letter  describes the amazement of the AdA and ACO scientist, when they saw VEP-1 and VEPP-2 under construction, and its indicates the wherabouts of the Dubna Proceedings, after 1963.  Like everything written by Marin, the letter also includes  instructive comments on the fate of historical documents.

\begin{figure}[htb]
\centering
\includegraphics[scale=0.135]{
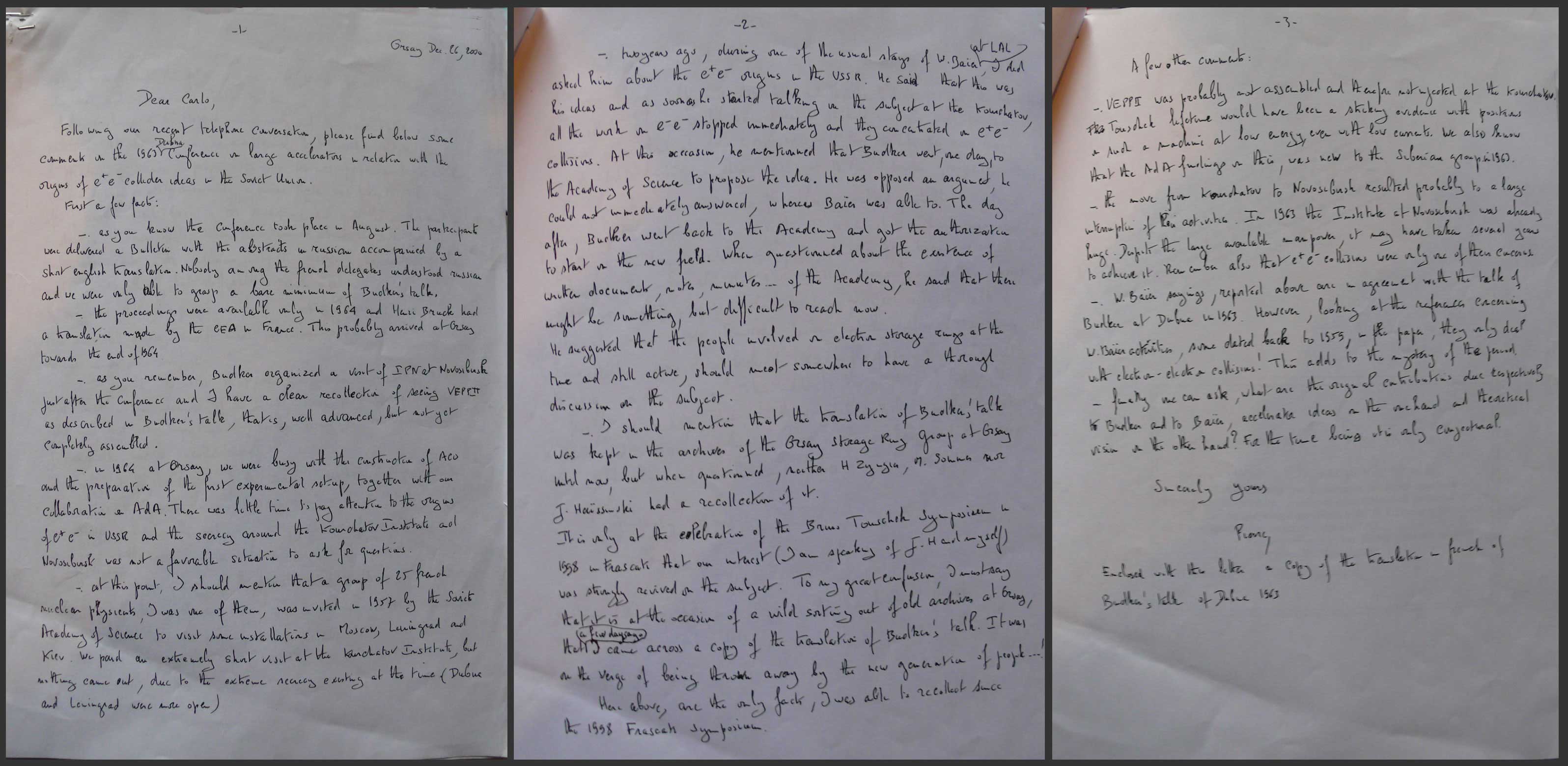}
\caption{Letter sent by \PM \ to \CB\ on December 26, 2000, about the French translation of the Dubna 1963 Conference Proceedings. Courtesy of \CB, personal papers.}
\label{fig:PM-toCB-2000}
\end{figure}
   
  Enclosed with the letter was a  copy of the translation in French  of Budker's talk of Dubna '63, whose first page we show in Fig.~\ref{fig:frenchbudker63}.
\begin{figure}
\centering
\includegraphics[scale=0.31]{
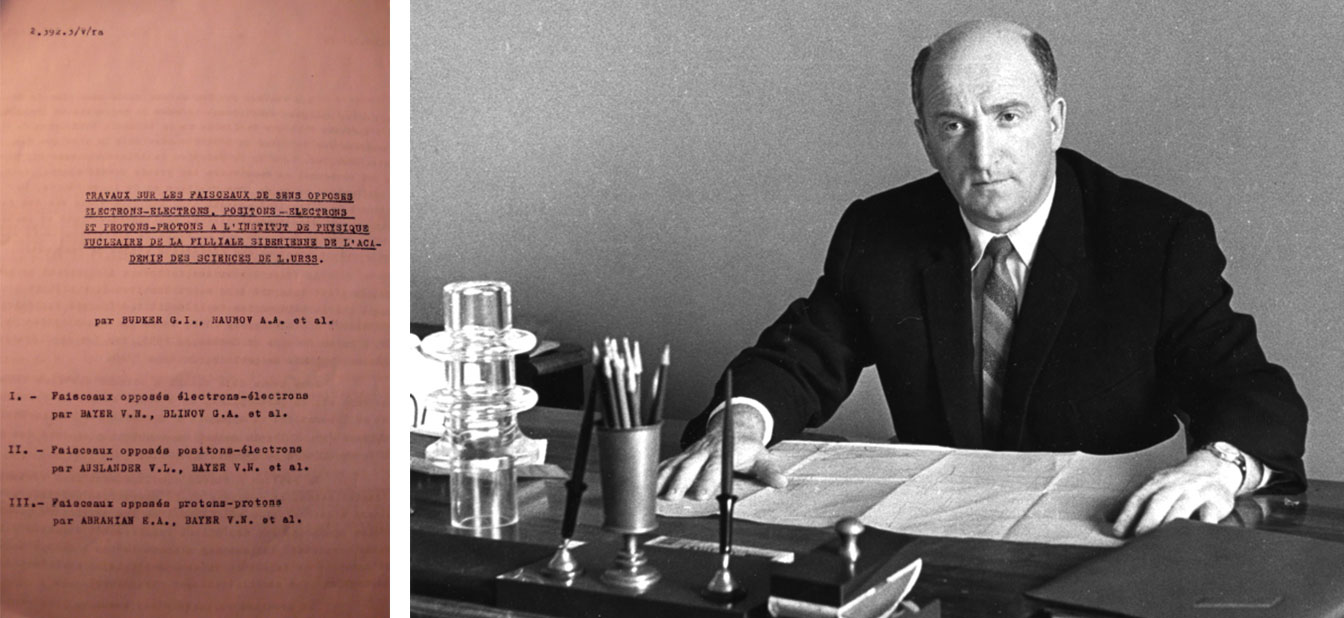}
\caption{At left, the first page of Budker's talk at Dubna '63, in the French translation, courtesy of \JH, and at right Gersh Budker's photo from \url{http://sesc.nsu.ru/famed/index.php?view=detail&id=83&option=com_joomgallery&Itemid=145}.} 
\label{fig:frenchbudker63}
\end{figure}

\section{Observing collisions}
\label{sec:collisions}
The summer conferences being over, back to Orsay, the work with AdA resumed in ernest. The report on the beam life time had been published and presented both 
at Brookhaven and in the USSR. It was now time to proceed to implement its consequences, and attack the next obstacle, the demonstration that although annihilation was out of reach, still AdA could prove that collisions were taking  place. 

Having grasped  the geometry and dynamics of the colliding beams of electrons and positrons, the team was now on the finishing line, but in order to  get the final prize, namely to be accepted   in the history of science as the first to witness, track and record electron-positron collisions  in a laboratory,   the team  had to  convince the world: only then, they could move ahead to continue their new dreams, ACO for the French and ADONE for the Italians.

{To complete the work,  the two teams had to start precise measurements of the process which could be observed with AdA, and confront them with theoretical expectations. Touschek had already a fair idea of the order of magnitude of the result, but precise formulas had to be given and numbers checked. Indeed, as the fall of 1963 started, the needed calculation for the process they could claim to have observed,  $e^+e^-\rightarrow  e^+e^- +\gamma$, became available.  The two students in Rome had spent the months of July and August, working on their thesis,  skipping almost entirely the usually sacred summer vacations, as Franco Buccella remembers. There were many obstacles, theoretical and computational, some of them almost intractable, but they persevered, and indeed  the article they would later write \citep{Altarelli:1964aa} is still a classic text, amply used and quoted.  They  worked through the summer of 1963, taking only one week vacations. Touschek helped them not to lose heart, giving them now and then some insight on why the calculation appeared so difficult, and indeed it was,   and, at times,  hinting at the  possible solution. Finally, they found their way to the result  through  an approximation which is still used and     they could then    start writing the thesis. The thesis   was defended, with full honours, at the University of Rome, in November 1963. Thus,   the theoretical results were ready for the time   the AdA team in Orsay would start  the final experimentation in December 1963,
when  the AdA team positioned themselves  in earnest, building statistics.}
\subsection{December 1963: Starting the final runs}
{ The first of four final   runs  took place in  December 1963. They had to observe the correlation of gamma rays with counting positrons and electrons, and check that the number of such events was in accordance with their acquired knowledge about the volume of the bunches and the calculation of the elementary cross-section done by Altarelli and Buccella. The second run took place in January 1964, a third one in February and the fourth one in April 1964.  In May, enough evidence had been accumulated, figures started to be drawn, the article was written.}
We show in Fig.~\ref{fig:graficorosa} the plot which showed the results of counting gamma rays when both electron and positron beams were circulating in the the AdA ring: the plot,  in Touschek's hand,  includes the results collected during  the  four data runs, and clearly indicated  correlation between the counts from both sides.
 \begin{figure}[htb]
 \centering
 \includegraphics[scale=0.18]{
 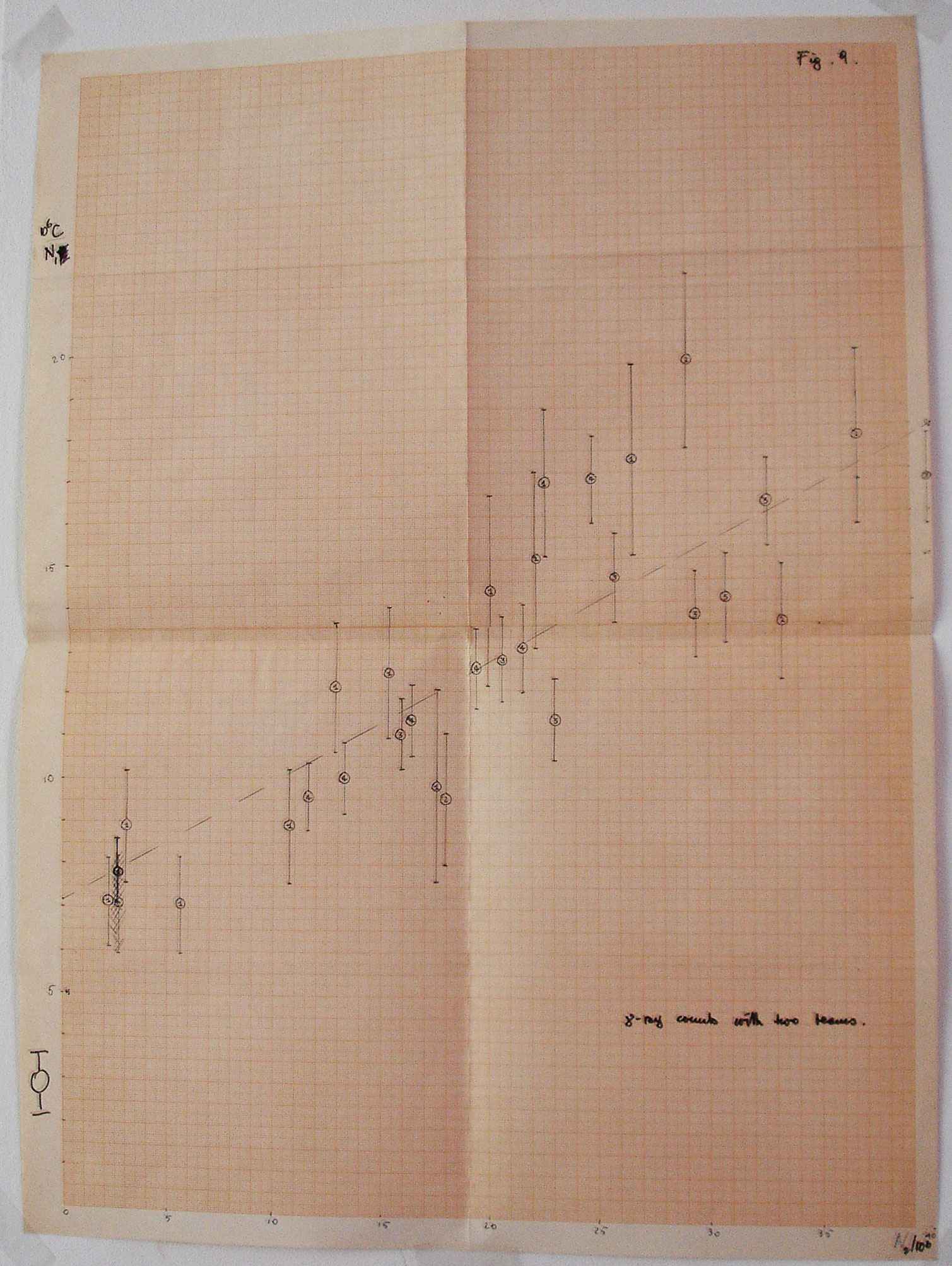}
 \caption{The synthesis of all the data for gamma rays count  collected when two beams were circulating in AdA, from Touschek's recorded plot, data  published as  (Fig. 9) in \citep{Bernardini:1964lqa}. {The original plot was photographed from  \CB's personal papers.}}
 \label{fig:graficorosa}
 \end{figure} 
{In Frascati and Rome,  the calculation of single hard bremsstrahlung in electron positron scattering  was finished,  and  the comparison between the theoretical calculation and AdA's data showed that AdA's observations were consistent with expectations that electrons and positrons had indeed made collisions. This result was  of course also based on the knowledge the AdA group had acquired of the beam dimensions and their internal dynamics, controlled by the Touschek effect. }

Thus the team did not lose heart and, in the end,    managed to demonstrate that the collisions were taking place and that the overlap between the electron and the positron bunches was complete. As {\JH} says in his 2013 interview:
\begin{quote}
\begin{small}
\dots this  was the most favourable situation to produce collisions and that was the    end point of the runs of AdA in Orsay.\\
Later on in my thesis what I did was to put together some kind of a synthesis of all the measurements which were  carried out by the Italian-French collaboration and so we performed with \PM \ and with Giuseppe Di Giugno some complementary experiments, so that all the pictures would be given in my thesis. The whole program lasted about two years and not more. It was very exciting time because the spirit in the team was really excellent. The personalities of the various people were complementary, I would say,  and so were the knowledge and the competences. We really had a great time. 
\end{small}
\end{quote}
\noindent
 {In  the first half of 1964, an article with the results from the four runs was prepared, first in {\it preprint}  format to circulate, by ordinary mail, among the interested community of accelerator physicists,    and was then submitted, in July 1964, 
 to the {\it Nuovo Cimento} \citep{Bernardini:1964lqa} with  the title {\it Measurement of the rate of interactions between stored electrons and positrons}.  
 
 We quote here \JH's words, in French, summarising the results of AdA's experimentation:\footnote{May 2013 interview with \JH \ in Orsay.}
 \begin{quote}
 \begin{small}
  Les mesures finales qui ont \'et\'e faites ont
\'et\'e des mesures qui ont port\'e sur le nombre de collisions par seconde
entre les \'electrons et les positrons. C'\'etait la premi\`ere fois au
monde que l'on montrait que les particules, effectivement, int\'eragissaient et
entraient en collision les unes avec les autres et donc, \c ca montrait que, on peut dire, que ces machines \'etaient utilisables pour faire de la physique
des tr\`es hautes \'energies
{et l'essentiel, pas tout, bien s\^ur, on a fait d'autres
d\'ecouvertes par la suite, mais l'essentiel, quand m\^eme, des caract\'eristiques de ce type de machines \'etait valid\'e 
 et permettait de penser que les g\'en\'erations ult\'erieures [de ce type de machine] s\'eraient utilisables pour faire de la physique des particules \`a tr\`es haute \'energie.}\footnote{
 The final measurements which were done  dealt  on  the rate  of collisions among electrons and positrons.   It was the first time in the world that one could show that the particles effectively  interacted and collided against each other, and this allowed to think  that these machines could be used to do [experiments] in  high energy physics. There were other discoveries, of course, but the essential point was that the characteristics of this type  of machines  were validated and that successive generations [of these machines] could  be used for  very high energy physics. }
\end{small}
  \end{quote}

\section{What was happening beyond the Iron Curtain}
\label{sec:ironcurtain}
With the publication of the results of the four runs, during 1963 and 1964, AdA's adventure was completed. It had proved that this type of machine could work, and the AdA team had been the first to show this. But had AdA been the first ever particle collider to function? 
And had been Touschek the first to put forward  the
idea of storing electrons and positrons  in a single ring in order to
make them collide and probe the QED vacuum? Had  he been the first to
realize the idea, together with a group of exceptional young scientists and technicians, from Orsay and Frascati? It seems to us, at the end of this chapter, that we can answer in the affirmative. This was an incredible feat, a victory of David against Goliath,  which sprang from the hard times of the war through which Touschek, and the rest of Europe, had survived. 

 Before going forward and describing what happened after AdA both to  our heroes and to the field of particle colliders, we should mention  here the story from the point of view of the scientists beyond the Iron Curtain,  but these further important aspects of the colliding beams developments will be described in a separate note. Here we will briefly recall that Budker's team at the Laboratory of New Acceleration Methods of the Institute of Atomic Energy of the USSR Academy of Sciences in Moscow (in 1960 named `Kurchatov Institute', after its founder Igor V. Kurchatov) had focused since 1957 on colliding beams, in particular on physics development and design   of the electron-electron collider VEP-1. After AdA, VEP-1, whose construction started in 1958-1959, was the first accelerator implementing colliding beams. As recalled by Alexander N. Skrinsky,  this activity had started ``just upon D. Kerst's proton collider suggestion and G. O'Neill's proposal to use radiation damping for electron beams storing and compression.'' \citep[14]{Skrinsky:1996aa}. Then, in 1958, Budker's Laboratory was transformed into the Institute of Nuclear Physics (INP) of the Siberian Branch of Academy of Sciences. Between 1961-1962 the whole INP moved to Novosibirsk, where the first circulating beam was obtained with VEP-1 in August 1963 and the first luminosity was detected in May 1964 \citep[407]{Levichev:2018aa}. In the meantime, they had begun to work on the idea of electron-positron collisions, which would later materialize in the VEPP-2 collider. The collider activity of INP was presented for the first time at the International Conference on High-Energy Accelerators held in Dubna in August 1963 \citep{Budker:1963aa}.
 \footnote{
 For an outline of accelerator developments in Novosibirsk and a presentation of research work with VEP-1 and VEPP-2 see \citep{Budker:1966ab} \citep{Budker:1966aa} \citep{Auslender:1966aa}. For an early account and presentation of the Russian projects made by Budker himself, see \citep{Budker:1967aa}.}


 We recall here some comments and impressions by Pierre Marin  about the Dubna Conference from the letter quoted in the previous section:
  
\begin{quote}
\begin{small}

  Dear Carlo, 
  
  following our recent telephone conversation, please find below some comments on the 1963 Dubna Conference on large accelerators in relation to the origin of $e^+ e^-$ collider ideas in the Soviet Union.
  
  First a few facts:
  \begin{description}
    \setlength\itemsep{0.1 em}
  \item  -- as you know the conference took place in August. The participants were delivered a Bulletin with the abstracts in Russian, accompanied by a short English translation. Nobody among the French delegates understood Russian and we were only able to grasp a bare minimum of Budker's talk
  \item -- the proceedings were available only in 1964 and Henri Bruck had a translation made by CEA in France. This probably arrived at Orsay towards the end of  1964
   \item -- as you remember, Budker organized a visit of IPN [Institute for Nuclear Physics] at Novosibirsk just after the Conference and I have a clear recollection of seeing VEPP-2, as described in Budker's talk, that is, well advanced but not yet assembled 
    \item -- in 1964 at Orsay, we were busy with the construction of ACO and the preparation of the first experimental set-up, together with our collaboration in AdA.\footnote{Efforts were  made in view of preparing experiences and detectors specifically dedicated to the new research topics to be explored with colliding beams \citep{marin:1965aa}. The group directed by A. Blanc-Lapierre included: 1) physicists and engineers of the Laboratoire de l'Acc\'el\'erateur Lin\'eaire d'Orsay and of the Centre d'\'Etudes Nucl\'eaires de Saclay who participated in the project and construction of ACO as well as in tests (M.M.G. Arzelier, J.E. Augustin, R.A. Beck, R. Belb\'eoch, M. Bergher, H. Bruck, G. Gendreau, P. Gratreau, J. Ha\"issinski, R. Jolivot, G. Leleux, P. Marin, B. Milman, F. Rumpf, E. Sommer, H. Zyngier); 2) Physicists participating to the realization of the experimental device at ACO (MM. J.E. Augustin, J. Buon, J. Ha\"issinski, F. Laplanche, P. Marin, F. Rumpf, E. Silva) \citep{Marin:1966}.}
     There was little time to pay attention to the origin of $e^+e^-$ in the USSR and the secrecy about the Kurchatov Institute at Novosibirsk was not a favourable situation to ask for questions
         \item -- at this point I should mention that a group of 25 French nuclear physicists,  I was one of them, was invited in 1957 by the Soviet Academy of Science to visit installations in  Moscow, Leningrad and Kiev. We paid an extremely short visit to the Kurchatov Institute, but nothing came out due to the extreme secrecy existing at the time (Dubna and Leningrad were not open)
       \item -- two years ago, during one of usual stays of W. Ba\u ier at LAL, I had asked about the $e^+e^-$ origins in the USSR. He said that this was his ideas and as soon as started talking on the subject at the Kurchatov, all the work on $e^-e^-$ stopped immediately and they concentrated on $e^+ e^-$ collisions. At that occasion,   he  mentioned that  Budker had gone, one day, to the Academy [of Sciences]  to propose   the idea. He was opposed an argument he could not immediately answer whereas Ba\u ier was able to. The day after, Budker went back to  the Academy and got the authorization to start on the new field. When questioned about the existence of written documents, notes, minutes \dots  of the Academy, he said there might be something but difficult to reach now.
A year after, Ba\u ier suggested that people involved at that time in electron storage rings  and still active should meet and have a thorough discussion of the subject
         \item -- I should mention that the translation of Budker's talk was kept in the archives of the Orsay Storage Rings Group at Orsay until now, but when questioned neither  H. Zyngier, M. Sommer nor \JH\ had any recollection of it. It is only at the celebration of the Bruno Touschek Symposium in 1998, at Frascati that our interest (I am talking of J.H. and myself) was strongly revived on the subject. To my great confusion, I must say that it is at the occasion of a wild sorting out of old  archives  at Orsay, that a few days ago I came across  a copy of the translation of Budker's talk. It was in the verge of being  thrown away by the new generation of people \dots ! 
    \end{description}
    
    Here above, are the only facts I was able to recollect since the 1998 Symposium.
    
\noindent     A few other comments:
\begin{description}
  \setlength\itemsep{0.1 em}
\item -- VEPPII was probably not assembled and not injected at the Kurchatov. Touschek life time would have been a striking evidence with positrons at such a machine at low energy, even with low currents. We also know that the AdA findings on this were new to the  Siberian group in 1963.
\item -- the move from Kurchatov to Novosibirsk resulted probably to a large interruption of their activities. In 1963 the institute at Novosibirsk was already huge. Despite the large available manpower  it may have several years to achieve it. Remember also that electron-positron collisions were only one of their concerns. 
\item -- Ba\u ier's sayings reported above are in agreement with Budker's talk.  However, looking at the references concerning the reported activities of Ba\u ier, some dated back to 1959, they only deal with electron electron collisions. This adds to the mystery of the period. 
\item -- finally one can ask what are the original contributions due respectively  to Budker and to Ba\u ier, accelerator ideas on the one hand  and theoretical vision on the other hand? For the time being  
it is only conjectural
\end{description}
Sincerely yours
\noindent
\\Pierre 
\end{small}
 \end{quote}
 \noindent
In the year 2006, the well known theoretical physicist  Vladimir N. Ba\u ier published a memoir which tells his  side of the electron-positron collider adventure. In his memoir, Vladimir Ba\u ier tells of the difficulties and delays encountered  by the Russian group led by Budker, delays due also to the moving to Novosibirsk, including  people and instrumentation. He also recalled that ``Because of pathological secrecy adopted at that time in USSR, all activity in Kurchatov Institute was considered as `for restricted use only' and the special permission for publication in open journals or proceedings was necessary for each article'' \citep[5]{Baier:2008aa}.} Ba\u ier  included an interesting date for the day in which he started working on electron-positron colliders:  28 October 1959, around the same time we can ascribe to Panofsky's seminar in Rome, which was held around 26th of October. As a matter of fact, this would have been just over a month after the CERN conference, in September, which had been well attended by many Russian scientists. In a sense, one month appears as a physiological time for the ideas to spread and be absorbed, but the coincidence of  dates, between Italy  and Russia,  is striking.

\section{After AdA}
\label{sec:afterAdA}
{When the AdA team published their fourth and last work on AdA's experimentation  \citep{Bernardini:1964lqa}, four and a half years had passed since the day when Touschek had proposed to build AdA. This had occurred on the occasion of a meeting about the future of the Frascati Laboratories,   on  Wednesday, February 17th, 1960. From the transcripts, we read:\footnote{L.N.F. Report N. 62, December 1960.}
\begin{quote}
\begin{small}
[\dots] From   minutes of previous meetings, he [\BT] has received an  impression of uncertainty about  the Frascati Laboratories future activities. He thus has started thinking about what  to do next: something such as a future goal.\footnote{{\it Una meta futura}, in Italian.}
\noindent  

[\dots] Now, an experiment really worth doing, an experiment which would  really be frontier physics [\dots]  would be an experiment for the   study of electron-positron collisions. It is an experiment that Panofschy [sic] plans to do: one has to do it first.\footnote{{\it Si tratta di arrivare prima di lui}, literally ``It is a matter to arrive there ahead of him."}
\end{small}
\end{quote} 
\noindent
And so they did, and the world of particle physics accelerators  dramatically changed.
While  AdA's last article was being published in December 1964, everywhere in the world  new projects for colliding beam facilities were already in place, in Italy with ADONE, in the USA with SPEAR, in the USSR with VEPP-2, in France with ACO, at CERN with the Intersecting Storage Ring for protons.\footnote{CERN's interest in storage rings for protons had an early start. Proposals for tangential rings for protons  were presented at the end of 1960, and it is in 1962 that the actual proposal for the ISR colliding  protons against protons was presented. The   first proton-proton collisions were recorded on 27 January 1971. CERN's  interest in storing antiprotons appeared  also very early.  In 1983, Peter Bryant wrote  a brief account   of the events leading to antiprotons in the ISR, on behalf  of then dissolved  CERN ISR division \citep{Bryant:1983jh}. Bryant recounts that the storage of anti-protons in the ISR was discussed as early as 1962, even before the ISR were constructed. Physics runs  with proton on antiprotons  in the ISR took place starting in  April 1981 until  December 1982, at  maximum 62.7  GeV c.m. energy.}
 Never again, in recent times, would  such a revolution take place in accelerator physics in such a short time.


{Since  1963, and more so after 1964, Bruno Touschek had focused on ADONE, the more beautiful and powerful collider under construction in Frascati. After 1964  \JH \  finished writing his thesis, and then joined \PM\  on  the ACO project, the  \ACO. }
{Both ADONE and ACO  had been envisioned and planned before AdA moved to Orsay. In December 1960, as AdA had been constructed and was near to enter in operation in February 1961, Touschek had proposed to build an electron-positron collider which could reach such energy to explore annihilation into pairs of particles and anti-particles of all the kinds  known at the time: $\mu's$ and $\pi's$, Kaons and protons and neutrons. ADONE would need to be much bigger, and costlier, but it was rapidly approved and, in 1963, construction of the ADONE building  had started, in an empty lot across the street from the land where the electron synchrotron and AdA had been built \citep{Amman:1963aa}.} 
It became operational in 1968. ADONE's experimentation brought the discovery of multiparticle production, and the confirmation, in 1974, of the American discovery of a new state of matter, made of charmed quarks, the $J/\Psi$ \citep{Aubert:1974js,Augustin:1974xw,Bacci:1974za}.

In Orsay, the idea of an electron-positron collider of an energy about three times bigger than AdA's had arisen  in Orsay, shortly after \PM \ had returned from Frascati in July 1961 and mentioned AdA's potential to the new laboratory director \ABL.
In Fig.~\ref{fig:dec1961letter} we show copy of the letter sent by \ABL\ to the Frascati director Italo Federico Quercia, which mentions the possible French project. This letter started the collaboration which 7 months later would bring AdA to Orsay. 
\begin{figure}[htb]
\centering
\includegraphics[scale=0.6]{
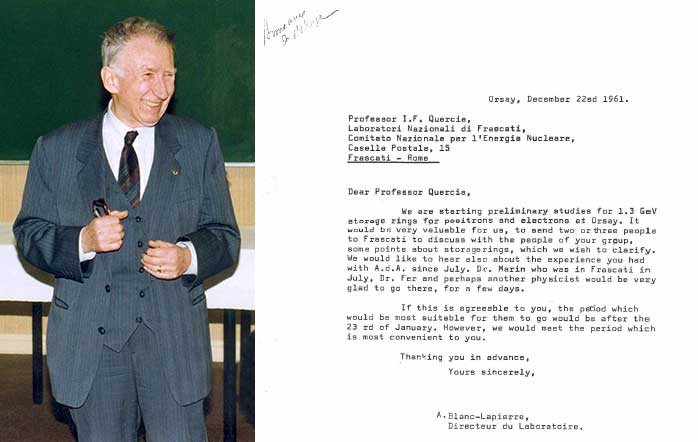}
\caption{
Copy of the letter sent by LAL's director \ABL, shown in the left panel,\ to the Frascati director,  to propose a collaboration, with mention of ``starting preliminary studies for 1.3 GeV storage rings for positron and electrons at Orsay". Letter is courtesy of \JH, from \LAL's Archives. }
\label{fig:dec1961letter}
\end{figure}

As described in \citep{Bonolis:2018gpn},  Edoardo Amaldi, then President of the Istituto Nazionale di Fisica Nucleare,  discouraged the French director \ABL\ from starting a European competition, since ADONE, at the time, had already been included in the European planning for future accelerators, and its funding was soon to come through the Italian government. 
Amaldi's reservation about the LAL storage ring project with a 1.3 GeV energy per beam was taken into account by the LAL management. In fact, the machine energy was downsized to 450 MeV per beam as presented by H. Bruck the year after (1963) at the Dubna conference on high energy accelerators \citep{Bruck:1963aa}. The choice of this energy value was (very likely) due to the newly discovered particles, the vector mesons, whose dynamics and properties could be optimally studied through such energy.
Thus,  in Orsay, the experience with AdA opened the way to the development of the large French accelerators \citep{Marin:2009},   and to their exploitation as producers of synchrotron light radiation.\footnote{A highlight of Orsay's road to sinchrotron light radiation studies and applications can be found in the docu-film \href{https://webcast.in2p3.fr/video/60-ans-dexploration-de-la-matiere}{Soixante ann\'ees d'exploration de la mati\`ere avec des acc\'el\'erateurs de particules}, by E. Agapito and G. Pancheri, with the collaboration of \JH.}

 If the Italians thought that taking AdA to the LINAC had been a good move in the game they played at a distance with the other  collider builders of the world, the French scientists also understood and  exploited the  added value that AdA had brought to Orsay. The construction of ACO was approved in 1963 \citep{Bruck:1963aa} and the collider became operational in 1967. 
 As presented in Dubna, ACO's original value of $E_{max}/beam$ was 450 MeV, but later on    this energy was pushed up
to 540 MeV, probably  to ensure comfortable conditions
to produce the $\phi$ meson and study its properties \citep{Augustin:1973ep}. 

{In Fig.~\ref{fig:ADONE-ACO-AdA}, we show 
  a recent, 2013, photograph of \JH\  with 
 the 1:1/4 copy of AdA,  in Orsay on the occasion of the 50th year anniversary of first observation of electron positron collisions, courtesy of \JH, and, at right Bruno Touschek discussing ADONE's construction in Frascati  around 1964.}

\begin{figure}[ht]
\centering
\includegraphics[scale=0.43]{
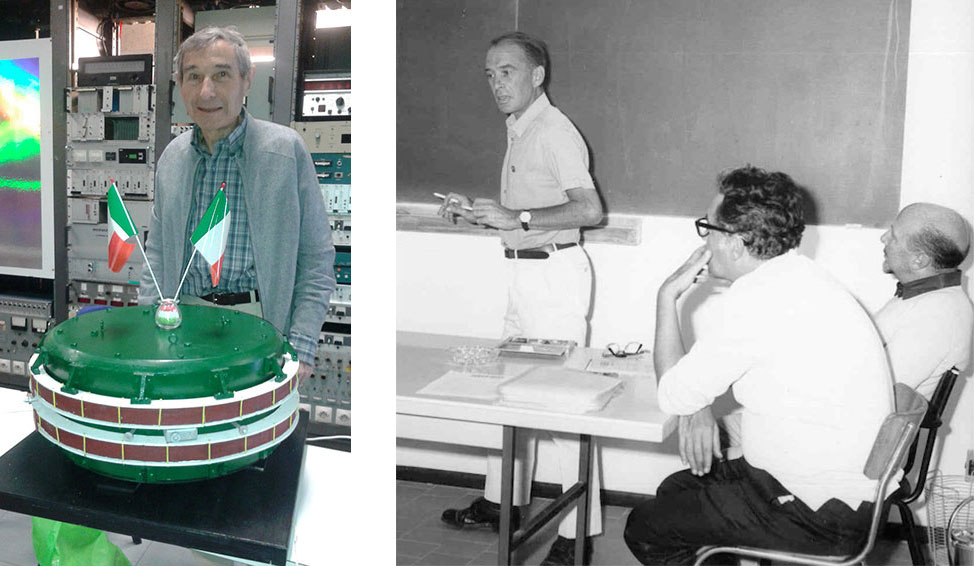}
\caption{At left,  \JH\ is shown  in the ACO control room, kept as it was then,  with the 1:1/4 copy of AdA,  on the occasion of  the Bruno Touschek Memorial Lectures, September 9-13, 2013, celebrating fifty years since the 1963 observation of electron-positron collisions. At right one can see Bruno Touschek in Frascati in 1964, discussing ADONE's project. In this photo,  the Frascati Laboratories director Italo Federico Quercia is sitting  at the far right, Ruggero Querzoli next to him.}
\label{fig:ADONE-ACO-AdA}
\end{figure} 

\

Following  AdA's experimentation, everywhere in the world  hopes for the physics which such machines could unearth, were high, and, indeed, in the next ten years,  these machines would make great discoveries which   challenged  current thinking of particle physics. Touschek witnessed part of it, not all. In 1978, barely 57 years old, he passed away, but 
the many sleepless nights in Orsay were not spent in vain, as his  legacy survives  \citep{Amaldi:1981} through the enormous modern day colliders and their discoveries  \citep{Aad:2012tfa}.


\section{Conclusions}
Before stating our conclusions, we show in Fig.~\ref{fig:varenna69} the photograph of the participants to the XLVI Summer Course held in Varenna, in June 1969. ADONE had just started functioning, and we can see in this picture all the major actors of colliding beam physics, until then: Matthew Sands,  Bruno Touschek, Gerard O'Neill, \JH, Vladimir N. Ba\u ier, together with many younger students and scholars.

\begin{figure}
\centering
\includegraphics[scale=0.68]{
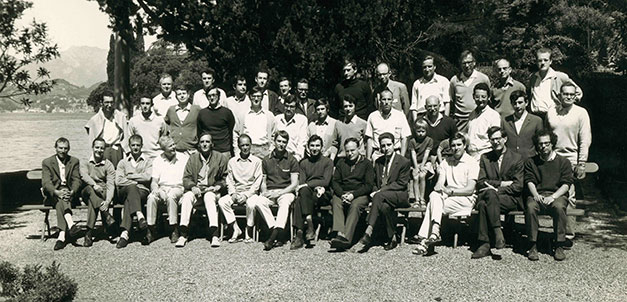}
\caption{Group photo of the XLVI Summer Course ``Physics with intersecting storage rings", of the Societ\`a Italiana di Fisica, 16-26 June 1969, Varenna, Italy \citep{Touschek:1971aa}. In front row,  fifth from left Matthew Sands, then \BT, Gerard O'Neill, \JH, Vladimir Ba\u ier. In first row, we also see, second  from left, Rinaldo Baldini, who worked on ADONE, and is now in the   experimental team at the Beijing Electron Positron Collider, BEPCII, and collaborates in preparations  for  China's next electron positron collider, \href{http://cepc.ihep.ac.cn}{CEPC}.  In second  row, third from right, one can see Claudio Pellegrini, who collaborated with Touschek on ADONE  and was later (2015) awarded the  \href{https://www6.slac.stanford.edu/news/2015-10-21-slacs-claudio-pellegrini-receives-fermi-award-white-house.aspx}{US Fermi Award}. Pedro Waloschek, Wider\o e's biographer, at the time senior scientist at DESY, is in the last row, first from left.}
\label{fig:varenna69}
\end{figure}

In this note, we have presented how events  unfolded during the final leg of AdA's journey   in the history of physics. In particular we have described   how AdA's transfer from Frascati to Orsay    proved the feasibility of electron-positron colliders as the tool {\it par excellence} for the future of particle physics.

 It was not a feat which could have been accomplished by a single nation or a few researchers:
  two national laboratories contributed their technical know-how,  the \LAL  \ d'Orsay with the LINAC, and the Frascati National Laboratories, first with the electron synchrotron,  and then with AdA. The physicists and technicians who had built these machines and were familiar with the ways of elementary particles such as the electrons, provided the intellectual  resources needed for success in a totally new field. 
  
  The story moves back and forth   through different locations:  in Orsay, at  the  \LAL\  between the LINAC and the Salle-500, some 100 meters away where  AdA had been installed, in Frascati, 
  { where the \v Cerenkov detectors and the photomultipliers} were built  and sent to Orsay to be  next to AdA, and in the University of Rome, where some of  the theoretical problems were worked out. 
  The actors were many, and with different roles, as needed for a prototype of a high energy physics experiment: the machine builders, the detector developers, the theoreticians to propose and interpret the results.  The peculiarity of such a group, was that the machine builders had also been the users of the machine, at the point that they had a very strong ``mental representation'' of the processes taking place inside the collider, as Carlo Bernardini has continued to stress during the years \citep{bernardini2006fisica}. 
  
  Experimental physicists came from both sides, and invented new ways to receive and interpret the signals from AdA.  Ruggero Querzoli and Giuseppe Di Giugno built the detectors in Frascati, \PM \  was the main mover behind AdA's transfer to Orsay and was later  to transfuse AdA's experience into the first French  modern day accelerator, ACO, 
  \FL \ collaborated during the early delicate phase, from AdA's arrival and instalment at LAL up to first research activities, \JH, initially a doctoral student, was  the {\it liason officer}  linking  AdA and the LINAC. He  collected the   data and, after most of the  Italian group went back to Italy, completed  many  measurements, and masterfully described  in his doctoral thesis all which would be known at the time about colliding beam machines, and how AdA's success was achieved. 
  
On the Italian side,   the theorists were Bruno Touschek,  Carlo Bernardini, Nicola Cabibbo and Raoul Gatto. On the French side, important work on the Touschek effects was later done by H. Bruck and J. Le Duff \citep{Bruck:1965eba}. 

 Sometimes, one of the scientists took on a different role: such was the case of Bruno Touschek,  a first rate theorist with an experimental vision, who had in his past the experience of  building  an electron machine.
  { Touschek's genius was that he was not only a theorist. In 1942, when he was still  a physics student, he had  left  Vienna, where his Jewish origin was well known,  for Germany, hoping  to continue his physics studies  {\it incognito} \citep{Bonolis:2011wa}. There, between Hamburg and Berlin, during some of the darkest  times of WWII, in Hamburg, he had learnt the art of making electron accelerators   from one of the  masters of European  accelerator science, \RW \ \citep{Wideroe:1994}.} He had further honed his knowledge and understanding   when after the war he   worked  at the Glasgow electron synchrotron \citep{Amaldi:1981}. As for the other theorist of the AdA group, Carlo Bernardini, he also  had a long experience with the design of the Frascati electron synchrotron,  going back to the early 1950s.

And what happened to the heroes of this story after 1964? As always,  the roads followed by the protagonists of a great adventure are different, sometimes diverging. Both Giorgio Ghigo and Ruggero Querzoli passed away not long after the end of AdA's adventure. 
Bruno Touschek,  returned   to Italy, fully immersed in planning for ADONE and on how to extract physics results from this machine,  which imply   radiative corrections  work for  ADONE, beyond the single bremsstrahlung process calculated by Altarelli and Buccella \citep{Altarelli:1964aa,Etim:1967}.  Carlo Bernardini joined  the ADONE project as well. Giuseppe di Giugno had to leave for his military service, and, during his absence, the
  second AdA log book, kept in his office in University of Naples, was thrashed. Di Giugno is the only one of the AdA team who left the academic life of particle physics: he profused his genius for electronics into music and was one of the early founders of today's digital world in music.\footnote{See \url{https://medias.ircam.fr/x7b9990}, \url{http://www.lucianoberio.org/node/36288} for some memories from Di Giugno's first approach to electronics in modern musics. See also \url{http://120years.net/sogitec-4x-synthesiser-giuseppe-di-giugno-france-1981/}.}

As for the French team,  \JH \ finished his thesis and started working on ACO, joining \PM, already head of the project.
Thus,  in Orsay,  through experimentation with AdA,  the world of colliders had been opened and with it the  road to the great French accelerators, and their future exploration with synchrotron light, through  SUPERACO, and, in modern days,  the 
ESFR in Grenoble and SOLEIL on the Saclay plateau, south of Paris, above the Vall\'ee de Chevreuse.

  What we have presented here  is only a small part of a larger story, which encompasses all of Europe in both space and time. In this note,  we have shown the hopes and efforts of the Franco-Italian team which in   Orsay opened the  road to high energy colliders. This road was opened  by AdA,  the  dream machine   built in Frascati in 1960  by Bruno Touschek and his group of top class collaborators. AdA  was indeed a {\it vrai bijou}, as Marin wrote 40 years later. A small pancake shaped machine, placed at  the center  of a hall next to the  Frascati electron synchrotron,  AdA    had enchanted Pierre Marin when he had seen it on a hot summer day, in July 1962, so long ago. 
  
  In AdA, for the first time in the world, electrons and positrons had been injected and could be  seen circulating and, in so doing,  emitting light. 

AdA is still there, on a meadow, under a canopy, in the Frascati Laboratories, to remind the many visitors and scientists of how  a group of Italian and French physicists  decided to defy all odds and built an accelerator   who led the way to high energy particle physics and made the history of particle colliders.  
  
  \vspace{0.5cm}

\begin{figure*}[htbp]
\centering\includegraphics[scale=0.4]{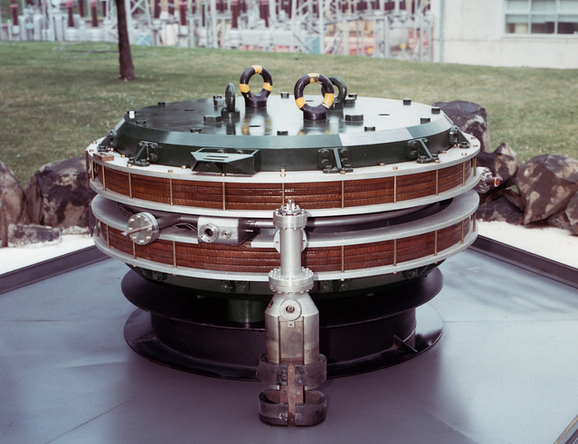}
\caption*{\href{http://www.lnf.infn.it/edu/AdA_EPSHistoricSite/gallery/}{AdA on the Frascati Laboratories grounds}, 2013 \copyright INFN.}
\end{figure*}
\vspace{0.3 cm}

\section*{Acknowledgments}
      We are very grateful   to our colleague and friend \JH\ for suggestions and comments.  Through the many  years  which preceded  this writing and those during which we worked on the Orsay period of AdA's story, he has given  us advice, providing  unpublished material, and correcting our misinterpretations or errors, with his memory of the times and knowledge of the subject.   We thank Giuseppe Di Giugno and Franco Buccella for sharing their memories of the period we have described, and Amrit Srivastava for translation of parts of \PM's book.  We thank Lluis Oliver,  Yogendra Srivastava  and Angelo Mainardi for encouragement
      and suggestions. 
      We also thank Achille Stocchi, Director of the \LAL  \ in Orsay, for hospitality during the period the interviews with French scientists took place,  and  acknowledge  courtesy of access to \LAL. From the Physics Department of  University of Rome La Sapienza,  we thank Gianni Battimelli and  are grateful to the library staff  for collaboration and access to the Archives.  From the Frascati National Laboratories we thank Antonino Cupellini  and Claudio Federici  for their help and collaboration with the photographic archive and bibliographic material. One of us, G. P. is grateful to Orlando Ciaffoni for advice and suggestions.  We thank the late Elspeth Ionge Touschek for sharing with us some of Touschek's photoes
 and other documents.     


      Finally we dedicate this work to the memory of our mentors, Bruno Touschek (1921-1978) and Carlo Bernardini, who sadly passed away in June 2018, 
      at 88 years of age. We deeply miss his friendship and advice, and would be happy if our writing  can make better  known to many  readers  the story of AdA's  wonderful adventure in particle physics.
   


\newpage
 \section*{Chronology of events from July 1962 to December 1964.}


 \begin{centering}
\hspace{-1cm}

 \begin{tabular}{|>{\scriptsize}l|>{\scriptsize}l|>{\scriptsize}l}

{\bf Date}&{\bf Event}&{\bf Source}\\  
  &  & \\
 June 28th 1962	&BT announces presumptive date of AdA's leaving & BT's letter to Francis Perrin \\
  & & \citep{Bonolis:2018gpn}\\
   July 4th 1962& (presumptive date) AdA leaves  Frascati&BT's letter to Francis Perrin\\
    & &  \citep{Bonolis:2018gpn} \\
$\approx $    July 6th 1962&AdA arrives in Orsay&Carlo Bernardini  in \citep{bernardini2006fisica}\\  
   On or after July 12th1962& AdA's lab outfits leave Frascati by plane&CB's letter to Fran\c cois Lacoste\\
    & &  in  \citep{Bonolis:2018gpn}\\
  On or after  July 12th, 1962& (presumptive date) Giorgio Ghigo with technicians flies to Orly&\\
  Fall 1962&Injection attempts, magnetic powder&\PM\ in \citep{Marin:2009}\\ 
   & & and \FL \ $^{(1)}$ \\
   November 1962& FB asks Gatto for a thesis, then postpones& personal communication to GP \\
   December 21st, 1962&Run for $e^+e^-\rightarrow 2 \gamma$&LogBook, as in \citep{Haissinski:1998aa}\\
   January 1963&(presumptive date) GA asks BT for a thesis, decides against&M. Greco and G. Pancheri $^{(2)}$\\
February 1963& Observation of Touschek effect&date on life time plot $^{(3)}$ \\
 March or late Feb.1963  & BT writes to O'Neill& O'Neill's letter to BT $^{(4)}$ \\
March 28 1963& O'Neill dismisses AdA's prospects&O'Neill's letter to BT $^{(4)}$ \\
April 1 1963& Submission to PRL&   \citep{Bernardini:1963sc}\\
March-April 1963 & BT envisages bremsstrahlung process and asks Gatto& presumptive,  log book\\
April 22-25 1963& O'Neill mentions ``Touschek effect'' at APS April Meeting, DC&See O'Neill's talk $^{(5)}$  \\
May 1 1963 & Touschek effect published& \citep{Bernardini:1963sc}\\
June 1963&GA and FB start thesis with Gatto& FB personal communication\\
Summer 1963&Work on thesis and check of calculations&FB personal communication\\
November 1963 & GA and FB graduate: calculations available  &FB personal communication\\
Dec 1963 & Use of small coupling quadrupole& \JH's thesis pag.80\\
Dec1963  & AdA first run for final publication&\citep{Bernardini:1964lqa}\\
Jan 1964 & AdA run and preparation of final AdA paper&\citep{Bernardini:1964lqa}\\
Feb 1964& AdA run and preparation of final AdA paper& \citep{Bernardini:1964lqa}\\
Apr 1964& AdA run and preparation of final AdA paper&\citep{Bernardini:1964lqa} \\
July 7th 1964&AdA 4th article is submitted to Nuovo Cimento&See Frascati internal note LNF-033 $^{(6)}$\\
December 1st 1964& Altarelli and Buccella calculation published& \citep{Altarelli:1964aa}\\
December 16th 1964& Final article is published in Nuovo Cimento&\citep{Bernardini:1964lqa} \\
\end{tabular}
\end{centering}

\vspace{0.5 cm}
\noindent 
\begin{scriptsize}
BT=Bruno Touschek, CB=Carlo Bernardini, GA=Guido Altarelli, FB=Franco Buccella\\
$^{(1)}$ In \url{http://www.lnf.infn.it/edu/materiale/video/AdA_in_Orsay.mp4}. \\
$^{(2)}$  \href{http://www.analysis-online.net/wp-content/uploads/2013/03/greco_pancheri.pdf}{Frascati e la fisica teorica: da AdA a DA$\Phi$NE}.\\
$^{(3)}$ 
See   Fig.~\ref{fig:graficorosa}.\\
$^{(4)}$ Bruno Touschek Archives (BTA) at Rome University Sapienza. \\
$^{(5)}$ Gerald O'Neill's \href{http://inspirehep.net/record/48292/files/C630610-p209.PDF}{Storage-Ring Work at Stanford} in 1963 Summer Study on Storage Rings, Accelerators and Experimentation at Super-High Energies 10 June - 19 July, 1963, Upton, NY.\\
$^{(6)}$LNF Internal notes are accessible at \url{http://www.lnf.infn.it/sis/preprint/search.php.}
\end{scriptsize}


   \bibliographystyle{abbrvnat}]
\bibliography{Touschek_book_Nov_6}
\end{document}